\documentclass[pra,notitlepage,twocolumn,superscriptaddress,floatfix,nofootinbib]{revtex4-2}

\usepackage{epsfig,color}
\usepackage{graphicx}
\usepackage{dcolumn}
\usepackage{bm}
\usepackage{amsmath, amsfonts, amssymb,mathrsfs}
\usepackage{pstricks}
\usepackage{amsxtra}
\usepackage{amsthm}
\usepackage{natbib}
\usepackage{qcircuit}
\usepackage{array}

\AtBeginDocument{\tabcolsep=7pt}
\usepackage{booktabs}

\newcommand{\ket}[1]{\mbox{$|#1\rangle$}}
\newcommand{\bra}[1]{\mbox{$\langle#1|$}}

\def\be{\begin{equation}}      
\def\ee{\end{equation}}
\def\beu{\begin{equation*}}   
\def\eeu{\end{equation*}}

\providecommand{\abs}[1]{\left\lvert#1\right\rvert}   
\DeclareMathOperator{\trace}{Tr}      

\providecommand{\mean}[1]{\langle#1\rangle}

\providecommand{\bk}{{\bm{k}}}

\newtheorem{proposition}{Proposition}[section]

\theoremstyle{definition}

 \definecolor{new}{rgb}{.08,.05,.8}

\newcommand{\delete}[1]{{}}

\begin{document}

\title{Quantum coding with low-depth random circuits}
\author{Michael J. Gullans}
	 \affiliation{Joint Center for Quantum Information and Computer Science, NIST/University of Maryland, College Park, Maryland 20742 USA}
	 \affiliation{Department of Physics, Princeton University, Princeton, New Jersey 08544, USA}
\author{Stefan Krastanov}
	\affiliation{John A. Paulson School of Engineering and Applied Sciences, Harvard University, Cambridge, Massachusetts 02138, USA}
    \affiliation{Department of Electrical Engineering and Computer Science, Massachusetts Institute of Technology, Cambridge, Massachusetts 02139, USA}
\author{David A. Huse}
    \affiliation{Department of Physics, Princeton University, Princeton, New Jersey 08544, USA}
\author{Liang Jiang}
    \affiliation{Pritzker School of Molecular Engineering, The University of Chicago, Illinois 60637, USA}
    \affiliation{AWS Center for Quantum Computing, Pasadena, California 91125, USA}
\author{Steven T. Flammia}
    \affiliation{AWS Center for Quantum Computing, Pasadena, California 91125, USA}

\begin{abstract}
Random quantum circuits have played a central role in establishing the computational advantages of near-term quantum computers over their conventional counterparts. 
Here, we use ensembles of low-depth random circuits with local connectivity in $D\ge 1$ spatial dimensions to generate quantum error-correcting codes. 
For random stabilizer codes and the erasure channel, we find strong evidence that a depth $O(\log N)$ random circuit is necessary and sufficient to converge (with high probability) to zero failure probability for any finite amount below the optimal erasure threshold, set by the channel capacity, for any $D$. 
Previous results on random circuits have only shown that $O(N^{1/D})$ depth suffices or that $O(\log^3 N)$ depth suffices for all-to-all connectivity ($D \to \infty$).
We then study the critical behavior of the erasure threshold in the so-called moderate deviation limit, where both the failure probability and the distance to the optimal threshold converge to zero with $N$. 
We find that the requisite depth scales like $O(\log N)$ only for dimensions $D \ge 2$, and that random circuits require $O(\sqrt{N})$ depth for $D=1$. 
Finally, we introduce an ``expurgation'' algorithm that uses quantum measurements to remove logical operators that cause the code to fail by turning them into either additional stabilizers or into gauge operators in a subsystem code. 
With such targeted measurements, we can achieve sub-logarithmic depth in $D\ge 2$ spatial dimensions below capacity without increasing the maximum weight of the check operators. 
We find that for any rate beneath the capacity, high-performing codes with thousands of logical qubits are achievable with depth 4--8 expurgated random circuits in $D=2$ dimensions.
These results indicate that finite-rate quantum codes are practically relevant for near-term devices and may significantly reduce the resource requirements to achieve fault tolerance for near-term applications.
\end{abstract}
\date{\today}
\maketitle


\section{Introduction}

Achieving reliable simulations of many-body quantum dynamics remains a central challenge across different areas of science.  
Quantum computers offer a natural computational advantage for such problems in near-term devices, as exemplified by recent experiments on random circuit sampling \cite{Arute19}.  
However, despite remarkable advances in quantum control and measurement \cite{Gaebler16,Debnath16,Corcoles15,Ofek16,Neill18,Arute19,Mi18,Huang19,Levine19,Egan20}, many platforms face daunting resource requirements when accounting for scalable quantum error correction \cite{Knill05,Bravyi05,Svore06,Fowler12}.  
On the other hand, it is now understood that one can significantly lower the resource requirements for fault-tolerance through e.g.~hardware efficient encodings \cite{Leghtas13,Mirrahimi14,Vlastakis13}, accurate noise estimation \cite{Martinis15,Harper20,Flammia2020}, noise-bias-preserving gates~\cite{Aliferis2008,Puri2020,Guillaud2019}, long-range interactions \cite{Tillich09,Hastings13,Gottesman13}, and better choices of codes with associated decoding algorithms  \cite{Tuckett18,Tuckett19,Tuckett20,BonillaAtaides20}. 
These developments suggest that fault-tolerance with much lower overhead is possible in near-term devices \cite{Gottesman13}.   

A common technique in classical error correction is to study random codes, which often nearly saturate the bounds for the optimal codes \cite{Shannon48,Gallager62,Gallager73}. 
Moreover, practical, near-optimal codes with efficient encoders and decoders are possible through random constructions of low-density parity check (LDPC) codes  \cite{Gallager62,Gallager73}.  In the quantum case, the decoding problem tends to be more difficult to solve (including for LDPC codes), but analogous random coding results have been obtained for stabilizer codes, where the decoding problem is similar to the classical case.  
Two-local random Clifford circuits with all-to-all connectivity have been shown to achieve an extensive code distance on $N$ qubits at a  depth upper bounded by $O(\log^3 N)$ \cite{Brown12,Brown13}.  
This scaling is comparable to provably optimal constructions for two-designs from Clifford gates at depth $O(\log N)$ with access to $O(N)$ additional ancillae \cite{Cleve15}.   
Spatial locality is often an important constraint in quantum computing architectures. 
In $D$ spatial dimensions, the expected  depth for local circuits to achieve an approximate two-design is upper bounded by $O( N^{1/D})$  \cite{Brandao16,Harrow18}.  
Such constructions are only required if the code needs to correct all errors up until threshold.   
Achieving optimal performance for local noise models requires fewer resources because the code only needs to correct typical errors in the thermodynamic limit.  

In this paper, we develop the general theory of optimal decoding with low-depth random encodings that include both unitaries and targeted measurements for one of the simplest error models given by the erasure channel.  
Many of the results apply for more general error channels, but optimal recovery probabilities are easy to compute for erasure errors, making it a useful error model for benchmarking quantum codes \cite{Delfosse16,Delfosse17}.   
We show that, in any spatial dimension, random Clifford encodings of finite-rate codes converge to zero failure probability below the optimal erasure threshold, set by the channel capacity, for depths $O(\log N)$; thus,  improving on the random circuit bounds described above.  
We then introduce an ``expurgation'' algorithm to surpass this logarithmic barrier and achieve convergence at a sub-logarithmic depth in $D>1$ dimensions.  
This method works by using quantum measurements to remove (expurgate) logical operators from the code that have a high-probability of failure until either a steady state code is reached or target coding parameters are obtained.  
These low-quality logicals are either turned into additional stabilizers or gauge operators to form a subsystem code.  
This expurgation process  monotonically increases the code distance and recovery probability of any  stabilizer subsystem code.   
At a practical level, one can use random coding and expurgation to generate high-performance, finite-rate codes for thousands of logical qubits with depth 4--8 circuits in two dimensions.

Our results also establish several connections between quantum error correction thresholds,  random matrix theory (RMT), and statistical physics.
Using an RMT ansatz, we develop a complete critical theory for optimal decoding of erasure errors for random stabilizer codes.  
We numerically benchmark this ansatz to a high degree of precision in the critical region of the erasure threshold. 
These scaling results guide our numerical analysis of optimal decoding for finite-depth encoders in finite-size systems. 
Focusing on the critical scaling theory of random codes at low depths, we find that random Clifford circuits can achieve the capacity of the erasure channel only at parametrically larger depth  $O( \sqrt{N})$ in 1D.  
In $D>1$ dimensions, however, random Clifford circuits retain the depth $\le O( \log N)$ scaling at capacity.  
The marginal dimension being 2D is consistent with Imry-Ma type arguments regarding the relevance of randomness in the error patterns at the optimal threshold \cite{Imry75}.   

We also analyze the  case of Haar random circuit encoders at high-depth $\ge O(N)$, where optimal decoding is likely exponentially hard.  
We find similar results  as for the high-depth Clifford encoders, but with small quantitative differences that indicate Haar random codes are slightly more optimal than random stabilizer codes.   
Through an approximate mapping to an Ising model, we argue that the erasure threshold with local random circuits can be generally understood as a type of first-order domain-wall pinning phase transition. 

\subsection{Relation to previous work} 

In this section, we discuss the relation of our results to some of the prior work on quantum error correcting codes and random quantum circuits. 

\subsubsection{Quantum error correcting codes}

Starting in the early days of quantum error correction, a common strategy for proving fault-tolerance was to study concatenated codes \cite{Shor96,Aharonov99,Gottesman00}. 
 These codes reduce decoherence by successively encoding quantum information in nested chains of small codes.  
 Unfortunately, this approach typically suffers from large space-time resource costs and low error thresholds \cite{Knill05,Svore06,Gottesman13}. 
A paradigmatic example of a code that, in balance, requires minimal resources is the 2D surface code \cite{Kitaev97,Dennis02}. 
This topological code saturates the  capacity for the erasure channel at zero code rate on a square lattice \cite{Dennis02}, is provably fault-tolerant under more general noise models \cite{Dennis02,Chubb18}, has highly efficient decoding algorithms \cite{Dennis02,Chubb18,Duclos10,Wootton12,Hutter14,Delfosse16,Delfosse17,Delfosse17b,Tuckett18,Tuckett19,Tuckett20,BonillaAtaides20}, and a large variety of fault-tolerant strategies for implementing gates  \cite{Bombin06,Bombin07,Bombin11,Horsman12,Bravyi05,Fowler12,Brown17,Webster19,Brown20}.  
However, despite its remarkable properties, the surface code requires a prohibitively large overhead in the number of physical qubits for applications on near term devices \cite{Fowler12}. 
With issues of this nature in mind, it remains a central goal to develop more space-efficient, ideally finite-rate, codes  that achieve similar levels of performance to the surface code \cite{Gottesman13}.  
 
Extending topological codes, or, more generally, low-density parity check (LDPC) codes,  to finite rate faces various theoretical obstructions in spatially local models \cite{Bravyi10b}.  
Two routes to overcome this are to use subsystem codes \cite{Bacon06} or remove the constraint of geometric locality, while keeping the LDPC condition.  
In all-to-all coupled systems, a large variety of finite-rate LDPC codes have been developed by extending the surface code to nonlocal geometries or adapting classical codes based on expander graphs \cite{Tillich09,Hastings13,Leverrier15,Bolt18}.  
Furthermore, several threshold theorems have been proved for a large family of these codes \cite{Gottesman13,Kovalev13,Fawzi18}.  
Maintaining all-to-all connectivity in the thermodynamic limit eventually runs into prohibitive resource constraints, but these codes are applicable to near-term ion trap quantum computers \cite{Monroe14,Landsman19} and quantum networks \cite{Kimble08}.   
Another interesting class of finite-rate codes that retain some locality structure, but are not of the LDPC type, are provided by holographic codes that originally arose in the study of quantum gravity and the AdS/CFT correspondence \cite{Pastawski15,Harris18}.  
The quasilocal codes considered here differ from these various classes of codes because their properties emerge from generic, local scrambling dynamics instead of concatenation, topology, expander graphs, or hyperbolic geometry.

\subsubsection{Random quantum circuits}

The theoretical methods in this work  draw from a variety of recent results in random quantum circuits, which have served as a powerful tool to examine quantum many-body dynamics in nonperturbative limits \cite{Hayden07,Lashkari11,Hosur16,Nahum16,Nahum17,vonKeyserlingk17,Nahum17b,Khemani17,Tibor17,Chan18}.  
Notable examples are their extensive applications to quantum gravity \cite{Hayden07,Lashkari11,Hosur16} and quantum chaos and equilibration \cite{Nahum16,Nahum17,Zhou19,vonKeyserlingk17,Nahum17b,Khemani17,Tibor17,Chan18}.  
Despite the intricate structure of a particular circuit, which forms the basis for average-case hardness of random circuit sampling \cite{Brandao13,Boixo18,Neill18,Arute19,Bouland18,Movassagh19} there is a notion of universality in the dynamics of extensive quantities such as the entanglement \cite{Nahum17} or the distance of the code generated by the circuit \cite{Brown13}.  
Such notions of universality build on the random matrix theory/eigenstate thermalization approach to describing late-time equilibration in closed quantum systems \cite{Deutsch91,Srednicki93,Nandkishore15,DAlessio16}.

More specifically, our results have direct relevance to a recently discovered phase transition that arises in monitored random circuits, where unitary gates are interspersed with random projective measurements \cite{Li18,Skinner18}.  
These models have attracted interest in condensed matter theory due to the potential connections to chaos, thermalization, conformal field theory, and the many-body localization phase transition \cite{Li19,Chan18b,Gullans19c,Choi20,Cao19,Szyniszewski19,Bao20,Jian19,Gullans19d,Zabalo20,Zhang20,Tang20,Li20,Fuji20,Lavasani20,Sang20,Ippoliti20,Alberton20,Lunt20,Rossini20,Goto20,LopezPiqueres20,Shtanko20,Vijay20,Lang20,Nahum20}.  
In the context of quantum information, their study has led to novel insights into emergent quantum error correction \cite{Gullans19c,Choi20,Fan20,Li20b,Fidkowski20}, as well as the sampling complexity of constant depth circuits in 2D \cite{Napp19}.  
Due to the repeated rounds of measurements acting on a code space density matrix, the dynamics during our expurgation algorithm display a similar phenomenology to the ``purification'' dynamics of a mixed state in the unitary-measurement models \cite{Gullans19c,Ippoliti20,Li20,Fidkowski20}; however, there are several important differences in the present case due to the nonrandom, targeted choice of measurements.   
Furthermore, since we show that logarithmic depth random circuits are sufficient to reliably encode quantum information, our results may provide  guidance for rigorous  existence proofs of the volume-law phase in some models.  
They may also help guide efforts in  developing fault-tolerant strategies for monitored random circuits that incorporate feedback.

\subsection{Structure of paper}

The paper is organized as follows: In Sec.~\ref{sec:approach}, we outline our theoretical approach for studying random quantum codes.  
We then summarize our main results and theoretical methods.  
In Sec.~\ref{sec:background}, we provide some  background on the basic concepts and terminology used to describe quantum error correction thresholds.   
In Sec.~\ref{sec:rmt}, we present the RMT solution to the erasure threshold for random stabilizer codes.  
In Sec.~\ref{sec:quasilocal}, we present our  results on the behavior of low-depth random circuit encoders for the erasure channel.  
In Sec.~\ref{sec:exp}, we present our expurgation algorithm to surpass the depth $O(\log N)$ barrier in $D>1$ dimensions.   
In Sec.~\ref{sec:Haar}, we present an analysis of the erasure threshold for general Haar random codes.  
In Sec.~\ref{sec:statmech}, we describe an approximate mapping of the erasure threshold to a first-order domain wall pinning transition that occurs in the ordered phase of the Ising model.  
We provide further discussions and present our conclusions in Sec.~\ref{sec:discussions}.  

We remark that the arguments in the paper use a combination of rigorous proofs, large-scale numerics,   conjectures, and some occasional heuristics.   
To test our ideas as strongly as possible with this approach, we analyze the problem from multiple perspectives and systematically compare our results across different spatial dimensions.  
What emerges from this analysis is a consistent framework to describe quantum coding with local random circuits.

\section{Summary}
\label{sec:approach}
  
 In this section, we outline the theoretical approach taken in this work and summarize our main results.
  
  \subsection{Theoretical approach}
  
  \begin{figure}[tb]
\begin{center}
\includegraphics[width = .48 \textwidth]{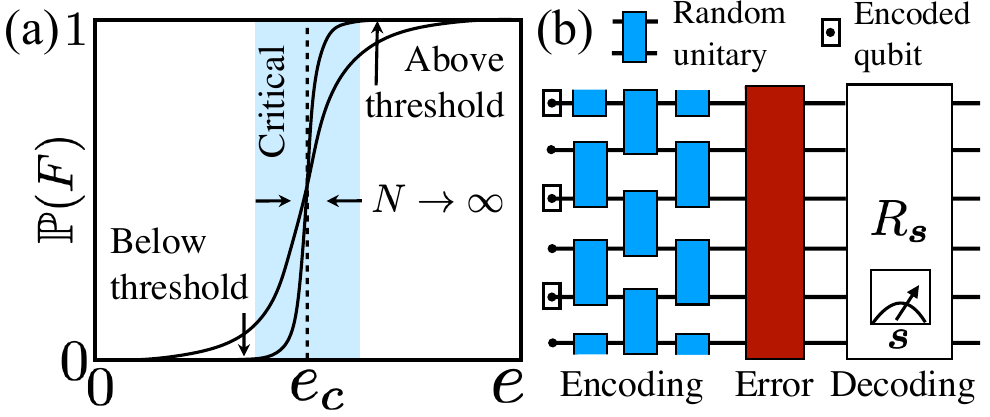}
\caption{(a) Probability of decoding failure $\mathbb{P}(F)$ vs. error rate $e$ through an error correction threshold for an optimal code. 
In this work, we probe the optimality of a given code ensemble by comparing the location and universality class of the critical point to random stabilizer codes.  
(b) Illustration of the models we study: the encoding circuit is a low-depth random unitary circuit and the error is an erasure of a fixed fraction $eN$ of $N$ qubits.  
The decoding proceeds via generalized measurements with outcomes $\bm{s}$ and recovery operators $R_{\bm{s}}$.  
We mostly focus on stabilizer codes, where optimal decoding of Pauli error channels like erasure errors is possible with stabilizer syndrome measurements followed by the conditional application of single-site Clifford gates.}
\label{fig:threshold}
\end{center}
\end{figure}

This paper is focused on developing a theory of optimal decoding for finite-rate codes generated by  random circuits.  
To approach this problem, we directly investigate the probability of successful recovery $\mathbb{P}(R)$ of the encoding and decoding scheme for the specific error model of erasures. 
This type of observable is complementary to other performance metrics that are agnostic to the error model, for example, the code distance. 
One advantage of studying recovery/failure probabilities is that it allows us to obtain a more detailed understanding of the code performance near the optimal threshold. 
For coding below the optimal threshold,  we have found that focusing on this observable often suggests methods to tailor the codes to the detailed properties of the noise, as we explore with our expurgation algorithm in Sec.~\ref{sec:exp}.

The  qualitative behavior of the optimal (minimal) failure probability $\mathbb{P}(F) = 1 - \mathbb{P}(R)$ is shown in Fig.~\ref{fig:threshold}(a).  
Here, $e$ is a parameter that characterizes the strength of noise in a given error model (or class of error models) and we assume that the implemented code is optimal for this error model.  
Below threshold, the failure probability converges to zero in the limit of large $N$.  
Past a critical error rate $e_c$ (set by the channel capacity limit for the optimal code), for $e>e_c$ the failure probability instead converges to one in the large-$N$ limit.   
This discontinuous behavior in the large-$N$ limit is characteristic of a phase transition.  
Motivated by results from classical error correction \cite{Gallager62,Gallager73}, we assume the  failure probability for an optimal code for large $N$ is well approximated by the average behavior of a high-depth random stabilizer code under optimal decoding \cite{Hayden07b}.   
One primary  question that we address is what minimal depth of a random encoding Clifford circuit is needed for large but finite $N$ to achieve near-optimal failure probability  for the specific case of erasure errors [see Fig.~\ref{fig:threshold}(b)].

More specifically, in finite-size systems, the failure probability for the optimal code will generically be a function of both the error rate $e$ and the number of (qu)bits $N$ per code block.  However, in the thermodynamic limit of large $N$, the failure probability will approach a scaling form in the vicinity of the critical error rate $e_c$ [see shaded region in Fig.~\ref{fig:threshold}(a)]
\be
- \log \mathbb{P}(F) = N^a f_{\rm opt}[ (e - e_c) N^{1/b} ],
\ee
where $a$ and $b$ are critical exponents and the corrections are assumed to be subleading in powers of $1/N$.  
We take the logarithm of the failure probability as it generally scales like a free energy, e.g. in the surface code \cite{Chubb18}.    
In coding theory, properties of the scaling functions for the optimal codes $f_{\rm opt}$ have been extensively studied  under optimal decoding of Markovian error channels (e.g., see Ref.~\cite{Chubb17}).   
The finite-size scaling behavior is important because it determines the rate of convergence to the ideal behavior  below threshold.
The underlying idea of this work is to use the scaling properties of the optimal codes for a given error channel as an ideal performance benchmark.  
We  effectively define a code as optimal if it achieves capacity at threshold and its threshold behavior lies in the same universality class as the  truly optimal codes for this error channel.

Of course, finding explicit and efficiently implementable representations for encoding and decoding maps of optimal codes is generally a difficult problem \cite{Tomamichel16}.  
To approximate this paradigm in a setting that allows for more theoretical progress and potential practical implications for quantum computing, we relax the benchmark critical behavior from that of optimal codes to the average behavior of high-depth \emph{random stabilizer} codes.  
As mentioned above, random codes typically achieve similar levels of performance as optimal codes.  
In quantum error correction,  even random stabilizer codes are often sufficient.
We present numerical evidence on small systems that the Haar random code transition is in the same universality class as the random stabilizer code transition. 
 However, we also see small quantitative differences between the scaling functions for the two cases, with slightly more optimal performance for the Haar codes.  
 Random stabilizer codes are, thus, better classified as ``near-optimal'' codes for the erasure channel, which is similar to a well-known result for the depolarizing channel   \cite{DiVincenzo98,Smith07}.

\subsection{Main results}
\label{sec:results}

As discussed in the introduction, our main results center around the resource requirements (in terms of encoding circuit depth) to achieve zero failure probability or approach capacity for finite-rate codes generated by random circuits.
In particular, we study stabilizer codes generated by two-local random Clifford circuits  on hypercubic lattices in $D$ spatial dimensions or on all-to-all coupled networks.  
The basic setup is illustrated in Fig.~\ref{fig:threshold}(b).  
In this example, every other qubit is mapped to an encoded or ``logical'' qubit at a code rate of $R = 1/2$ and the random circuit is implemented in 1D.  
We provide a summary of the scalings found in this work in Table~\ref{tab:depth}.

The error model is taken to be an erasure model where $eN$ sites of an $N$-qubit system are randomly selected and traced out of the system, with those sites heralded to the decoder but unknown to the encoder.  
The failure probability for the more physically relevant case of independent, identically distributed (iid) erasures at each site with probability $e$ can be determined from the failure probability for the fixed-fraction erasure model, which is why we mostly focus on the latter.\footnote{The failure probability $\mathbb{P}_{\rm iid}(F)$ for iid random erasures can be found from the fixed-fraction failure probability $\mathbb{P}(F)$ and the probability $p_{\rm iid}(n_e)$ of erasure number $n_e$ in the iid model.  Since $n_e$ is known to the decoder, $\mathbb{P}(F) = \mathbb{P}_{\rm iid}(F|n_e)$ is effectively a conditional distribution, i.e., $\mathbb{P}_{\rm iid}(F) = \sum_{n_e} \mathbb{P}(F) p_{\rm iid}(n_e)$.} 
For the random stabilizer codes, we show that the transition is in a certain sense first order, since for $e<e_c$ the logarithm of the failure probability is proportional to $-(e_c-e)N$ in the limit of large $N$.  
If we interpret this as a free energy, it is extensive and its density has a discontinuity in the first derivative with respect to $e$ at $e_c$, as is the case for first-order phase transitions.  
This first-order transition is rounded out for finite $N$.  
This finite-size rounding is minimal if we take an error model with erasures on a fixed fraction  $eN$ of sites.    
The finite-size rounding of the transition in the iid model is much stronger (by a factor of $\sqrt{N}$).

\begin{table}[tb]
    \centering
    \begin{tabular}{cccc}
    \toprule
    $D$ &$e<e_c$ & $e= e_c - O(\tfrac{1}{N^b})$ & Expur.\ $e<e_c$  \\
    \midrule
    1  & $(\tfrac{1}{e_c - e})\log N$ & $N^{1/2}$ &$\log N$ \\
    2 & $\log N$ & $\log N$ (conj.) & $(\log N)^c$,~$c < 1$ \\
    $>2$ & $\log N$ & $\log N$ & $(\log N)^c$,~$c < 1$ \\
    \bottomrule
    \end{tabular}
    \caption{Random Clifford circuit encoding depths required to reach zero failure probability for finite-rate codes under  erasure errors in $D$ dimensions.  
    Here, $e_c - O(1/N^b)$ denotes coding arbitrarily close to the critical region of the optimal erasure threshold in the thermodynamic limit.  
    We find $b = 1$ for the fixed-fraction erasure model and $b=1/2$ for the iid model. $D=2$ is the marginal dimension for the relevance of spatial randomness in the errors to the threshold behavior, which makes the scaling at the optimal threshold difficult to reliably determine from numerics or Imry-Ma arguments.  
    The last column shows the results upon expurgation of bad logical operators using quantum measurements (see Sec.~\ref{sec:exp}).}
    \label{tab:depth}
\end{table}

We find that the best known analytical bounds for the convergence rate to a two-design strongly overestimate the circuit depth $d$ required for convergence of the failure probability towards the high-depth $[d = O(N)]$ limit.  
Most notably, in any $D\ge 2$, at the critical point the depth required scales as $d \le O( \log N) $, which is comparable to the optimal depth for generating a two-design in systems without spatial locality constraints $O( \log N )$  \cite{Cleve15}.  
Even in $D=1$, we find that removing the randomness in erasure locations by taking regularly spaced erasures leads to a required depth to approach zero failure probability below the optimal threshold of $O(\log N)$.  
Spatial randomness in the erasure locations seems to only be a relevant perturbation to the finite-size scaling behavior in $D=1$, and not for $D\ge 2$.

To simplify the analysis, we will  fix the initial code rate at precisely $R = 1/2$ in most of our discussion and drop this argument from the scaling functions.  
Also fixing the initial spatial arrangement of the logical qubits to be every other site in the lattice, the failure probability has a four-parameter dependence
\be
-\log \mathbb{P}(F) = F(e,D,d,N).
\ee
We first consider the high-depth limit $d = O(N)$ of the failure probability, which does not depend on $D$.  
Through a RMT ansatz, we obtain an asymptotic form for the failure probability in the fixed fraction model that depends only on the total number of erasures $eN$ (an integer) relative to the  threshold number $e_c N$ (which does not have to be an integer): $\lim_{N\to\infty} \lim_{d \to \infty}F(e,d,D,N) = f_{\infty}[(e-e_c)N]$ for $e$ near $e_c$.  
Here, $e_c = (1-R)/2 $ coincides with the channel capacity limit for the erasure threshold \cite{Bennett97}.  For this fixed-fraction erasure error model, the scaling function $f_\infty(x_n)$ is only well-defined on a countably infinite set of values in the thermodynamic limit.    
The RMT solution  predicts a value for the critical failure probability $\mathbb{P}(F) = 0.38968\ldots $ that is independent of $R$ for $0<R<1$; for $R=1/2$ we verify this value numerically to a  precision of $10^{-4}$.   

To understand the scaling with depth we first consider a simple mean-field model for the below threshold behavior in which we break the system up into individual blocks of size $O(\log N)$.  
A simple analysis of this model based on the results of Ref.~\cite{Brandao16} shows that the convergence to the high-depth behavior of the failure probability in $D$ dimensions is  typically $O(\log N)$ for random Clifford encodings, but can be made as low as depth $ O[(\log N)^{1/D}]$ through the optimized encodings of Ref.~\cite{Cleve15}.  
In the latter case, there is a reduction in the effective rate of the code due to the use of $O(\log N)$ ancilla qubits per block in the encoding scheme.
  
Using an Imry-Ma type argument \cite{Imry75}, we then argue that the positional randomness of the erasures is irrelevant for the finite-size scaling in $D \ge 2$.    
As a result, we conjecture that the critical points for all $D\ge 2$ have the same leading order scaling with depth as the below threshold behavior predicted from the mean-field model
\begin{align*}
- \lim_{N\to\infty} &\log \mathbb{P}(F)|_{e = e_c,D \ge 2} = f_{Dc}( d - A \log N).
\end{align*}
Using this ansatz, we find a consistent scaling collapse in our numerics.

We also study the scaling behavior with depth $d$ in the 1D case $(D=1)$, which has to be treated separately.  
By studying the convergence of the critical failure probability $\mathbb{P}(F)|_{e=e_c}$ to the RMT prediction, we find numerical evidence for a leading order scaling behavior of the form
\be \label{eqn:1dscaling}
- \lim_{N\to\infty}\log \mathbb{P}(F)|_{e=e_c,D=1} = f_{1c}(d/\sqrt{N}).
\ee
   In contrast, below the critical error rate, we find that the failure probability for $d > O(\log N)$ exhibits exponential decay with the depth $ \mathbb{P}(F)|_{e < e_c,D=1} \sim e^{- d/A(e)}$ for some function $A(e)$ that diverges as $(e_c-e)^{-1}$ upon approaching $e_c$.  
This behavior leads to an overall $O(\log N)$ depth for convergence to zero failure probability below threshold, but with a rate that goes to zero at the optimal erasure threshold. 
We argue that the $\sqrt{N}$ scaling at $e_c$ has an intuitive explanation as arising from the Poisson fluctuations in the number of excess erasures in a given extensive region.

After establishing these scaling results, we introduce our expurgation algorithm based on measuring logical operators in the system that are likely to lead to failures.    
We prove that the code distance and recovery probability for Pauli error channels will monotonically increase with this expurgation strategy.  
We then numerically study the performance of the algorithm in 2D and all-to-all coupled systems. 
 In both cases, we see strong evidence that a given target failure probability can be achieved with a sub-log-$N$ depth circuit.

Finally, to test the generality of these results obtained for stabilizer codes, we study the erasure threshold for Haar random circuits.  
We first study the high-depth limit using small scale numerics.  
We find consistent critical behavior with the random stabilizer code threshold, but small quantitative differences in the scaling functions (as noted above).  
Using well-studied mappings of  two-local Haar random circuits to statistical mechanics models \cite{Harrow09,Lashkari11,Nahum17}, we describe an approximate mapping of the erasure threshold  to a first-order pinning transition for domain walls that occurs in the ordered phase of $D+1$-dimensional Ising models.  
Such transitions display similar phenomenology to our numerically observed results for random Clifford circuits.   

\section{Preliminaries}
\label{sec:background}

In this section, we introduce the basic terminology and concepts underlying  quantum error correcting codes, optimal decoding, and quantum error correction thresholds.  
We derive a formula used throughout the paper for the recovery probability of stabilizer subsystem codes under erasure errors.

\subsection{Optimal decoding}
\label{sec:optdec}

The general setup we consider follows the illustration in Fig.~\ref{fig:threshold}(b).
 Information is first mapped into a nonlocal code space, it is then subjected to local errors and decoded.  
 In the theory of fault-tolerance, one needs to consider errors in both the encoding and decoding steps; however, we will not address such issues here and assume both the encoding and decoding operations are implemented perfectly.  

In the quantum case, these three operations are typically described using the language of quantum channels, which are linear maps that are completely-positive and trace preserving \cite{NielsenChuang}.  
Denoting the encoding, error, and decoding channel by $\mathcal{E}$, $\mathcal{N}$ and $\mathcal{D}$, respectively, the central object of interest is the composite channel
\be
\mathcal{D} \circ \mathcal{N} \circ \mathcal{E}.
\ee
Error correction can be done perfectly when this composite channel acts as the identity on the allowed input states  $\mathcal{D} \circ \mathcal{N} \circ \mathcal{E}(\rho) = \rho$ or is unitarily equivalent to the identity $\mathcal{D} \circ \mathcal{N} \circ \mathcal{E}(\rho) = U \rho U^\dag$ for a known unitary $U$.  

When this map is not exactly unitarily equivalent to the identity, then it is convenient to use a fidelity metric to quantify its proximity to the identity. 
 One natural fidelity metric that we study in this work is the max-average state fidelity \cite{NielsenChuang}
\be
F_{\rm avg} = \max_{\mathcal{D}} \int d \psi \bra{\psi} \big[ \mathcal{D} \circ \mathcal{N} \circ \mathcal{E}(\ket{\psi} \bra{\psi}) \big] \ket{\psi},
\ee
where $d \psi$ is taken as a uniform measure over pure input states for $\mathcal{E}$ and the maximum is taken over all possible decoding maps $\mathcal{D}$.  
This fidelity metric quantifies the degree to which a randomly drawn code word can be recovered back to its initial state under optimal decoding.

A closely related fidelity metric to this average state fidelity is the entanglement fidelity, which measures the degree to which the map preserves entanglement with a reference system \cite{Schumacher96b}.  
Given an initial  density matrix $\rho_S$ on the system $S$, we purify it to the state $\rho_{SR} = \ket{\Psi_{SR}}\bra{\Psi_{SR}}$ by introducing entanglement with a reference system  $R$ such that  $\rho_S = \trace_R[ \rho_{SR}]$.  
Then, the entanglement fidelity under optimal decoding is
\be
F_{e}(\rho_S) = \max_{\mathcal{D}} \bra{\Psi_{SR}} \big[ \mathcal{D} \circ \mathcal{N} \circ \mathcal{E}(\ket{\Psi_{SR}} \bra{\Psi_{SR}}) \big] \ket{\Psi_{SR}},
\ee
where the maps act as the identity on the reference system and $F_e(\rho_S)$ is independent of the choice of purification.
Conveniently, the max-average fidelity is equivalent to the max-entanglement fidelity of the completely mixed state through the formula $F_{\rm avg} =  [q\, F_{\rm ent}(\mathbb{I}/q)+1]/(q+1)$, where $q$ is the dimension of the input space \cite{Horodecki99,Nielsen02}.  

In cases where the optimization over decoders is difficult to compute, we can still gain insight into the quantum error correction threshold by studying the coherent quantum information \cite{Schumacher96,Schumacher01}
\be
I_c(\rho_S,\mathcal{N}\circ \mathcal{E}) = S(\rho_{S'}) - S(\rho_{RS'}),
\ee
where $S(\rho) = - \trace[\rho \log_2 \rho]$ is the von Neumann entropy, $\rho_{S'} = \mathcal{N}\circ \mathcal{E}(\rho_S)$, and $\rho_{RS'}=\mathcal{N}\circ \mathcal{E}(\rho_{SR})$.
The coherent quantum information is closely related to the entanglement fidelity because when $|I_c(\rho, \mathcal{N}\circ \mathcal{E}) - S(\rho)| < \epsilon$, then it implies that $F_e(\rho) \ge 1 - 2 \sqrt{\epsilon}$ \cite{Schumacher01}.   
In our analysis of random stabilizer codes, we directly compute $F_{\rm avg}$, while for Haar random codes we use the coherent quantum information to bound  $F_e(\mathbb{I}/q)$.

The coherent quantum information is fundamentally related to the quantum channel capacity through  the limiting formula \cite{Lloyd97,Devetak05}
\be 
Q(\mathcal{N}) = \lim_{N \to \infty} \frac{1}{N} \max_{\rho} I_c(\rho,\mathcal{N}^{\otimes N}).
\ee 
In this work, we study erasure errors, which for a single qubit is defined by the channel
\be \label{eqn:eraseN}
\mathcal{N}(\rho) = (1-e) \rho \otimes \ket{0}\bra{0} + e/2\, \mathbb{I} \otimes \ket{e}\bra{e}.
\ee
  The states $\ket{0}/\ket{e}$ are orthogonal states that herald the absence/occurrence of the erasure on this site.  
  Note that, in many physically relevant  scenarios, the state of the system itself is mapped to an orthogonal state under an erasure error, which is an equivalent description of this channel for our purposes.  
  We choose the representation in Eq.~(\ref{eqn:eraseN}) to simplify the notation in later discussions.   
  
  The heralded nature of the erasure locations dramatically simplifies the decoding problem, as we discuss below  for stabilizer codes.  
  Furthermore, due to this classical register, the capacity of the erasure channel is additive [i.e., $Q(\mathcal{N}) = \max_\rho I_c(\rho,\mathcal{N})$] and is derivable from the no-cloning theorem \cite{Bennett97}.  
  It is also easy to compute the channel capacity from the maximization of the coherent quantum information $ Q(\mathcal{N})= (1- 2 e)$, where $e$ is the local erasure probability on each site.  
  Equivalently, for a code rate $R = k/N = Q$, the optimal erasure threshold in the thermodynamic limit is $e_c = (1-R)/2$. 

\subsection{Optimal decoding: stabilizer codes}
\label{sec:optdecstab}

These concepts of optimal decoding are illustrated more concretely by considering the example of qubit stabilizer codes.  
An $[N,k]$ qubit stabilizer code encodes $k$ logical qubits in $N$ physical qubits.  
The codewords are spanned by the set of $2^k$ stabilizer states that are the simultaneous eigenstates of a stabilizer group $S \subset \mathcal{P}_N$, which is an abelian subgroup of the Pauli group on $N$ qubits $\mathcal{P}_N$ such that $-\mathbb{I} \notin S$.  
 Given a generating set $\{ \bar{Z}_1, \ldots, \bar{Z}_{N-k} \}$ for $S$, optimal decoding is possible through projective measurements of these generators (called syndrome measurements) for Pauli error channels.  
 These are channels that have a Kraus representation of the form 
\be
\mathcal{N}(\rho) = \sum_{E,\bk} p(E,\bk) E \rho E^\dag \otimes \ket{\bk}\bra{\bk},
\ee
where $E$ is an element of $\mathcal{P}_N$, $\ket{\bk}$ are orthogonal quantum states that are used to store classical data (e.g., the erasure locations), and $p(E,\bk) \ge 0$ is a joint probability distribution over the allowed error operators $E$ and register indices $\bk$.    
Such quantum-classical channels are sometimes called a ``quantum instrument'' \cite{Wildebook}.  
  Erasure errors can be represented in this form because of the following identity for the partial trace operation on site $n$
\be \label{eqn:erase}
\trace_n[\rho] \otimes \mathbb{I}_n/2 = \frac{1}{4} \big( \rho + X_n \rho X_n +Y_n \rho Y_n + Z_n \rho Z_n \big),
\ee
where $\mathbb{I}_n,X_n,Y_n,$ and $Z_n$ are the four Pauli operators.  

The Pauli group operation can be represented by standard matrix multiplication of the $N$-qubit Pauli operators.  
For two Pauli group elements $P_{1,2}$, we use the notation $[[ P_1,P_2 ]] = \trace[ P_1P_2 P_1^{-1} P_2^{-1}]/2^N$ to denote their scalar commutator: if $P_1$ and $P_2$ commute, then $[[P_1,P_2]]=1$, otherwise, $[[P_1,P_2 ]] = -1$.  
We can extend the generating set for $S$ to a complete generating set for $\mathcal{P}_N$ by appending destabilizer operators $\{\bar{X}_1,\ldots,\bar{X}_{N-k}\}$ that satisfy $[\bar{X}_i,\bar{X}_j] = \delta_{ij} \mathbb{I}$ and $[[ \bar{Z}_i,\bar{X}_j ]] =  (-1)^{\delta_{ij}}$, a generating set for the logical operators $L_i$ (these are  Pauli group elements that commute with $S$, but are not contained in $S$ \cite{NielsenChuang}), and the Pauli group element $i \, \mathbb{I}$.
Since each $E$ is an element of the Pauli group, we can  decompose them based on the outcomes they produce in the syndrome measurements $E_{\bm{s} \bk} =  g_{\bm{s}\bk} L_{E_{\bm{s}\bk}}$, where $\bm{s} =(s_1,\ldots,s_{N-k})$ is a vector of syndrome bits  $(s_i = 0/1)$, $g_{\bm{s}\bk}$ is Pauli group element satisfying $[[ g_{\bm{s}\bk},\bar{Z}_i ]] =  (-1)^{s_i}   $,  and $L_{E_{\bm{s}\bk}}$ is a logical operator.  
The $g_{\bm{s} \bk}$ is non-unique and is allowed to be linearly dependent on elements of $S$, the destabilizers, logical operators, and $i \mathbb{I}$.  

After applying the error  channel $\mathcal{N}$ and performing a projective measurement of the syndrome bits $\bm{s}$ and register state $\bk$, the state is mapped to
\be
M_{\bm{s}\bk} \circ \mathcal{N}\circ \mathcal{E}(\rho) =  \sum_{E_{\bm{s}\bk}} p(E_{\bm{s}\bk},\bk) g_{\bm{s}\bk} L_{E_{\bm{s}\bk}} \rho L_{E_{\bm{s}\bk}}^\dag g_{\bm{s}\bk}^\dag,
\ee
where we traced over the classical register for notational convenience.
Applying the correction operator $g_{\bm{s}\bk}^\dag$, which is a product of single-site Clifford gates, the state becomes a mixture of states in the code space
\be
g_{\bm{s}\bk}^\dag [ M_{\bm{s}\bk} \circ \mathcal{N}\circ \mathcal{E}(\rho) ]g_{\bm{s}\bk} = \sum_{E_{\bm{s}\bk} } p(E_{\bm{s}\bk},\bk) L_{E_{\bm{s}\bk}} \rho L_{E_{\bm{s}\bk}}^\dag .
 \ee
 Below threshold  in the thermodynamic limit, all the $L_{E_{\bm{s}\bk}}$ must converge in probability to the same logical operator $ L_{\bm{s} \bk}$ up to multiplication by elements of the stabilizer group $S$, i.e., $L_{E_{\bm{s}\bk}} =  g_{ E_{\bm{s}\bk}} L_{\bm{s}\bk}$ for some $g_{ E_{\bm{s}\bk}}$ in $S$.  
 Since the operators $g_{ E_{\bm{s}\bk} }$ act trivially in the code space, the initial state can then be perfectly recovered by applying the additional unitary correction operator $L_{\bm{s}\bk}^{\dag}$.  
 
 In finite-size systems, where perfect decoding is not generally possible,  an optimal decoding strategy is any maximum-likelihood decoder based on the observed  $\bm{s}$ and $\bk$ \cite{Chubb18}.  
 In this approach, we further break up the set of all $E_{\bm{s}\bk}$ into logical operator classes $E_{\bm{s}\bk}^i = g_{\bm{s}\bk}g_{ E_{\bm{s}\bk}^i }L_{\bm{s}\bk}^{i}$, such that the $g_{ E_{\bm{s}\bk}^i }$ are in $S$ and the $L_{\bm{s}\bk}^i$ cannot be related (modulo a phase) through multiplication by elements of $S$.  
 Conditioned on $\bm{s}$ and $\bk$, the decoder applies a unitary correction operator $R_{\bm{s}\bk}^\dag =g_{\bm{s}\bk} L_{\bm{s}\bk}^{i_m}$ to the state $R_{\bm{s}\bk} [M_{\bm{s}\bk} \circ \mathcal{N}\circ \mathcal{E}(\rho)] R_{\bm{s}\bk}^\dag$ with $L_{\bm{s}\bk}^{i_{m}}$ corresponding to the most \emph{likely} logical error equivalence class.  
 This operator can be computed by finding the value of $i$ that maximizes the probability 
 \be \label{eqn:psi}
 Z_{i}(\bm{s},\bk) = \sum_{E_{\bm{s}\bk}^i} p(E_{\bm{s}\bk}^i,\bk),
 \ee
 i.e., $Z_{\max}(\bm{s},\bk) = \max_i Z_{i}(\bm{s},\bk)$.  
 The probability of a perfect recovery for all input states under optimal decoding is then given by
 \be \label{eqn:pperf}
\mathbb{P}(R) = \sum_{\bm{s},\bk} Z_{\max}(\bm{s},\bk).
 \ee
 
For general codes and Pauli error channels, finding $L_{\bm{s}\bk}^{i_{m}}$ is likely to be exponentially hard, but, in the case of erasure errors, an efficient optimal decoding strategy has been derived by Delfosse and Zemor \cite{Delfosse17}.  
Briefly reviewing their argument: one can use the fact that erasure error locations are heralded, which implies that the  decoder only needs to find the most likely error operator that acts on the erased sites.  
Once the sites are known, according to Eq.~(\ref{eqn:erase}) all error operators are equally likely, which implies that the probabilities in Eq.~(\ref{eqn:psi}) are all equal for a fixed $\bk$; therefore, a maximum-likelihood strategy is to simply choose $R_{\bm{s}\bk}$ to be any Pauli operator that lives on the erased sites and produces the observed syndrome measurement.  
Using the standard representation for stabilizer states from the Gottesman-Knill theorem \cite{Gottesman98,Aaronson04}, such a Pauli operator can be found given the syndrome check operators, erasure locations, and syndrome measurement outcomes using Gaussian elimination in a time at most $O(N^3)$.

Here, we derive an explicit formula for the recovery probability under erasure errors that is convenient for our purposes.  
We make use of a generating matrix for the error operators that act in the erased region $\bm{e}$
\be \label{eqn:ms}
M(S,L,\bm{e}) = \left( \begin{array}{c | c}
\bm{s}_{z_{i_1}} & \bm{\ell}_{z_{i_1}}  \\
\bm{s}_{x_{i_1}} & \bm{\ell}_{x_{i_1}}  \\
 \vdots & \vdots \\
 \bm{s}_{z_{i_{n_e}}} & \bm{\ell}_{z_{i_{n_e}}}  \\
 \bm{s}_{x_{i_{n_e}}} & \bm{\ell}_{x_{i_{n_e}}}  
 \end{array}
 \right).
 \ee
The first $n_s = N-k$ columns $\bm{s}_E$ are vectors of syndrome bits for a local basis of error operators defined by the relation $(-1)^{s_{E,i}} = [[\bar{Z}_i, E ]]$. 
The last $2k$ columns  $\bm{\ell}_E$ similarly encode the scalar commutator of the local errors with a generating set for the logical operators.  
Crucially, stabilizer  codes are additive codes, which implies that if $E = E_1 + E_2$, then $\bm{s}_{E} = \bm{s}_{E_1}+\bm{s}_{E_2}$ and $\bm{\ell}_{E} = \bm{\ell}_{E_1}+\bm{\ell}_{E_2}$.  
As a result, the row vectors $(\bm{s}_{\mu_i} | \bm{\ell}_{\mu_i})$ act as a generating set for all possible syndromes and their associated logical errors in the erased region.
 
 We now show how to compute the recovery probability from the matrix $M$.  
 Performing row reduction on $M$ identifies all errors that map to the all zero syndrome, but have a nontrivial logical operator content.  
 Errors of this type can be used to enumerate all uncorrectable errors for that set of erasure locations.  
 For each matrix $M(S,L,\bm{e})$, we define 
 \be
 r_M(S,L,\bm{e}) = {\rm rank}(M) - {\rm rank}(M_S),
 \ee
 where $M_S$ is the submatrix of $M$ consisting of the first $n_s$ columns of $M$.  $r_M$ counts the number of basis vectors for errors that have a zero syndrome, but act nontrivially on the logical subspace. 
 
  For each syndrome, the decoder  can only apply a single  recovery operator; however, this recovery strategy will always fail with some probability if the error is linearly dependent on one of the $r_M$ basis vectors with trivial syndrome and nontrivial logical operator content.  
  Since all the errors occur with equal probability for erasures, the optimal recovery probability  can then be directly computed as
 \be
 \mathbb{P}(R| S,L,\bm{e}) = \frac{1}{2^{r_M(S,L,\bm{e})}}.
 \ee
Incidentally, $k - r_M$ is also equal to the coherent quantum information of this encoding scheme under erasure errors.  
As we take advantage of in Sec.~\ref{sec:exp}, these formulas directly generalize to stabilizer subsystem  codes by removing columns of $M$ associated with generators for the gauge group.

 \section{Random Stabilizer Code Threshold}
\label{sec:rmt}

In this section, we present a solution to the critical theory of the erasure threshold for random stabilizer codes based on an RMT ansatz.  
Establishing the basic phenomenology of the random stabilizer erasure threshold transition is standard material in quantum information theory \cite{Preskill}.  
Our main contribution is to derive analytic predictions for the code-averaged probability of perfect recovery $\bar{\mathbb{P}}(R|n_e)$ according to Eq.~(\ref{eqn:pperf}), where $n_e$ is the number of erased sites in the fixed-fraction model and $k$ is the number of logical  encoded qubits.
In Appendix \ref{app:favg}, we further show that $\bar{\mathbb{P}}(R | n_e)$ is equal to the code-averaged max-average fidelity $\bar{F}_{\rm avg}$ in the thermodynamic limit.  
We use this result to argue that the Haar random erasure threshold, where we  only approximate $\bar{F}_{\rm avg}$, is in the same universality class as the random stabilizer erasure threshold.  

The encoding circuit $U$ for a random stabilizer code is a random Clifford unitary on $N$ qubits.  
Since spatial locality is irrelevant in this discussion, we take the initial unencoded logical qubits to be given by the last block of $k$ qubits, which implies that the stabilizer group has generators
\be
\bar{Z}_{i}= U Z_{i} U^\dag,~i=1,\ldots, n_s,
\ee
where $n_s = N-k$ is the number of stabilizer generators.
We use the optimal decoding strategy for erasure errors described in Sec.~\ref{sec:optdecstab}:  given a set of erased sites $\bm{e}$ and syndromes $\bm{s}$, the  decoder applies any Pauli operator $R_{{\bm{s}\bm{e}}}$ that lives on $\bm{e}$ and flips each stabilizer generator $\bar{Z}_i$ to have the same sign as $s_i$ \cite{Delfosse17}.  
 The circuit-averaged probability for perfect recovery under optimal decoding satisfies
\be
\bar{\mathbb{P}}(R| n_e) =  \mathbb{E}_U \sum_{\bm{e}} \mathbb{P}(R| U, \bm{e}) p(\bm{e}) =  \mathbb{E}_U \mathbb{P}(R| U, \bm{e}), 
\ee
because $\mathbb{E}_U \mathbb{P}(R| U, \bm{e})$ depends on $\bm{e}$ only through $n_e = |\bm{e}|$ for a fully random Clifford circuit.

Since random stabilizer codes are nondegenerate in the large $N$ limit, every correctable error needs to map to a unique syndrome.  
We can find the number of unique syndromes for a given $U$ and $\bm{e}$ by determining the $\mathbb{F}_2$-rank $n_{se}$ of the syndrome matrix $M_S(S,L,\bm{e})$ formed from the first $n_s$ columns of $M$ from Eq.~\eqref{eqn:ms}. 
Intuitively, $2^{n_{se}}$ is simply the total number of unique syndromes available to the decoder for this erasure pattern. 
Thus, the average recovery probability is
\be \label{eqn:pbarr}
\bar{\mathbb{P}}(R | n_e) = \mathbb{E}_U[2^{n_{se} - 2 n_e}] = 2^{\bar{n}_{se} - 2 n_e},
\ee  
where $\bar{n}_{se} \equiv \log_2 \mathbb{E}_U[ 2^{n_{se}(U,\bm{e})}]$ is just a function of $n_e$, $n_s=N-k$, and $N$.

We can approximate the behavior of $n_{se}$ for a random stabilizer code in the two limits  $n_{s} \ll 2 n_{e}$ or $n_s \gg 2 n_e$ \cite{Preskill}.  
In the former case, the number of possible errors is exponentially larger than the number of available syndromes.  
For a random $U$, each syndrome occurs with nearly equal probability; thus, each syndrome will be occupied with high probability and result in the scaling $\bar{n}_{se} = n_s$.  
In the opposite limit, the number of available syndromes is exponentially larger than the number of errors.  
As a result, there is a high probability that each error gets mapped to a distinct syndrome, resulting in the scaling $\bar{n}_{se} = 2 n_e$.    
These estimates show that $\bar{\mathbb{P}}(F) = 1- \bar{\mathbb{P}}(R)$ has a discontinuity at the channel capacity bound $ n_s/N = 1 - R = 2 n_e/N = 2 e $ in the large-$N$ limit.  

In the RMT approach described below, we can explicitly calculate $\bar{\mathbb{P}}(F)$ for all values of $n_e$ and $n_s$, including arbitrarily close to threshold,
  \be \label{eqn:pre}
 \bar{\mathbb{P}}(F)  \approx \left\{ \begin{array}{cc}
   2^{-\abs{\delta}-1}  & 2n_e \ll n_s, \\
  1- r_c,~& 2n_e = n_s, \\
  1-2^{-\abs{\delta}}  & 2n_e \gg n_s, 
  \end{array}\right. 
  \ee
  where $\delta = 2 n_e - n_s = 2(e-e_c)N$ is the distance from the critical point and $r_c$ is the recovery rate at the critical point. 
  As we show in the section below, $  r_c = 0.610322 \ldots$ in the RMT model. 
  From this RMT solution, we also find that the higher order corrections to this formula are exponentially suppressed  in the distance from the critical point $O(2^{-2 |\delta|})$.  

\subsection{ RMT solution}

The exact formula for the code-averaged recovery probability is given by
\be
\bar{\mathbb{P}}(R|n_e) = \mathbb{E}_{U} \sum_{m=0}^{2 n_e} \mathbb{P}\big[ M(S,L,\bm{e})~\textrm{has}~r_M = m  \big] \frac{1}{2^{m}} .
\ee
  There is no need to average over $\bm{e}$ for a fixed erasure number because the circuit average removes the dependence on the spatial locations of errors. 
In the RMT approach, we  assume that the syndrome matrices $M(S,L,\bm{e})$ and $M_S(S,L,\bm{e})$ are given by  random $2 n_e \times (n_s+2k)$ and $2 n_e \times n_s$ matrices, respectively.  
We do not expect this result to be true exactly because it ignores the constraint that the time-evolution preserves commutation relations; however, we conjecture that it is accurate up to exponentially small corrections in $N$.  
The reason it is a probable hypothesis is that $M$ and $M_S$ can be constructed by taking  submatrices of a much larger $2N\times 2N$ tableau representation for $U$ \cite{Gottesman98,Aaronson04}. 
The RMT ansatz is based on the assumption that these  submatrices  are insensitive to the ``global'' constraint on the otherwise random $U$ that it preserves commutation relations of Pauli group elements. 

In the RMT anstaz and for $e<1/2$,  the matrix $M$ will  have full rank $2 n_e$ with a probability that converges to one exponentially in $N$ because the number of columns is much greater than the number of rows.  
The average recovery probability then reduces to a combinatorial formula regarding the rank distribution of the $M_S$ matrix
\beu
\bar{\mathbb{P}}(R|n_e) =_{\rm RMT} \sum_{m} \frac{ \textrm{\#~$2 n_e \times n_s$ matrices of rank $m$}}{\textrm{\# $2 n_{e}\times n_s$ matrices}} \frac{2^{m}}{2^{2 n_e}}.
\eeu
The denominator is the number of matrices over $\mathbb{F}_2$ of size $2 n_e \times n_s$, which is equal to $2^{2 n_e n_s}$ since each entry can take one of two independent values.    
Finding the number of $2 n_e \times n_s$ matrices of rank $m$ is a less trivial, but familiar, result in combinatorics that also has applications to classical error correction \cite{Abdel12}.  
For completeness, we provide a derivation in Appendix \ref{app:combinatorics}.

\begin{figure}[tb]
\begin{center}
\includegraphics[width = .48 \textwidth]{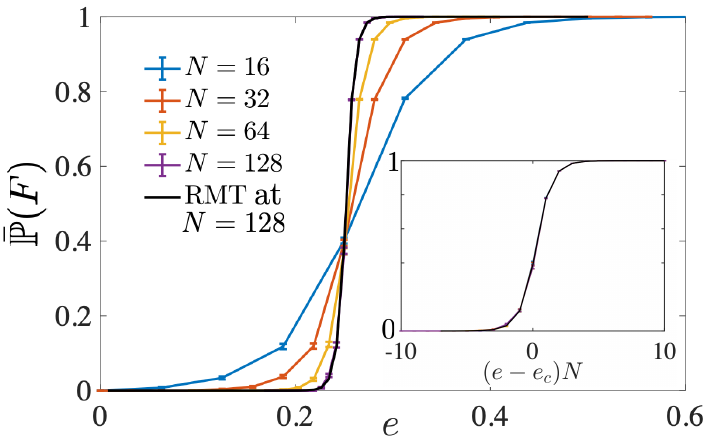}
\caption{Failure probability for 1D circuit of depth $N$ with periodic boundary conditions obtained via numerically sampling $10^3$ random codes.  
We took a fixed fraction erasure error at a code rate $R=1/2$, for which the critical erasure rate is $e_c = N/4$.  
Black line shows the RMT prediction computed for $N = 128$, which agrees with the numerical results to within the statistical error bars. 
The inset shows an excellent collapse for this full range of sizes according to the predicted scaling.}
\label{fig:RMT}
\end{center}
\end{figure}

Using this formula, the probability of successful recovery has the analytic expression
\beu
\bar{ \mathbb{P}}(R|n_e)  =_{\rm RMT} \sum_{m=0}^{n_{se}^{m}} \frac{\prod_{\ell=0}^{m-1}(2^{2 n_e} - 2^\ell) (2^{n_s} - 2^\ell)}{2^{2 n_e n_s} \prod_{\ell=0}^{m-1} (2^{m} - 2^\ell)} \frac{2^m}{2^{2 n_e}},
 \eeu
 where $n_{se}^{m} = \min(n_s,2 n_e)$. $\bar{\mathbb{P}}(F) = 1 - \bar{\mathbb{P}}(R)$ has the asymptotic behavior given by Eq.~(\ref{eqn:pre}) with the  critical parameter
 \be
 r_c =_{\rm RMT}  \sum_{m=0}^{\infty}  \frac{\prod_{\ell=1}^{\infty}\Big(1 - \frac{1}{2^{m+\ell}}\Big)^2 }{2^{m(m+1)} \prod_{\ell=1}^{\infty} \Big(1 - \frac{1}{2^{\ell}}\Big)},
 \ee
 which is approximately $r_c \approx 0.610322 \ldots$.
 By numerically sampling  $M_S$ matrices generated by depth $N$ 1D local circuits with periodic boundary conditions, we have verified that these RMT predictions accurately approximate the true failure probability on these sizes.  
 The results are shown in Fig.~\ref{fig:RMT} up to size $N = 128$ for $R=k/N=1/2$.  
 We find excellent agreement between the exact numerical results and the RMT prediction throughout the critical region, even for sizes down to $N=16$.  
 To obtain a more precise comparison, we  estimate the success probability with higher precision at the critical point. 
 Randomly generating $10^8$ $M_S$ matrices from a depth $2 N$ circuit ($N = 40$) in 1D with periodic boundary conditions provides our current best  estimate 
 \be
 r_c \approx 0.61029 \pm 3.7 \cdot 10^{-5} = 0.61029(4) ,
 \ee
which agrees with the RMT value at a precision of $ 10^{-4}$.   
In Appendix \ref{app:selfavg}, we further show that the recovery probability is self-averaging at $e_c$ in the sense that a typical random code has a recovery probability that converges to $r_c$ in the large-$N$ limit.

\section{Quasilocal Random Stabilizer Code Threshold}
\label{sec:quasilocal}

In this section, we investigate the erasure threshold for random stabilizer codes generated by finite-depth quantum circuits in finite-size systems.

\subsection{Block model: mean-field limit}
\label{sec:block}

\begin{figure}[tb]
\begin{center}
\includegraphics[width = 0.42 \textwidth]{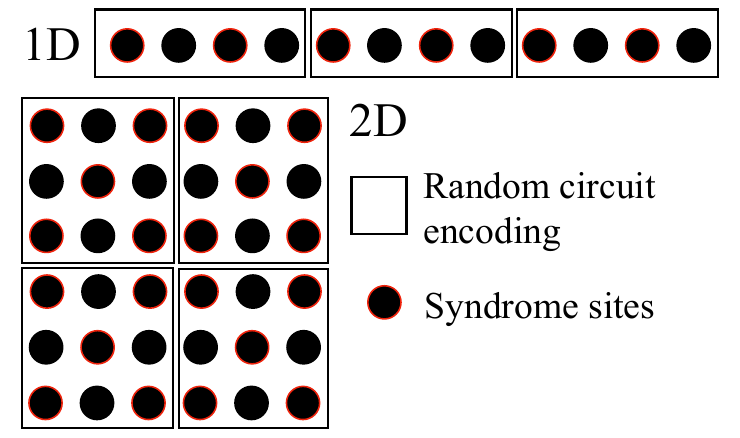}
\caption{Toy model for the below threshold behavior of finite-depth random unitary-encoding circuits in which we remove gates that couple different blocks of qubits.
 }
\label{fig:block}
\end{center}
\end{figure}

We can gain a surprising amount of insight into this local random coding problem by first considering a toy  model with the simplified block encoding scheme illustrated in Fig.~\ref{fig:block}.  
Furthermore, the basic arguments in this section are not specific to the erasure channel. 
In this model, we remove gates that couple different blocks of qubits such that each block undergoes completely independent random unitary dynamics.  
Intuitively, this model can be interpreted as a type of mean-field model for the random code transition.  
At large depth, the average failure probability for this model becomes an upper bound on the average failure probability of the random code transition. 

Specifically, we break up a system of $N$ qubits into cubic blocks of size $N_b = L^D$ where $D$ is the space dimension of the encoding Clifford circuit in each block and $L$ is the linear size of the block.  
Each block has approximately $(1-R) N_b$ stabilizers and $R N_b$ logical qubits.     
Running a high-depth $[d=O(N_b)]$ random Clifford circuit on each block results in a rate $R$ random stabilizer code on this block of qubits.  
If we apply an erasure error below the  random code threshold, then the average recovery probability is just the product of the average recovery probability for each block (since the codes between blocks are uncorrelated)
\be
\bar{\mathbb{P}}(R | n_e) = \prod_{i=1}^{N/N_b} \big(1 - \mean{2^{-\delta_i - 1}}_{\bm{e}} \big) + O(N 2^{-N_b}/N_b),
\ee
where $\delta_i = (1-R) N_b - 2 n_{e i}$ is the distance from the critical point in block $i$ with $n_{e i}$ erased sites.  
In order for our approximations to be valid we require that $N_b$ grows as $O( \log N)$ or faster. 
We make use of the fact that the fluctuations in the number of erasures in each region are determined by the central limit theorem
\be
n_e^b = e N_b + \Delta_b,~ \Delta_b \sim \mathcal{N}(0, \sigma_b^2),~\sigma_b = \sqrt{e(1-e)N_b}
\ee
As a result, the average failure probability is given by
\begin{align}
\bar{\mathbb{P}}_{\rm RMT}(F) &\le \bar{\mathbb{P}}(F)  \approx \frac{N}{2N_b} \mean{ 2^{-\delta_i} }_{\bm{e}},\\
\mean{ 2^{-\delta_i } }_{\bm{e}} &\approx 2^{2(e-e_c) N_b} \int_{-\infty}^0 \frac{dx}{\sqrt{2\pi}} 2^{2 \sigma_b  x} \exp(-x^2/2) \\ \nonumber
&= \frac{2^{2 ( e - e_c ) N_b}}{\sqrt{2\pi e(1-e) N_b} \ln 4},
\end{align}
where $x =(e - e_c)\sqrt{N} /\sqrt{e(1-e)}$.
Thus, for $e<e_c$, the failure probability converges to zero when 
\be
\lim_{N\to \infty} \frac{\log_2 (N/N_b^{3/2})}{N_b} <  2 (e_c - e) 
\ee
which is satisfied for $N_b = O( \log N)$, i.e., when the block length scales as $L =O[ (\log N)^{1/D}]$.  

To achieve a random code on each block, we naively need to apply a depth $d = O(L)$ circuit \cite{Brandao16,Harrow18}; however, this neglects the fact that there are rare Clifford circuits where Pauli operators remain localized to a given site.  
In particular, to preserve commutation relations every two-qubit Clifford gate has to map at least one single-site Pauli operator for each site to another single-site Pauli operator.  
The probablity of such localized logicals appearing in a system of size $N$ scales as $N/A^d$ for a constant $A$ that depends on the ensemble of gates used in the random circuit.  
Note, that even if all two-qubit gates are entangling, $A$ will still be finite.  
This constraint implies that one needs to apply a depth $O(\log N)$ random Clifford circuit regardless of dimensionality to avoid these rare localized operators.  
As a result, the block model only converges to zero failure probability for depth $d = O(\log N)$ for all spatial dimensions.  
Interestingly, after our work appeared, a similar type of argument was used in Corollary 4 of Ref.~\cite{Dalzell20} to prove a lower bound of $O(\log N)$ on the depth required to achieve a form of anticoncentration in random circuits.  
The ensemble was formed from two-local circuits with the gates drawn randomly from a two-design.   
Developing a more complete understanding of the relation between encoding properties of  low-depth random circuits and  other observables, e.g., anticoncentration or sampling complexity, is an interesting subject for future work.  

In the case of the channel coding problem considered here, there are two routes to overcome the $O(\log N)$ lower bound on the depth required to achieve zero failure probability below capacity.  
One simple approach within the block model picture is to apply an optimized implementation of a two-design following Ref.~\cite{Cleve15}, but including SWAP gates to map the all-to-all circuit to a local geometry.  This approach also requires the use of $O(\log N)$ ancilla qubits per block, which, by our conventions, would effectively reduce the overall rate of the code.  
With such optimized circuits, one can deterministically encode each block into a high-performance code in depth $O( N_b^{1/D})$; thereby, allowing convergence of the full system to zero failure probability at depth $O[(\log N)^{1/D}]$.  
This argument shows that, in principle, one can surpass the $O(\log N)$ scaling by introducing long-range correlations into the encoding and allowing for additional ancilla qubits.  
In practice, however, the block model will always have a relatively weak convergence with depth because it is not taking advantage of correlations that can build up between blocks.  
To achieve the sub-logarithmic scaling in practice, we therefore use the expurgation strategy described in Sec.~\ref{sec:exp} below.  
In this approach, these rare localized logical operators are directly removed from the code by the expurgation process.

\subsection{Critical scaling }

\begin{figure}[tb]
\begin{center}
\includegraphics[width = 0.48 \textwidth]{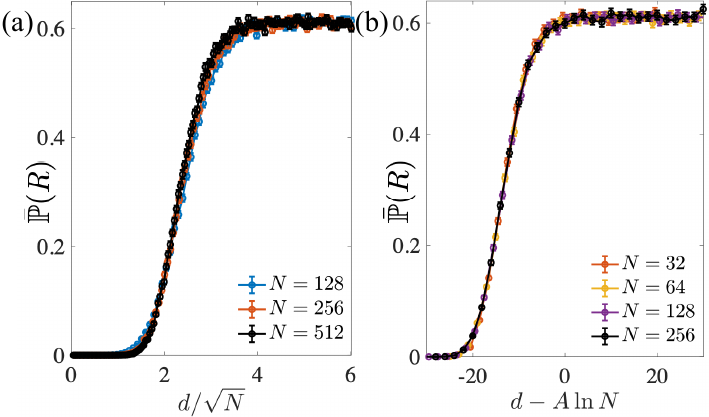}
\caption{(a)  Recovery probability vs. scaled depth $d/\sqrt{N}$ ($d=$ number of two-qubit gates per site) for a 1D random circuit in a brickwork arrangement at the channel capacity limit  $(R,e)=(1/2,e_c)$.  
(b) Recovery probability vs. scaled depth $(A=6.5)$ for  a nonrandom erasure error in which every fourth site is erased from the system.  
In this case, the $d=0$ failure probability is $1/2$.  
Each two-qubit gate in these circuits is a random Clifford gate. }
\label{fig:regerase}
\end{center}
\end{figure}

In the vicinity of the critical point for the random stabilizer code, it is clear that the block encoding scheme fails because each individual block fails with a large probability.    
As mentioned in the previous section, we expect the original model to achieve better performance because the ``blocks'' formed by the finite depth circuit are effectively correlated with each other.  
This implies that the error correction in regions with excessive numbers of erasures can be assisted by nearby regions.  
As shown in Fig.~\ref{fig:regerase}(a), we numerically observe that the convergence to the critical properties of the random code behavior occurs at depth $O( \sqrt{N})$ in 1D.  
On the other hand, for $D \ge 2$,  the convergence, even at the critical point, occurs at depth $O(\log N)$. 
As we  show below, this distinction between $ D =1$ and $D\ge 2$ can be traced to the familiar fact that the boundary of a contiguous region in 1D is effectively zero dimensional.  
In the discussion below, we assume we are working at depth greater than $O(\log N)$ so that large inhomogeneities in the quality of the random code are smoothed out, while what is left over is the randomness in the error pattern. 

We first give an argument for the $\sqrt{N}$ scaling in 1D based on a mapping to a random walk for the iid model.  
If we sum up the number of erasures relative to the critical number along the length of the system, this is a biased random walk that travels a certain distance on summing around the full system.  
The random walker's time is the system's space, while the random walker's space is an excess number of erasures in that segment of the system's space.  
The failures occur where this random walk does a ``backtrack'' of distance $d$.  
So the characteristic $d=d^*$, where the failure probability converges towards its high-depth $[d = O(N)$] value,  is the $d$ where these backtracks become rare in the system of length $N$.  
From the statistics of random walks,  this has a probability of occurring that falls off as  $\exp(-A N/d^2)$  for some constant $A$, but the region that is dense can be at of order  $d^2/N$  distinct locations \cite{Fisher84}.  
By considering only regions where these local fluctuations are above threshold, we arrive at the scaling form $d^* = N^{1/2} g[(e_c-e)N^{1/2}]$    with  $g(x) \sim (1/x) \log(x)$  at large $x$  and  $g(0)$ of order one.  
Thus, well below threshold   in 1D ($x \gg 1$), the scaling for the critical depth is $d^* \sim (e_c-e)^{-1} \log N$.  
The depth required to converge to zero failure probability is always $O(\log N)$ in 1D, but the prefactor diverges as one approaches the optimal threshold.

A related argument that connects more directly to the syndrome matrix $M(S,L,\bm{e})$  proceeds as follows:   Imagine we apply a fixed number of erasures at the critical point $n_e = n_s/2$, but distributed randomly throughout the system. 
If we cut the system into two halves, then one half of the system will effectively be above threshold with $\sim \sqrt{N}$ extra erasures, while the other half will be below threshold.  
In order to correct the above threshold region, we need to ``borrow'' a sufficiently large number of error syndrome basis elements in $M(S,L,\bm{e})$ from the region that is below threshold.  
This requires that the minimum support of our error syndrome basis elements is $\sim \sqrt{N}$ to satisfy this condition; thus, we need to run a depth $\sim \sqrt{N}$ circuit to generate sufficiently long-range error syndromes in the syndrome matrix $M$.

To  test our argument that it is only the local fluctuations in the erasure number that determine the required depth, we compare the convergence to the RMT prediction for random erasures in Fig.~\ref{fig:regerase}(a)  against regularly arranged erasures in Fig.~\ref{fig:regerase}(b).  
In the spatially nonrandom case, the error is chosen randomly from one of the four regularly spaced erasure patterns with $n_e = e_c N = N/4$.  
In contrast to the random error model, we see convergence to the large depth limit with an $O(\log N)$ scaling.  

We remark that the recovery probability for a depth zero circuit with this nonrandom error and our layout of logical qubits is equal to $1/2$.  
Thus, the recovery probability is nonmonotonic with depth: it is $1/2$ at $d=0$, then drops close to zero for $0<d \ll A \log{N}$ and then improves to $\cong 0.6$ for $d> A \log{N}$;  the coefficient is found to be $A \cong 6.5$.

\begin{figure}[tb]
\begin{center}
\includegraphics[width = .48 \textwidth]{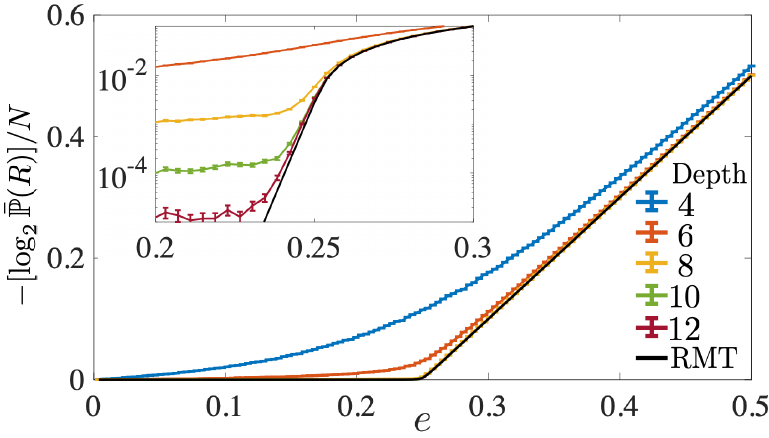}
\caption{Recovery probability vs. erasure fraction for a two-dimensional random circuit in a brickwork arrangement of gates with periodic boundary conditions for different depths $d$ for $N=256$  and $R=1/2$. 
Different sizes collapse to the same curve for this way of scaling except within a region of width $|e-e_c| \sim 1/N$ near the critical point. 
We sequentially cycle through 4 layers so that each site interacts with its north, east, south, west neighbor for each 4 units of depth.     
The scaling behavior converges to the RMT prediction exponentially with depth throughout the critical region.  
The inset shows the same data on a logarithmic scale, illustrating the scaling $\mathbb{P}(F) \sim e^{-d/A}$ for $e <e_c$.  
Each two qubit gate in the circuit consists of an iSWAP gate followed by a random single-site Clifford on each site. }
\label{fig:2D}
\end{center}
\end{figure}

The situation changes dramatically in higher integer dimensions where the prefactor of the $\log N$ scaling of $d^*$ does not need to diverge as one approaches the optimal erasure threshold.  
In this case, the random fluctuations in erasure number within a given region can be overcome by the overlapping syndromes near the boundary whenever
\be
L^{D-1}d \sim \sqrt{N} \sim L^{D/2} \to d \sim L^{1-D/2}.
\ee
This tension between random fluctuations and ordering tendencies  is familiar from  Imry-Ma arguments.    
This scaling  indicates that $D = 2 $ is the marginal dimension for the relevance of random erasure locations.  
For $D > 2$, at the depth $d \ge A \log N$ needed to produce a near-optimal code, the effect of this erasure-location randomness is subdominant.  
This appears to remain true in the marginal dimension $D=2$, where the subdominance is only by factors of $\log{N}$.   
In Fig.~\ref{fig:2D} we show the numerical results for the recovery probability through the erasure threshold at different values of the depth in two dimensions.  
We clearly see the exponential convergence to the RMT prediction throughout the critical region.

In the case of intermediate dimensions $ 1 < D <2$, such as can be realized in fractal lattices and critical percolation clusters, the perimeter of a region with excess erasures may have a nontrivial scaling with $N$ that is also not spatially uniform.  
As a result, it would be an interesting subject for future work to precisely determine the  fate of the critical scaling on particular real space lattices with these intermediate dimensions.

\subsection{Spatial correlations of uncorrectable errors}

When used as a toy model for the low depth regime $\log N \ll d \ll N^{1/D}$, the block model  suggests that errors will generally be bunched in space.  
In particular, this model leads to the intuition that regions with excess erasures will fail first with an uncorrectable error of  weight $\sim d^D$.   
To test this argument we consider a setup inspired by the entanglement fidelity: two of the logical qubit sites are initially entangled with external reference qubits and the other logical qubits are in a random pure product state.  

Using these reference qubits as local probes, we define an error as occurring in the vicinity of location $i$, if reference qubit $i$ loses its entanglement with the system following the full encoding, error, and decoding procedure.  
Specifically, we study the change in mutual information between each probe and the system   
\be
\Delta I(R_i:S) = I(R_i:S) - I(R_i:S'),
\ee
where $I(A:B)=S(\rho_A)+S(\rho_B)-S(\rho_{AB})$, $\rho_{R_iS'}= \mathcal{D}\circ \mathcal{N} \circ \mathcal{E}(\rho_{R_iS})$ is the  density matrix of the system and reference probe $i$ following the decoded error channel, $\rho_{R_i} = \trace_{S'}[\rho_{ R_i S'}]$, and $\rho_{S'} = \trace_{R_i}[\rho_{R_iS'}]$. 
Initially, the mutual information $I(S:R_i) = 2$.   
In these stabilizer code models with Pauli error channels, the mutual information changes in discrete integer steps.  
For two reference qubits entangled with the system at sites $x_{1}$ and $x_2$, we then define code-averaged local error profiles
\begin{align}
P_{i}(x_{12},d) &= \bar{ \mathbb{P}}[\Delta I(R_i:S) > 0], \\
P_{12}(x_{12},d) &= \bar{ \mathbb{P}}[\Delta I(R_1:S) +  \Delta I(R_2:S)>0],
\end{align}
where  $\bar{\mathbb{P}}(\cdot) =  \mathbb{E}_U \mathbb{P}(\cdot)$ and $x_{12} = |x_1 - x_2|$ is the distance between the probes.

\begin{figure}[tb]
\begin{center}
\includegraphics[width = .47 \textwidth]{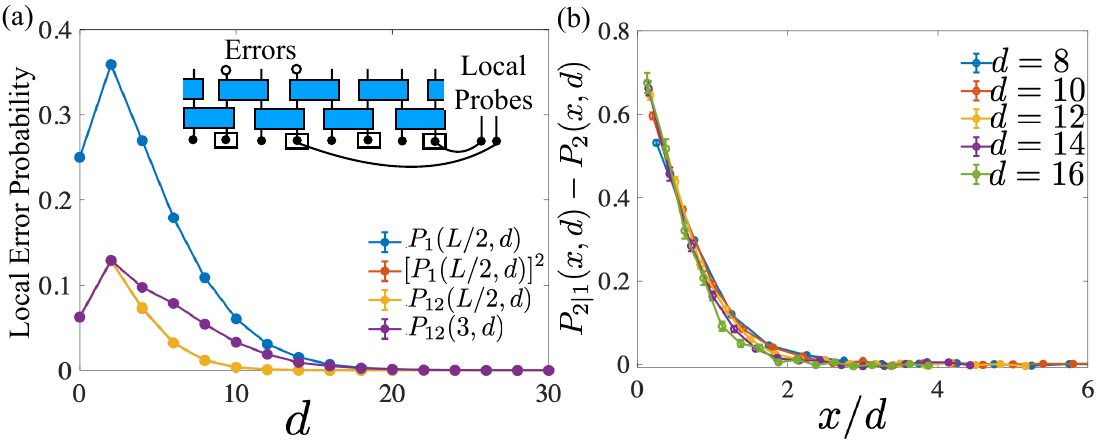}
\caption{(a) Local uncorrectable error probability of one  $P_1(x_{12},d)$ or both $P_{12}(x_{12},d)$  reference probe qubits  entangled with the system vs. $d$.   
Here, we took $D=1$, $R=1/2$, $n_e/N = e_c =1/4$, and $N= 128$. 
Each two qubit gate in the circuit is a random Clifford gate.  
The local error probability is defined as the probability that the mutual information of a reference qubit is less than maximal after the optimal decoding. 
Note, the red curve almost perfectly coincides with the yellow curve, indicating an absence of connected correlations for these far separated uncorrectable errors. 
(b) Conditional error probability of probe $2$ when an error effects probe 1 vs. scaled distance for different $d$. 
When a probe fails in a given region it implies that the second probe a distance $\sim d$ also fails with high probability. }
\label{fig:cor}
\end{center}
\end{figure}

Numerical results for these error profile functions are shown in Fig.~\ref{fig:cor}(a) for $D=1$ with length $L = N = 128$.  
We took $R=1/2$ and  $n_e = e_c N = N/4$ so that uncorrectable errors occur relatively frequently.  
We see convincing evidence that spatial locality plays an important role for these low depth codes, despite the potential for nonlocal effects induced by the syndrome measurements.  
In particular,  when $x_{12} = L/2$, then the joint failure distribution $P_{12}(L/2,d)$ factorizes into a product distribution $[P_{1}(L/2,d)]^2$ at low depths $d$.  
On the other hand, when $x_{12} < d$, there is a clear bunching effect whereby $P_{12}(x_{12},d) > [P_{i}(L/2,d)]^2$.  
We study this more quantitatively in Fig.~\ref{fig:cor}(b) in terms of the conditional failure probability of reference probe $2$ given that reference probe $1$ failed: $P_{2|1}(x,d) = P_{12}(x,d)/P_1(x,d)$.  
Rather intuitively, we see a collapse of the curves for different depths when this conditional profile is plotted as a function of $x/d$.

  These spatial correlations in the uncorrectable errors are an indication that these low-depth codes retain features associated with spatial locality despite achieving the critical behavior of fully random or high-depth codes.  
  Thus, in many respects, they are a truly distinct class of codes from fully random stabilizer codes.

  \section{Expurgation algorithm}
  \label{sec:exp}

  As discussed in Sec.~\ref{sec:block}, there are strategies in higher dimensions to overcome the $\log N$ depth scaling found for random Clifford circuits.  
  In this section, we introduce a natural method to improve the performance of these low-depth codes based on the fact that the dominant failure mode at depths $[\log N]^{1/D} \le d \le \log N$ are rare regions with bad logical qubits.

  The basic ingredient in our algorithm is the efficient implementation of quantum measurements of stabilizer code-space density matrices \cite{Aaronson04}.  
  We assume we are given a single logical operator $g$.  
  We can update a generating set for the stabilizer code and its logical operators by making a projective measurement of the code space density matrix following the tableau rules outlined by Aaronson and Gottesman \cite{Aaronson04}
\be
\begin{split}
\rho_{S} & = \frac{1}{2^N} \prod_{i =1}^{N-k} (\mathbb{I}+\bar{Z}_i) \to (1\pm g) \rho_S (1\pm g)/2 \\
&= \frac{1}{2^N} \prod_{i =1}^{N-k} (\mathbb{I}+\bar{Z}_i) (\mathbb{I} \pm g),
\end{split}
\ee
where the sign of the measurement outcome is randomly chosen.  
This projective measurement operation will not affect the original generating set for $S$ except to add $g$ to the list of generators; however, it will modify the logical operators to ensure that all of the remaining logical operators commute with $g$.  
This implies that the ``destabilizer'' operator $\bar{g}$ associated with $g$ is no longer a logical operator.  
As a result, we can form an $[N,k-1]$ stabilizer code or stabilizer subsystem code by converting  $g$ or  $\bar{g}$  into a stabilizer or gauge operator, respectively.  
This procedure can be iterated to successively convert logical operators into additional stabilizers or gauge operators, while leaving the original syndrome stabilizers unaffected.  

Specifically, in our expurgation algorithm we begin with an $[N,k]$ stabilizer code with stabilizer group $S$ and logical operators $L$.  
We then randomly generate an erasure pattern $\bm{e}$ and compute the matrix $M(S,L,\bm{e})$.  
Performing row reduction allows us to form a basis $\{g_i\}$ of linearly independent errors that map to the zero syndrome, but have nontrivial logical operator content.  
We then perform a sequence of projective measurements of these operators as described above to form a new stabilizer or subsystem code.  
This procedure is iterated many times until either the rate of the  code  approaches a specified target value, the failure probability reaches a certain threshold, or  the number of logical operators goes to zero (i.e., expurgation fails). 

To put this algorithm on firmer mathematical footing, we prove the  following two simple propositions:
\begin{proposition}
Let $S$ be the stabilizer group for  a stabilizer subsystem code with logical operators $L$ and gauge group $G$.  
For every $g \in L \otimes G$ that acts nontrivially on $L$, the distance of $S$ after expurgating $g$ into $S$ or $G$ monotonically increases.
\end{proposition}
Let us order all $4^N-1$ Pauli group operators by their Hamming weight (number of nontrivial sites) and compute the anticommutator of every Pauli group element with a generating set for $S$ and $L$
\be
\left( \begin{array}{c | c}
\bm{s}_{E_1} & \bm{\ell}_{E_1}  \\
 \vdots & \vdots \\
 \bm{s}_{E_{4^N-1}} & \bm{\ell}_{E_{4^N-1}}  
 \end{array}
 \right).
 \ee
We assume  $g$ and an anticommuting logical $\bar{g}$ are two of the generators and they commute with all other generators for $L$.  
The distance of the subsystem code can be found by finding the first Pauli group element $E_D$ in this list that has $\bm{s}_{E_D} = 0$ and a nontrivial anticommutator vector $\bm{\ell}_{E_D}$.  
If we expurgate $g$, then this removes $g$ and $\bar{g}$ from the list of generators, which amounts to removing two of the columns from $\bm{\ell}_{E_D}$ and adding one column to $\bm{s}_{E_D}$ or not (depending on the expurgation strategy). 
 If $\bm{\ell}_{E_D}$ becomes trivial, then the distance might increase depending on what happens to the next Pauli group element in the list ordered by the Hamming weight.  
 If $\bm{s}_{E_D}$ becomes nontrivial, then the distance might also increase.  
 If $\bm{s}_{E_D}$ remains trivial and $\bm{\ell}_{E_D}$ remains nontrivial, then the distance stays the same. 
 Therefore, the distance is monotonic. \qed

Essentially, we use the following two properties of expurgation: (i) The stabilizer group never shrinks in size (it can  even grow depending on the strategy) and (ii) the number of logical operators only decreases.
Hence, the relevant set of operators that commute with stabilizers, and anticommute with some logical operator never grows. 
Therefore, the code distance -- defined as the minimum Hamming weight of elements in the relevant set of operators  -- never decreases.

A related proposition that follows a similar line of reasoning is:
\begin{proposition}
Let $S$ be the stabilizer group for  a stabilizer subsystem code with logical operators $L$ and gauge group $G$.  
For every $g \in L \otimes G$ that acts nontrivially on $L$, the optimal decoding recovery probability of $S$ after expurgating $g$ into $S$ or $G$ monotonically increases for all Pauli error channels.
\end{proposition}
For each Pauli group element $E$ in the list from Proposition 1,  we let their probability of appearing in the error channel be $p(E)$.  
We  then  group this list of anticommutator vectors into subsets with the same syndrome vector $\bm{s}_i$, which each occur with total probability $\mathbb{P}(\bm{s}_i)$.  
We further break up these groups into error classes $\bar{E}_{ij}$ of errors with identical values of $\bm{\ell}_{E_{ij}}$. 
The conditional recovery probability  is the probability of the most likely error class 
\be
p_i = \max_j\sum_{E \in \bar{E}_{ij}} p(E) 
\ee
divided by $\mathbb{P}(\bm{s}_i)$, such that the total total recovery probability is 
$ \mathbb{P}(R) = \sum_{i} p_i.$
Expurgation of $g$ will never decrease the total value of this sum.  
In the case where $g$ is turned into a gauge operator, then the syndrome classes and their total probabilities are unchanged, while the logical equivalence classes for that syndrome can only combine with each other or stay the same.  
As a result, $p_i$ is monotonically increasing for each $i$, which makes $\mathbb{P}(R)$ monotonically increase under expurgation. 
A similar argument holds when $g$ is turned into a check operator. \qed

The dynamics during this expurgation process bears close resemblance to the purification dynamics of $\rho_S$ for random circuit models with measurements  studied by two of the authors \cite{Gullans19c} and developed further in Refs.~\cite{Li20,Ippoliti20,Fidkowski20}.  
In that case, though, the measurements are not selectively chosen to project out certain logical operators, but rather they are chosen as random, few-site projective measurements.  
In both dynamics, however, we observe a similar trend that the entropy of the code-space density matrix progressively decreases with measurements until it reaches a plateau value.  
The plateau can either be at a subextensive value (a ``pure'' phase) or at a finite entropy density (a ``mixed'' phase).   
What is common between both types of dynamics is that, whenever there is residual entropy in the code-space density matrix, then the expurgated code is able to better protect the remaining logical qubits against future errors in the system that are statistically independent from the errors that helped form the code.

  \begin{figure}[tb]
\begin{center}
\includegraphics[width = .47 \textwidth]{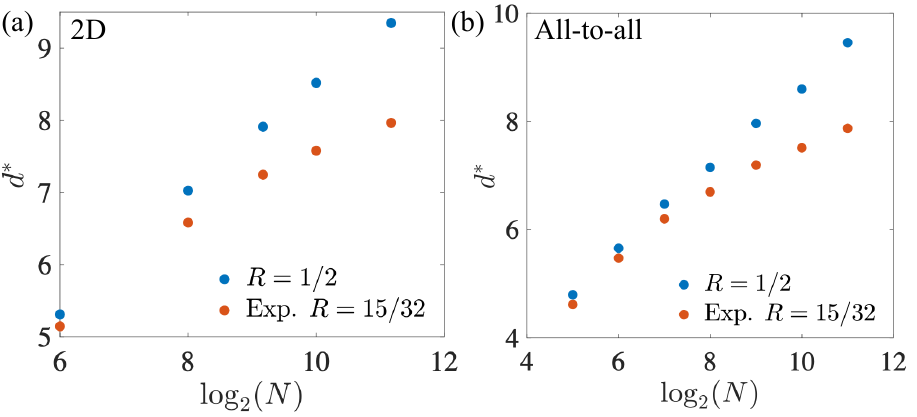}
\caption{(a) Interpolated depth $d^*$ to reach 50 \% failure probability for a 2D random Clifford circuit with periodic boundary conditions vs log-system size.  
All logical operators were turned into gauge degrees of freedom during expurgation.  
We removed the small amount of $N/32$ to aid in extracting the scaling vs  $\log_2 N$ to large sizes and small depths.    
We took an erasure fraction $n_e/N = 1/8$ for both the expurgation algorithm and the calculation of the failure probability.  
(b) Same as (a), but for an all-to-all circuit in which $N/2$ pairs of sites are randomly selected to apply a two-qubit gate for each unit of depth.  
Each two-qubit gate in both geometries is a random Clifford gate. }
\label{fig:exp}
\end{center}
\end{figure}

In Fig.~\ref{fig:exp}, we provide an illustrative example of the performance improvements that are possible with this expurgation strategy for 2D and all-to-all random circuit encodings.  
In both cases, all expurgated logicals were turned into gauge qubits, which has the advantage that the syndrome check operators are unchanged. 
 In this case, the support of each check operator is determined by the initial encoding circuit.  
 Maintaining low-weight check operators has advantages for fault-tolerance by limiting the effects of measurement errors.  
 For both geometries, we see nearly linear scaling of $d^*$ with $\log N$ before expurgation. 
 After expurgation, $d^*$ has a strongly sublinear scaling with $\log N$.

We have also studied the performance of these expurgated codes in 1D, but we do not find improvement of the $\log N$ depth scaling upon expurgation. 
 It is an interesting subject for future work to better characterize the full range of possibilities that result from this type of targeted expurgation process for quantum codes that begin with many logical qubits.

\section{Haar random code threshold}
\label{sec:Haar}

In this section, we study the Haar random erasure threshold.  
We find a similar threshold erasure rate and critical scaling behaviors as the random stabilizer erasure thresholds; however, we observe small quantitative differences in the scaling functions near the critical point for the two codes.  
These results indicate that  Haar random codes are more optimal than random stabilizer codes for erasure errors.

\begin{figure}[tb]
\begin{center}
\includegraphics[width = .48 \textwidth]{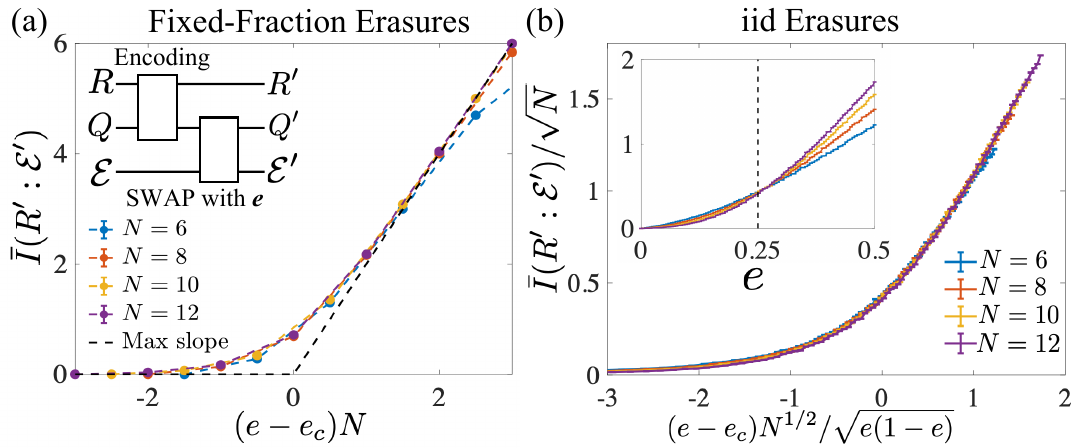}
\caption{(a)  $I(R':\mathcal{E}')$ for the Haar random encoding following a fixed-fraction erasure error at $R = 1/2$.   
For  $e<e_c = (1-R)/2$, $I(R':\mathcal{E}')$ rapidly decays to zero, whereas it grows to an extensive value for $e>e_c$. 
(b) Finite-size scaling in the iid erasure model.    
(inset) Scaled $I(R':\mathcal{E}')$ with an unscaled erasure rate.  
The curves for different system sizes cross near the channel capacity bound $e_c = 1/4$.}
\label{fig:haarcode}
\end{center}
\end{figure}

In contrast to our analysis of the stabilizer codes, we do not perform an optimal decoding analysis and only test for the existence of an erasure threshold.  
We consider the coherent quantum information of an initial state in the code space after application of the erasure channel in Eq.~(\ref{eqn:eraseN}) on a random set $\bm{e}$  of the sites 
\begin{align}
I_c = S(\rho_{Q'}) - S(\rho_{\bm{e}}), \rho_{Q'}  = \trace_{\bm{e}} [\rho_Q]  , 
\rho_{\bm{e}}  = \trace_{\bar{\bm{e}}} [\rho_Q] ,
\end{align}
where $\rho_Q$ is an initial encoded density matrix, and $\rho_{\bm{e}}$ is the reduced density matrix on $\bm{e}$.
  We study the purified channel where a reference system $R$ is used to purify $\rho_Q$ and an environment $\mathcal{E}$ purifies the error operation [see inset to Fig.~\ref{fig:haarcode}(a)].  
  In the case of the erasure error, the interaction of the system with the environment is through a SWAP operation of each erased qubit in $\bm{e}$ with a qubit in $\mathcal{E}$.  
  The mutual information between the reference and the fictitious environment is equal to
\be
\begin{split}
I(R':\mathcal{E}')& = S(\rho_{R'}) +S(\rho_{\mathcal{E}'}) - S(\rho_{R' \mathcal{E}'})  \\
&=  S(\rho_Q) - I_c = R N  - I_c \ge 0,
\end{split}
\ee
where $R$ is the rate of the code.  
As we discussed in Sec.~\ref{sec:optdec}, when $I(R':\mathcal{E}') = |S(\rho_Q) - I_c|< \epsilon$, then the max-entanglement fidelity for that input state satisfies $F_e(\rho_Q) \ge 1- 2 \sqrt{\epsilon}$, i.e., for $\epsilon$ sufficiently small, the error channel can be approximately decoded.

In Fig.~\ref{fig:haarcode}, we show the results of numerical simulations for $\bar{I}(R':\mathcal{E}')=\mathbb{E}_U I(R':\mathcal{E}')$ for the Haar random code with a $\rho_Q$ that acts trivially on the code space.  
We show the results for both fixed fraction and iid erasure errors.  
We see consistent scaling results with the random stabilizer code: the fixed-fraction error model leads to a finite-size rounding of the transition over a region scaling as $|e-e_c|\sim 1/N $.  
The random fluctuations in the total number of erasures in the iid model then round out the threshold even more, producing  a ``critical'' region of width $|e-e_c|\sim 1/\sqrt{N}$ and amplitude $\bar{I} \sim \sqrt{N}$ at $e_c$.

\begin{figure}[tb]
\begin{center}
\includegraphics[width = .48 \textwidth]{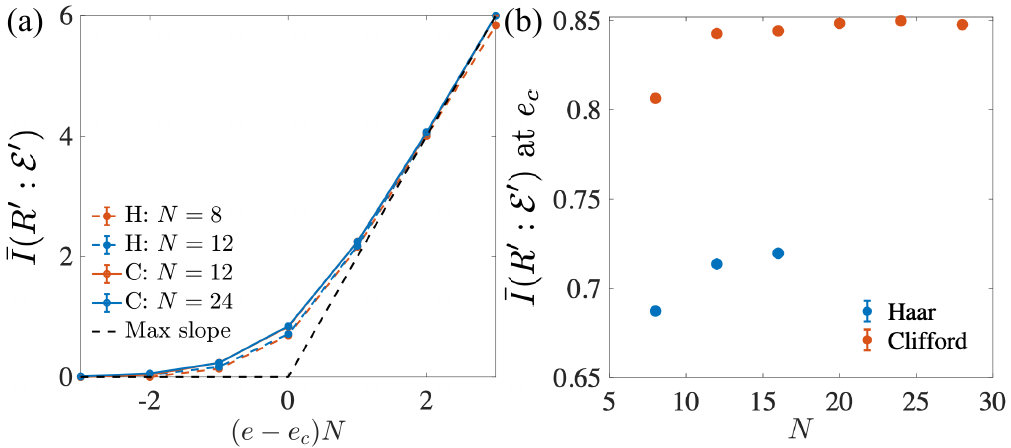}
\caption{(a)  $\bar{I}(R':\mathcal{E}')$ for the fixed fraction erasure errors in the vicinity of the critical point for the Haar random (H) and random stabilizer (C - Clifford) codes.  
(b) Comparison of the code performance at the critical point. 
The Haar codes have slightly better performance than the random stabilizer codes with $\bar{I}(R':\mathcal{E}') = 0.720(5)$ and $0.848(5)$, respectively. }
\label{fig:haar_vs_cliff}
\end{center}
\end{figure}

To obtain a more direct comparison between the Haar random and random stabilizer codes, we show the average coherent quantum information  of each code ensemble in Fig.~\ref{fig:haar_vs_cliff}.  
We see remarkably close quantitative agreement between the code performance; however, there are significant differences that appear at the critical point.  
In particular, in Fig.~\ref{fig:haar_vs_cliff}(b), we see that the two codes appear to be converging to substantially different values of $\bar{I}(R':\mathcal{E}')$ in the large-$N$ limit of $0.720(5)$ (Haar) and $0.848(5)$ (Clifford).  
Thus, a Haar random code is slightly more optimal than a random stabilizer code in this region where the code fails.  
These quantitative differences in the scaling function indicate that the random stabilizer code does not necessarily saturate the performance of an optimal code, even at leading order in the large-$N$ limit.  
In the case of the depolarizing channel, the optimal decoding threshold for a random stabilizer code is expected to be smaller than the channel capacity limit \cite{DiVincenzo98,Smith07}.

\section{Statistical mechanics mapping: domain wall pinning}
\label{sec:statmech}

In this section, we present an approximate mapping of the erasure threshold to a first-order domain-wall pinning transition in a related statistical mechanics description.  
This discussion applies to both Clifford and Haar models.

\begin{figure}[tb]
\begin{center}
\includegraphics[width = .45 \textwidth]{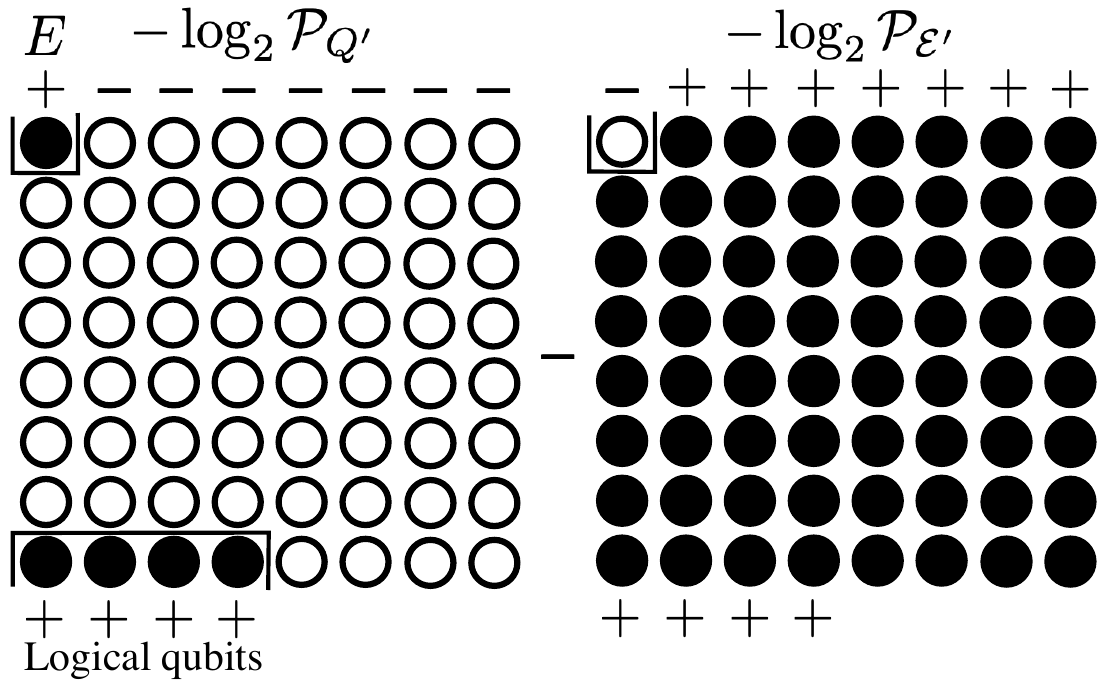}
\caption{Approximate description of below threshold behavior in Ising model.  
The coherent quantum information is the free energy cost of flipping the top boundary condition on $\bm{e}$ and $\bar{\bm{e}}$ in the ordered phase of the Ising model.  
Below threshold, the top boundary condition on $\bar{\bm{e}}$ polarizes the system into the ordered phase aligned along the same direction.  
The error correction threshold occurs when the bulk of the system on the left flips to be polarized in the $+$ direction as the size of $\bm{e}$ is increased, which is a first-order domain-wall pinning phase transition.}
\label{fig:wetting}
\end{center}
\end{figure}

We consider the quenched average of the purity of a subregion $A$
\be 
\begin{split}
-\log_2 {\mathcal{P}}_{A} &\equiv -\log_2 [ \mathbb{E}_U \trace[\rho_{A}^2] ]\\
& \le - \mathbb{E}_U \log_2  \trace[\rho_{A}^2 ] \le   \mathbb{E}_U S(\rho_{A}) .
\end{split}
\ee
A natural approximation to the  coherent quantum information is the difference in log-average purity of each subregion
\be
I_p = -\log_2 {\mathcal{P}}_{Q'} + \log_2 {\mathcal{P}}_{\bm{e}}.
\ee
Although this quantity does not have a clear significance for error correction in general systems, we expect that, for deep Haar random circuits, the fluctuations in $I_c$ over circuits are small enough that it is well approximated by $I_p$ \cite{Nahum16,Nahum17,Zhou19}.
For any $U$ constructed of local two-qubit gates distributed according a two-design, we can compute $I_p$ after circuit averaging using a well-studied mapping between the average purity of subregions of a $D$ dimensional random circuit to a $D+1$ dimensional partition function of an Ising model with certain boundary conditions at late times \cite{Nahum17}.  
The condition that the initial state is mixed on the logical qubit degrees of freedom corresponds to a spin polarized bottom boundary condition on the logical qubit sites \cite{Bao20}. 
In this mapping, $I_p$ becomes the free energy cost of flipping the polarization of the top erased boundary condition in the presence of the polarized boundary condition due to the logical qubits (see Fig.~\ref{fig:wetting}).

The  temperature of the  effective Ising model is well below the transition temperature, which implies that the free energy is minimized primarily through energy minimization.  
Using a minimal energy surface approximation, we obtain a direct estimate for the analog of the mutual information between the reference and the environment for the log-average purity
\be
\begin{split}
I_p(R': & \mathcal{E}') =  -\log_2 {\mathcal{P}}_{Q} - I_p = R N - I_p \\
& \approx \left\{ \begin{array}{c c}
0, & 2 n_e \ll n_s, \\
 2 n_e - n_s, & n_s \ll 2 n_e \ll (1+R)N, \\
2 RN, & (1+R)N \ll 2 n_e
\end{array} \right. 
\end{split}
\ee
where $n_s = (1-R)N$.
This quantity undergoes a phase transition at the same point as the optimal erasure threshold; thus, we suspect it captures some essential features of the threshold for the optimal code.   
 In particular, the point $2 n_e = n_s$ corresponds to a transition in the left half of Fig.~\ref{fig:wetting} where the top boundary condition is no longer sufficiently strong to polarize the bulk of the system.  
 In this case, the middle domain flips to align with the logical qubits. 

Although it is clear that our codes  will not be robust against erasure errors that occur during the encoding circuit, we can gain some additional insight into the breakdown of the threshold using this statistical mechanics model.  
In the Ising model mapping, erasures in the bulk correspond to fixing a finite density of spins in the bulk to point along the $+$ direction, which will overcome the surface pinning effect and prevent the formation of the ordered $-$ phase in the left of Fig.~\ref{fig:wetting}.  
As a result, in order to have a fault-tolerant encoding, some form of error correction should be applied during the evolution itself.

\section{Conclusions}
\label{sec:discussions}

In this paper, we revisited the study of quantum error correcting codes generated by low depth random circuits.  
In any spatial dimension, we found that a depth $O(\log N)$ random circuit is necessary and sufficient to achieve high-performance coding against erasure errors below the optimal erasure threshold, set by the channel capacity.  
However, in 1D, coding arbitrarily close to the optimal threshold requires a depth $O(\sqrt{N})$ circuit due to the relevance of spatial randomness in errors near code capacity.  
The marginal dimension for high-performance, low-depth coding at capacity is 2D where spatial randomness becomes an irrelevant perturbation. 

Although spatial randomness in the errors becomes irrelevant above 1D, there are still large inhomogeneities in the quality of the random code due to random circuit fluctuations.  
Using a simple block model, we showed that the effects of code randomness in $D>1$ can be mitigated through  correlated coding and the use of additional ancilla qubits that effectively reduce the rate of the code.  
An alternative strategy, that works better in practice, is to expurgate low-weight logical operators from the code using quantum measurements.  
With these methods, we found that good coding becomes possible  at sub-log-$N$ depths.  
Codes with rates near 1/2 generated by our random coding algorithms can achieve high performance at depth 4--8 in 2D for large erasure rates and block sizes of thousands of qubits.

The results in this work open up many directions for future research. 
To  develop these codes for use on near-term devices, a more general theory of optimal decoding for Pauli error channels should be developed. 
 Efficient optimal decoding can likely be implemented for these low-depth codes by taking advantage of their strongly local nature.  
 For example, a brute force method is sufficient in the block encoding model with logarithmic block sizes.  
 It will also be interesting to consider the performance of these codes in conventional threshold theorems, including strategies for achieving full fault-tolerance, e.g., as can always be achieved with concatenation.  
 
Another promising avenue of research is to further develop the expurgation algorithm, which we used to significantly reduce the required depth to achieve successful decoding of erasure errors.  
It has now been well established that fault-tolerant thresholds can be significantly improved by tailoring codes to the detailed properties of the noise \cite{Tuckett18,Tuckett19,Tuckett20,BonillaAtaides20}.  
The expurgation algorithm provides a wide variety of additional techniques to tailor codes to specific noise models. 
In addition, it may be possible to further improve the expurgation by using quantum measurements that explicitly implement entanglement swapping, similar to techniques used for the measurement based preparation of the surface code states \cite{Raussendorf05,Han07}.

As mentioned in the introduction, developing more concrete connections between the results here and measurement-induced phase transitions is also promising to explore. 
 Unitary-measurement models that include both errors and active error correction may realize a different universality class of these transitions that might be more resilient in near-term quantum computing devices.

\begin{acknowledgements}
We thank Steve Girvin, Pradeep Niroula, and Sarang Gopalakrishnan for helpful discussions. M.J.G and D.A.H were supported in part by the DARPA Driven and Nonequilibrium Quantum Systems (DRINQS) program. L.J acknowledges support from the ARO (W911NF-18-1-0020, W911NF-18-1-0212), ARO MURI (W911NF-16-1-0349), AFOSR MURI (FA9550-19-1-0399), DOE (DE-SC0019406), NSF (EFMA-1640959, OMA-1936118, EEC-1941583), and the Packard Foundation (2013-39273).
\end{acknowledgements}

\appendix

\section{Max-average fidelity for random stabilizer erasure threshold}
\label{app:favg}

In this appendix, we prove that the  max-average fidelity  converges to the  perfect recovery probability  for the random stabilizer erasure threshold in the thermodynamic limit.

For an initial random pure state $\ket{0} \ket{\psi}$ on the unencoded logical qubits at the $k$ sites $i = n_s+1,\ldots,N$ for $n_s = N-k$, the probability of successful error correction following the encoding by the Clifford unitary $U$, erasure at sites $\bm{e}$, syndrome measurements with outcome $\bm{s}$, and maximum-likelihood recovery is given by
\be \label{eqn:p1}
\begin{split}
 {\mathbb{P}}(R|U &,\psi,\bm{s},\bm{e}) \mathbb{P}(\bm{s}|U,\psi,\bm{e}) = \\
 \bra{0}  \bra{\psi}  &U^\dag R_{{\bm{s}\bm{e}}}  \prod_{i=1}^{n_s} P_i^{s_i} \trace_{\bm{e}} [U\ket{0}  \ket{\psi} \bra{0}  \bra{\psi}  U^\dag ] \otimes \frac{\mathbb{I}_{\bm{e}}}{2^{n_e}} \\
& \times \prod_{i=1}^{n_S} P_i^{s_i} R_{{\bm{s}\bm{e}}}^\dag U \ket{0}  \ket{\psi} \sum_{\bm{s}_{\bm{e}}} \delta_{\bm{s}  \bm{s}_{\bm{e}}},
 \end{split}
\ee
where $P_i^{s_i} = (\mathbb{I} - (-1)^{s_i} U Z_i U^\dag)/2$ is a syndrome projector for sites $i = 1,\ldots,n_s$ and $R_{\bm{s}\bm{e}}$ is the conditional recovery operator.   
We include a sum over Kroenecker delta functions  $\delta_{\bm{s}   \bm{s}_{\bm{e}} }$, where $\{ \bm{s}_{\bm{e}} \}$ are the set of possible syndrome outcomes for $\bm{e}$.  
This term is nonzero only when the observed syndrome is allowed for a given $\bm{e}$, thus, it serves as a projector onto the space of allowed syndromes.   
The maximum possible size of $\{ \bm{s}_{\bm{e}} \}$ is $2^{2 n_e}$ as this is the number of Pauli group elements with support only on $\bm{e}$ (modulo a phase).  

The precise form of $R_{\bm{s}\bm{e}}$ depends on the encoding circuit $U$ in addition to $\bm{s}$ and $\bm{e}$, therefore, it cannot be calculated in general without completely specifying  $U$.  
On the other hand, we can use the fact that it can be moved past the syndrome projectors by turning this into a projector onto the perfect syndrome outcome.  
Since it has its support entirely on $\bm{e}$, we can then cancel the product of the two recovery operators to arrive at the much simpler formula
\be 
\begin{split}
 {\mathbb{P}}(R&|U,\psi,\bm{s},\bm{e}) \mathbb{P}(\bm{s}|U,\psi,\bm{e})  = \sum_{\bm{s}_{\bm{e}} }  \delta_{\bm{s}  \bm{s}_{\bm{e}}} \\
\times &\bra{0}  \bra{\psi}  U^{\dag}  \trace_{\bm{e}} [U \ket{0}  \ket{\psi}   \bra{0}  \bra{\psi}   
 U^\dag ]  \otimes \frac{\mathbb{I}_{\bm{e}}}{2^{n_e}} U \ket{0}  \ket{\psi}   .
\end{split}
\ee
Summing over syndrome measurements gives the recovery probability 
\begin{align} \label{eqn:pru} 
\mathbb{P}(R|U,\psi,\bm{e}) & = \frac{1}{2^{2 n_e - n_{se}}} +  \frac{1}{2^{2 n_E - n_{se} }} \\ \nonumber
& \times \sum_{P_{\bm{e}} \ne \mathbb{I}} |   \bra{0}  \bra{\psi}  U^\dag P_{\bm{e}} U \ket{0}  \ket{\psi}  |^2, \\
n_{se} & = \log_2 | \{ \bm{s}_{\bm{e}} \}| \le 2 n_e.
\end{align}
Here, we used the identity 
\be
\trace_{\bm{e}}[\rho]\otimes \frac{\mathbb{I}_{\bm{e}}}{2^{n_e}}  = \frac{1}{2^{2n_e}} \sum_{P_{\bm{e}}} P_{\bm{e}} \rho P_{\bm{e}}^\dag, 
\ee
where $P_{\bm{e}}$ runs over a basis of Pauli group elements that act on sites in the subset ${\bm{e}}$.
Since the Clifford group forms a two-design, we have the  identity from random matrix theory
\be \label{eqn:rmt}
\begin{split}
\mathbb{E}_{U }& U_{a'a} U^*_{b'b}  U_{c'c} U^*_{d'd} = \mathbb{E}_{\mu } U_{a'a} U^*_{b'b}  U_{c'c} U^*_{d'd}\\
& \frac{1}{4^N-1}\bigg[\delta_{a'b'}\delta_{c'd'}\delta_{ab} \delta_{cd}  + \delta_{a'd'} \delta_{b'c'} \delta_{ad} \delta_{bc} \\
&- \frac{1}{2^N}(\delta_{ab} \delta_{cd} \delta_{a'd'} \delta_{b' c'} + \delta_{a'b'}\delta_{c'd'} \delta_{ad} \delta_{bc}) \bigg],
\end{split}
 \ee
 where $\mathbb{E}_{\mu }$ is an average over the Haar measure on the unitary group on $N$ qubits.
 This formula can be used to bound the average of the second term in Eq.~(\ref{eqn:pru}) as 
\begin{align}
\mathbb{E}_{U,\psi} &\bigg[\frac{1}{2^{2 n_e - n_{se} }}  \sum_{P_{\bm{e}} \ne \mathbb{I}} |  \bra{0}\bra{\psi} U^\dag P_{\bm{e}} U \ket{0}  \ket{\psi} |^2 \bigg] \\
& \le \frac{1}{2^{2 n_e - n_{se}^{m} }}  \sum_{P_{\bm{e}} \ne \mathbb{I}} \mathbb{E}_{\mu,\psi} | \bra{0}\bra{\psi}  U^\dag P_{\bm{e}} U \ket{0}  \ket{\psi} |^2 ,\\ \label{eqn:a9}
&=\frac{1}{2^{2 n_e - n_{se}^{m} }}  \sum_{P_{\bm{e}} \ne \mathbb{I}} \mathbb{E}_{\mu} | \bra{0}\bra{0}  U^\dag P_{\bm{e}} U \ket{0}  \ket{0} |^2 \\
& = \frac{(2^{2 n_e}-1)(2^N -1)}{2^{2 n_e - n_{se}^{m }}(4^N -1)} = O(2^{-N + n_{se}^{m}}),
\end{align}
where $n_{se}^{m} = {\min} (n_{s},2 n_e)$ and in Eq.~(\ref{eqn:a9}) we used the fact that $U \ket{\psi} = U U_{\psi} \ket{0}$ for $U_{\psi}$ distributed according to the Haar measure.
As a result, up to corrections that decay exponentially with $N$ for any erasure rate $e < 1/2$, we find the formula for the code-averaged max-average gate fidelity
\begin{align}
\bar{F}_{\rm avg} &=  \mathbb{E}_{U,\psi}  [ \mathbb{P}(R|U,\psi,\bm{e})]  = 2^{\bar{n}_{se}-2 n_e}, \\
2^{\bar{n}_{se}}& =  \mathbb{E}_U [2^{n_{se}}].
\end{align}
This expression for $\bar{F}_{\rm avg}$ is equal to $\bar{\mathbb{P}}(R)$ from Eq.~(\ref{eqn:pbarr}).

\section{Counting rank-$m$ matrices over $\mathbb{F}_2$}
\label{app:combinatorics}

In this appendix, we reproduce the standard formula for the number of rank-$m$ $2 n_e \times n_s$ matrices over $\mathbb{F}_2$.
  To find the formula, we first use the fact the number of $m \times n_s$ matrices of rank $m$ is given by
\be \label{eqn:1}
\prod_{k=0}^{m-1} (2^{n_s} - 2^k) = (2^{n_s} - 1) (2^{n_s} - 2) \cdots (2^{n_s} - 2^{m-1})
\ee
because we have $2^{n_s}-1$ choices for the first row and $2^{n_s}- 2^{i-1}$ choices for row $i$ to ensure that it is linearly independent from the first $i-1$ rows.  
When $2 n_e > m$, then we have to account for  linear dependence between rows of the matrix, which leads to a degeneracy that is equal to the number of $m$-dimensional subspaces of a $2 n_e$ dimensional vector space over $\mathbb{F}_2$
\be \label{eqn:2}
\frac{\prod_{k=0}^{m-1}(2^{2 n_e} - 2^k)}{\prod_{k=0}^{m-1} (2^{m} - 2^k)}.
\ee
Here, the numerator counts the total number of bases of an $m$-dimensional subspace of a vector space of dimension $2 n_e$ and the denominator counts the number of bases for each subspace of dimension $m$.  
The number of $2 n_e \times n_S$ matrices of rank $m$ is given by the product of Eq.~(\ref{eqn:1}) and Eq.~(\ref{eqn:2})
\be
\frac{\prod_{k=0}^{m-1}(2^{2 n_e} - 2^k)  (2^{n_s} - 2^k)}{\prod_{k=0}^{m-1} (2^{m} - 2^k)}.
\ee

\section{Independent-identically distributed erasure errors}
\label{app:iid}
    
    In this appendix, we derive the leading order RMT solution for the recovery probability for independent-identically distributed (iid) erasure errors with error rate $e$.
As noted in Sec.~\ref{sec:results}, the failure probability for iid  errors  can be obtained from the failure probability for the fixed-fraction model with $n_e = e N$.  
After averaging over all possible $n_e$, there is additional rounding of the transition due to Poisson fluctuations in the total number of erasures.     
To evaluate the associated  finite-size scaling, we make use of the fact that the total number of erasures is an extensive variable whose fluctuations are governed by the central limit theorem
\be
n_e = e N + \Delta,~ \Delta \sim \mathcal{N}(0, \sigma^2),~\sigma = \sqrt{e(1-e)N}
\ee
In averaging over the erasure errors, we can ignore the critical region for fixed $n_e$ because it  has a width $\sim 1$ that is much less than the typical fluctuations in $n_e \sim \sqrt{N}$
\be
\mean{\log_2  \mathbb{P}(F)}_{\bm{e}} \approx  \int_{\Delta_0}^{\infty} d\Delta \frac{  \exp ( -\Delta^2/2 \sigma^2) }{\sqrt{2\pi }\, \sigma }  2 (   \Delta_0 - \Delta) ,
\ee
where  $e_c = (1-R)/2$ and $\Delta_0 = (e-e_c) N$. 
After introducing the scaling variable $x =(e - e_c)\sqrt{N} /\sqrt{e(1-e)}$, we find 
\begin{align}
-\mean{\log_2 & {\mathbb{P}(F)}}_{\bm{e}}   = \sqrt{N}  f(x,e) , \\
f(x,e)  &=  \sqrt{e (1-e)} \bigg[ \frac{\exp(-x^2 /2 )}{\sqrt{ \pi/2}} -x\, \mathrm{erfc}\Big(\frac{x}{\sqrt{2}}\Big)\bigg]  ,
\end{align}
where $\mathrm{erfc}(\cdot)$ is the complementary error function and $f(x,e)$ is the scaling function for this random code transition.  
At the critical point, $f(0,e_c) =  \sqrt{2 e_c(1-e_c)/\pi}$.   
This analysis implies that the critical region after averaging over $n_e$ has a width scaling as $|e-e_c| \sim 1/\sqrt{N}$ that arises from the width of the probability distribution of $n_e$.  
Similarly, the average log-failure probability at the critical erasure rate $e_c$  scales as $\sqrt{N}$. 

\section{Self-averaging of random stabilizer code transition}
\label{app:selfavg}

\begin{figure}[tb]
\begin{center}
\includegraphics[width = 0.42 \textwidth]{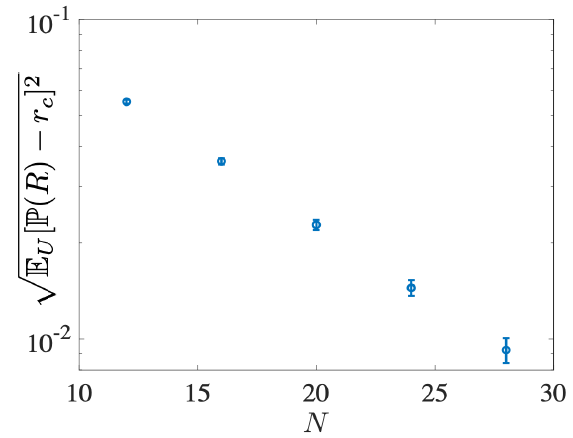}
\caption{Fluctuations in the recovery probability over random codes vs. $N$ for the fixed-fraction erasure  model with $R=1/2$ at $e=e_c$.  
The recovery probability for a random code appears to be self-averaging towards the RMT prediction $r_c$ at large $N$.  
We took a depth $2N$ encoding circuit in 1D with a brickwork arrangement of gates.  
Each two-site gate in the circuit was a random Clifford gate. }
\label{fig:selfavg}
\end{center}
\end{figure}

One of the central assumptions in this work is that the finite-size scaling behavior of random stabilizer codes near threshold well approximates the behavior of the optimal codes.  
A necessary condition for this to be true is that the random codes are self-averaging in the sense that a single realization of a random code has the same properties as the average over codes in the large $N$ limit.  
To test this self-averaging condition, we investigate the convergence with $N$ towards the RMT prediction for the critical recovery probability $r_c$ for single realizations.  
 Numerical Monte Carlo results for the standard deviation are shown in Fig.~\ref{fig:selfavg}.  
 We fix a random Clifford unitary $U$ generated by a high-depth circuit (depth $2N$ in 1D).  
 For that circuit, we then estimate $\mathbb{P}(R)$ at the critical point of the optimal codes for the fixed-fraction erasure  model.   
 By generating many codes, we can then estimate the variance $\mathbb{E}_U [\mathbb{P}(R)-r_c]^2$ through sampling.  
 Over the range of sizes shown in the figure, we see clear exponential decay of the standard deviation with $N$, indicating that $\mathbb{P}(R)$ self-averages to the RMT prediction $r_c$ in the large $N$ limit.
\\

\bibliographystyle{apsrev-nourl-title}
\bibliography{RandCodes}

\begin{thebibliography}{144}
\expandafter\ifx\csname natexlab\endcsname\relax\def\natexlab#1{#1}\fi
\expandafter\ifx\csname bibnamefont\endcsname\relax
  \def\bibnamefont#1{#1}\fi
\expandafter\ifx\csname bibfnamefont\endcsname\relax
  \def\bibfnamefont#1{#1}\fi
\expandafter\ifx\csname citenamefont\endcsname\relax
  \def\citenamefont#1{#1}\fi
\expandafter\ifx\csname url\endcsname\relax
  \def\url#1{\texttt{#1}}\fi
\expandafter\ifx\csname urlprefix\endcsname\relax\def\urlprefix{URL }\fi
\providecommand{\bibinfo}[2]{#2}
\providecommand{\eprint}[2][]{\url{#2}}

\bibitem[{\citenamefont{Arute et~al.}(2019)\citenamefont{Arute, Arya, Babbush,
  Bacon, Bardin, Barends, Biswas, Boixo, Brandao, Buell et~al.}}]{Arute19}
\bibinfo{author}{\bibfnamefont{F.}~\bibnamefont{Arute}},
  \bibinfo{author}{\bibfnamefont{K.}~\bibnamefont{Arya}},
  \bibinfo{author}{\bibfnamefont{R.}~\bibnamefont{Babbush}},
  \bibinfo{author}{\bibfnamefont{D.}~\bibnamefont{Bacon}},
  \bibinfo{author}{\bibfnamefont{J.~C.} \bibnamefont{Bardin}},
  \bibinfo{author}{\bibfnamefont{R.}~\bibnamefont{Barends}},
  \bibinfo{author}{\bibfnamefont{R.}~\bibnamefont{Biswas}},
  \bibinfo{author}{\bibfnamefont{S.}~\bibnamefont{Boixo}},
  \bibinfo{author}{\bibfnamefont{F.~G. S.~L.} \bibnamefont{Brandao}},
  \bibinfo{author}{\bibfnamefont{D.~A.} \bibnamefont{Buell}},
  \bibnamefont{et~al.}, \emph{\bibinfo{title}{{Quantum supremacy using a
  programmable superconducting processor}}}, \bibinfo{journal}{Nature}
  \textbf{\bibinfo{volume}{574}}, \bibinfo{pages}{505} (\bibinfo{year}{2019}).

\bibitem[{\citenamefont{Gaebler et~al.}(2016)\citenamefont{Gaebler, Tan, Lin,
  Wan, Bowler, Keith, Glancy, Coakley, Knill, Leibfried et~al.}}]{Gaebler16}
\bibinfo{author}{\bibfnamefont{J.~P.} \bibnamefont{Gaebler}},
  \bibinfo{author}{\bibfnamefont{T.~R.} \bibnamefont{Tan}},
  \bibinfo{author}{\bibfnamefont{Y.}~\bibnamefont{Lin}},
  \bibinfo{author}{\bibfnamefont{Y.}~\bibnamefont{Wan}},
  \bibinfo{author}{\bibfnamefont{R.}~\bibnamefont{Bowler}},
  \bibinfo{author}{\bibfnamefont{A.~C.} \bibnamefont{Keith}},
  \bibinfo{author}{\bibfnamefont{S.}~\bibnamefont{Glancy}},
  \bibinfo{author}{\bibfnamefont{K.}~\bibnamefont{Coakley}},
  \bibinfo{author}{\bibfnamefont{E.}~\bibnamefont{Knill}},
  \bibinfo{author}{\bibfnamefont{D.}~\bibnamefont{Leibfried}},
  \bibnamefont{et~al.}, \emph{\bibinfo{title}{High-fidelity universal gate set
  for ${^{9}\mathrm{Be}}^{+}$ ion qubits}}, \bibinfo{journal}{Phys. Rev. Lett.}
  \textbf{\bibinfo{volume}{117}}, \bibinfo{pages}{060505}
  (\bibinfo{year}{2016}).

\bibitem[{\citenamefont{Debnath et~al.}(2016)\citenamefont{Debnath, Linke,
  Figgatt, Landsman, Wright, and Monroe}}]{Debnath16}
\bibinfo{author}{\bibfnamefont{S.}~\bibnamefont{Debnath}},
  \bibinfo{author}{\bibfnamefont{N.~M.} \bibnamefont{Linke}},
  \bibinfo{author}{\bibfnamefont{C.}~\bibnamefont{Figgatt}},
  \bibinfo{author}{\bibfnamefont{K.~A.} \bibnamefont{Landsman}},
  \bibinfo{author}{\bibfnamefont{K.}~\bibnamefont{Wright}}, \bibnamefont{and}
  \bibinfo{author}{\bibfnamefont{C.}~\bibnamefont{Monroe}},
  \emph{\bibinfo{title}{{Demonstration of a small programmable quantum computer
  with atomic qubits}}}, \bibinfo{journal}{Nature}
  \textbf{\bibinfo{volume}{536}}, \bibinfo{pages}{63} (\bibinfo{year}{2016}).

\bibitem[{\citenamefont{C{\'o}rcoles et~al.}(2015)\citenamefont{C{\'o}rcoles,
  Magesan, Srinivasan, Cross, Steffen, Gambetta, and Chow}}]{Corcoles15}
\bibinfo{author}{\bibfnamefont{A.~D.} \bibnamefont{C{\'o}rcoles}},
  \bibinfo{author}{\bibfnamefont{E.}~\bibnamefont{Magesan}},
  \bibinfo{author}{\bibfnamefont{S.~J.} \bibnamefont{Srinivasan}},
  \bibinfo{author}{\bibfnamefont{A.~W.} \bibnamefont{Cross}},
  \bibinfo{author}{\bibfnamefont{M.}~\bibnamefont{Steffen}},
  \bibinfo{author}{\bibfnamefont{J.~M.} \bibnamefont{Gambetta}},
  \bibnamefont{and} \bibinfo{author}{\bibfnamefont{J.~M.} \bibnamefont{Chow}},
  \emph{\bibinfo{title}{{Demonstration of a quantum error detection code using
  a square lattice of four superconducting qubits}}}, \bibinfo{journal}{Nature
  Communications} \textbf{\bibinfo{volume}{6}}, \bibinfo{pages}{2152}
  (\bibinfo{year}{2015}).

\bibitem[{\citenamefont{Ofek et~al.}(2016)\citenamefont{Ofek, Petrenko, Heeres,
  Reinhold, Leghtas, Vlastakis, Liu, Frunzio, Girvin, Jiang et~al.}}]{Ofek16}
\bibinfo{author}{\bibfnamefont{N.}~\bibnamefont{Ofek}},
  \bibinfo{author}{\bibfnamefont{A.}~\bibnamefont{Petrenko}},
  \bibinfo{author}{\bibfnamefont{R.}~\bibnamefont{Heeres}},
  \bibinfo{author}{\bibfnamefont{P.}~\bibnamefont{Reinhold}},
  \bibinfo{author}{\bibfnamefont{Z.}~\bibnamefont{Leghtas}},
  \bibinfo{author}{\bibfnamefont{B.}~\bibnamefont{Vlastakis}},
  \bibinfo{author}{\bibfnamefont{Y.}~\bibnamefont{Liu}},
  \bibinfo{author}{\bibfnamefont{L.}~\bibnamefont{Frunzio}},
  \bibinfo{author}{\bibfnamefont{S.~M.} \bibnamefont{Girvin}},
  \bibinfo{author}{\bibfnamefont{L.}~\bibnamefont{Jiang}},
  \bibnamefont{et~al.}, \emph{\bibinfo{title}{{Extending the lifetime of a
  quantum bit with error correction in superconducting circuits}}},
  \bibinfo{journal}{Nature} \textbf{\bibinfo{volume}{536}},
  \bibinfo{pages}{441} (\bibinfo{year}{2016}).

\bibitem[{\citenamefont{Neill et~al.}(2018)\citenamefont{Neill, Roushan,
  Kechedzhi, Boixo, Isakov, Smelyanskiy, Megrant, Chiaro, Dunsworth, Arya
  et~al.}}]{Neill18}
\bibinfo{author}{\bibfnamefont{C.}~\bibnamefont{Neill}},
  \bibinfo{author}{\bibfnamefont{P.}~\bibnamefont{Roushan}},
  \bibinfo{author}{\bibfnamefont{K.}~\bibnamefont{Kechedzhi}},
  \bibinfo{author}{\bibfnamefont{S.}~\bibnamefont{Boixo}},
  \bibinfo{author}{\bibfnamefont{S.~V.} \bibnamefont{Isakov}},
  \bibinfo{author}{\bibfnamefont{V.}~\bibnamefont{Smelyanskiy}},
  \bibinfo{author}{\bibfnamefont{A.}~\bibnamefont{Megrant}},
  \bibinfo{author}{\bibfnamefont{B.}~\bibnamefont{Chiaro}},
  \bibinfo{author}{\bibfnamefont{A.}~\bibnamefont{Dunsworth}},
  \bibinfo{author}{\bibfnamefont{K.}~\bibnamefont{Arya}}, \bibnamefont{et~al.},
  \emph{\bibinfo{title}{{A blueprint for demonstrating quantum supremacy with
  superconducting qubits}}}, \bibinfo{journal}{Science}
  \textbf{\bibinfo{volume}{360}}, \bibinfo{pages}{195} (\bibinfo{year}{2018}).

\bibitem[{\citenamefont{Mi et~al.}(2018)\citenamefont{Mi, Benito, Putz, Zajac,
  Taylor, Burkard, and Petta}}]{Mi18}
\bibinfo{author}{\bibfnamefont{X.}~\bibnamefont{Mi}},
  \bibinfo{author}{\bibfnamefont{M.}~\bibnamefont{Benito}},
  \bibinfo{author}{\bibfnamefont{S.}~\bibnamefont{Putz}},
  \bibinfo{author}{\bibfnamefont{D.~M.} \bibnamefont{Zajac}},
  \bibinfo{author}{\bibfnamefont{J.~M.} \bibnamefont{Taylor}},
  \bibinfo{author}{\bibfnamefont{G.}~\bibnamefont{Burkard}}, \bibnamefont{and}
  \bibinfo{author}{\bibfnamefont{J.~R.} \bibnamefont{Petta}},
  \emph{\bibinfo{title}{{A coherent spin{\textendash}photon interface in
  silicon}}}, \bibinfo{journal}{Nature} \textbf{\bibinfo{volume}{555}},
  \bibinfo{pages}{599} (\bibinfo{year}{2018}).

\bibitem[{\citenamefont{Huang et~al.}(2019)\citenamefont{Huang, Yang, Chan,
  Tanttu, and Hensen}}]{Huang19}
\bibinfo{author}{\bibfnamefont{W.}~\bibnamefont{Huang}},
  \bibinfo{author}{\bibfnamefont{C.~H.} \bibnamefont{Yang}},
  \bibinfo{author}{\bibfnamefont{K.~W.} \bibnamefont{Chan}},
  \bibinfo{author}{\bibfnamefont{T.}~\bibnamefont{Tanttu}}, \bibnamefont{and}
  \bibinfo{author}{\bibfnamefont{B.}~\bibnamefont{Hensen}},
  \emph{\bibinfo{title}{{Fidelity benchmarks for two-qubit gates in silicon}}},
  \bibinfo{journal}{Nature} \textbf{\bibinfo{volume}{569}},
  \bibinfo{pages}{532} (\bibinfo{year}{2019}).

\bibitem[{\citenamefont{Levine et~al.}(2019)\citenamefont{Levine, Keesling,
  Semeghini, Omran, Wang, Ebadi, Bernien, Greiner, Vuleti{\'c}, Pichler
  et~al.}}]{Levine19}
\bibinfo{author}{\bibfnamefont{H.}~\bibnamefont{Levine}},
  \bibinfo{author}{\bibfnamefont{A.}~\bibnamefont{Keesling}},
  \bibinfo{author}{\bibfnamefont{G.}~\bibnamefont{Semeghini}},
  \bibinfo{author}{\bibfnamefont{A.}~\bibnamefont{Omran}},
  \bibinfo{author}{\bibfnamefont{T.~T.} \bibnamefont{Wang}},
  \bibinfo{author}{\bibfnamefont{S.}~\bibnamefont{Ebadi}},
  \bibinfo{author}{\bibfnamefont{H.}~\bibnamefont{Bernien}},
  \bibinfo{author}{\bibfnamefont{M.}~\bibnamefont{Greiner}},
  \bibinfo{author}{\bibfnamefont{V.}~\bibnamefont{Vuleti{\'c}}},
  \bibinfo{author}{\bibfnamefont{H.}~\bibnamefont{Pichler}},
  \bibnamefont{et~al.}, \emph{\bibinfo{title}{{Parallel Implementation of
  High-Fidelity Multiqubit Gates with Neutral Atoms}}}, \bibinfo{journal}{Phys.
  Rev. Lett.} \textbf{\bibinfo{volume}{123}}, \bibinfo{pages}{170503}
  (\bibinfo{year}{2019}).

\bibitem[{\citenamefont{Egan et~al.}(2020)\citenamefont{Egan, Debroy, Noel,
  Risinger, Zhu, Biswas, Newman, Li, Brown, Cetina et~al.}}]{Egan20}
\bibinfo{author}{\bibfnamefont{L.}~\bibnamefont{Egan}},
  \bibinfo{author}{\bibfnamefont{D.~M.} \bibnamefont{Debroy}},
  \bibinfo{author}{\bibfnamefont{C.}~\bibnamefont{Noel}},
  \bibinfo{author}{\bibfnamefont{A.}~\bibnamefont{Risinger}},
  \bibinfo{author}{\bibfnamefont{D.}~\bibnamefont{Zhu}},
  \bibinfo{author}{\bibfnamefont{D.}~\bibnamefont{Biswas}},
  \bibinfo{author}{\bibfnamefont{M.}~\bibnamefont{Newman}},
  \bibinfo{author}{\bibfnamefont{M.}~\bibnamefont{Li}},
  \bibinfo{author}{\bibfnamefont{K.~R.} \bibnamefont{Brown}},
  \bibinfo{author}{\bibfnamefont{M.}~\bibnamefont{Cetina}},
  \bibnamefont{et~al.}, \emph{\bibinfo{title}{{Fault-Tolerant Operation of a
  Quantum Error-Correction Code}}} (\bibinfo{year}{2020}),
  \eprint{arXiv:2009.11482}.

\bibitem[{\citenamefont{Knill}(2005)}]{Knill05}
\bibinfo{author}{\bibfnamefont{E.}~\bibnamefont{Knill}},
  \emph{\bibinfo{title}{{Quantum computing with realistically noisy devices}}},
  \bibinfo{journal}{Nature} \textbf{\bibinfo{volume}{434}}, \bibinfo{pages}{39}
  (\bibinfo{year}{2005}).

\bibitem[{\citenamefont{Bravyi and Kitaev}(2005)}]{Bravyi05}
\bibinfo{author}{\bibfnamefont{S.}~\bibnamefont{Bravyi}} \bibnamefont{and}
  \bibinfo{author}{\bibfnamefont{A.}~\bibnamefont{Kitaev}},
  \emph{\bibinfo{title}{{Universal quantum computation with ideal Clifford
  gates and noisy ancillas}}}, \bibinfo{journal}{Phys. Rev. A}
  \textbf{\bibinfo{volume}{71}}, \bibinfo{pages}{022316}
  (\bibinfo{year}{2005}).

\bibitem[{\citenamefont{Svore et~al.}(2007)\citenamefont{Svore, DiVincenzo, and
  Terhal}}]{Svore06}
\bibinfo{author}{\bibfnamefont{K.~M.} \bibnamefont{Svore}},
  \bibinfo{author}{\bibfnamefont{D.~P.} \bibnamefont{DiVincenzo}},
  \bibnamefont{and} \bibinfo{author}{\bibfnamefont{B.~M.}
  \bibnamefont{Terhal}}, \emph{\bibinfo{title}{{Noise Threshold for a
  Fault-Tolerant Two-Dimensional Lattice Architecture}}},
  \bibinfo{journal}{Quant. Inf. Comp.} \textbf{\bibinfo{volume}{7}},
  \bibinfo{pages}{297} (\bibinfo{year}{2007}).

\bibitem[{\citenamefont{Fowler et~al.}(2012)\citenamefont{Fowler, Mariantoni,
  Martinis, and Cleland}}]{Fowler12}
\bibinfo{author}{\bibfnamefont{A.~G.} \bibnamefont{Fowler}},
  \bibinfo{author}{\bibfnamefont{M.}~\bibnamefont{Mariantoni}},
  \bibinfo{author}{\bibfnamefont{J.~M.} \bibnamefont{Martinis}},
  \bibnamefont{and} \bibinfo{author}{\bibfnamefont{A.~N.}
  \bibnamefont{Cleland}}, \emph{\bibinfo{title}{Surface codes: Towards
  practical large-scale quantum computation}}, \bibinfo{journal}{Phys. Rev. A}
  \textbf{\bibinfo{volume}{86}}, \bibinfo{pages}{032324}
  (\bibinfo{year}{2012}).

\bibitem[{\citenamefont{Leghtas et~al.}(2013)\citenamefont{Leghtas, Kirchmair,
  Vlastakis, Schoelkopf, Devoret, and Mirrahimi}}]{Leghtas13}
\bibinfo{author}{\bibfnamefont{Z.}~\bibnamefont{Leghtas}},
  \bibinfo{author}{\bibfnamefont{G.}~\bibnamefont{Kirchmair}},
  \bibinfo{author}{\bibfnamefont{B.}~\bibnamefont{Vlastakis}},
  \bibinfo{author}{\bibfnamefont{R.~J.} \bibnamefont{Schoelkopf}},
  \bibinfo{author}{\bibfnamefont{M.~H.} \bibnamefont{Devoret}},
  \bibnamefont{and}
  \bibinfo{author}{\bibfnamefont{M.}~\bibnamefont{Mirrahimi}},
  \emph{\bibinfo{title}{{Hardware-Efficient Autonomous Quantum Memory
  Protection}}}, \bibinfo{journal}{Phys. Rev. Lett.}
  \textbf{\bibinfo{volume}{111}}, \bibinfo{pages}{120501}
  (\bibinfo{year}{2013}).

\bibitem[{\citenamefont{Mirrahimi et~al.}(2014)\citenamefont{Mirrahimi,
  Leghtas, Albert, Touzard, Schoelkopf, Jiang, and Devoret}}]{Mirrahimi14}
\bibinfo{author}{\bibfnamefont{M.}~\bibnamefont{Mirrahimi}},
  \bibinfo{author}{\bibfnamefont{Z.}~\bibnamefont{Leghtas}},
  \bibinfo{author}{\bibfnamefont{V.~V.} \bibnamefont{Albert}},
  \bibinfo{author}{\bibfnamefont{S.}~\bibnamefont{Touzard}},
  \bibinfo{author}{\bibfnamefont{R.~J.} \bibnamefont{Schoelkopf}},
  \bibinfo{author}{\bibfnamefont{L.}~\bibnamefont{Jiang}}, \bibnamefont{and}
  \bibinfo{author}{\bibfnamefont{M.~H.} \bibnamefont{Devoret}},
  \emph{\bibinfo{title}{{Dynamically protected cat-qubits: a new paradigm for
  universal quantum computation}}}, \bibinfo{journal}{New J. of Phys.}
  \textbf{\bibinfo{volume}{16}}, \bibinfo{pages}{045014}
  (\bibinfo{year}{2014}).

\bibitem[{\citenamefont{Vlastakis et~al.}(2013)\citenamefont{Vlastakis,
  Kirchmair, Leghtas, Nigg, Frunzio, Girvin, Mirrahimi, Devoret, and
  Schoelkopf}}]{Vlastakis13}
\bibinfo{author}{\bibfnamefont{B.}~\bibnamefont{Vlastakis}},
  \bibinfo{author}{\bibfnamefont{G.}~\bibnamefont{Kirchmair}},
  \bibinfo{author}{\bibfnamefont{Z.}~\bibnamefont{Leghtas}},
  \bibinfo{author}{\bibfnamefont{S.~E.} \bibnamefont{Nigg}},
  \bibinfo{author}{\bibfnamefont{L.}~\bibnamefont{Frunzio}},
  \bibinfo{author}{\bibfnamefont{S.~M.} \bibnamefont{Girvin}},
  \bibinfo{author}{\bibfnamefont{M.}~\bibnamefont{Mirrahimi}},
  \bibinfo{author}{\bibfnamefont{M.~H.} \bibnamefont{Devoret}},
  \bibnamefont{and} \bibinfo{author}{\bibfnamefont{R.~J.}
  \bibnamefont{Schoelkopf}}, \emph{\bibinfo{title}{{Deterministically Encoding
  Quantum Information Using 100-Photon Schr{\"o}dinger Cat States}}},
  \bibinfo{journal}{Science} \textbf{\bibinfo{volume}{342}},
  \bibinfo{pages}{607} (\bibinfo{year}{2013}).

\bibitem[{\citenamefont{Martinis}(2015)}]{Martinis15}
\bibinfo{author}{\bibfnamefont{J.~M.} \bibnamefont{Martinis}},
  \emph{\bibinfo{title}{{Qubit metrology for building a fault-tolerant quantum
  computer}}}, \bibinfo{journal}{npj Quantum Inf} \textbf{\bibinfo{volume}{1}},
  \bibinfo{pages}{1} (\bibinfo{year}{2015}).

\bibitem[{\citenamefont{Harper et~al.}(2020)\citenamefont{Harper, Flammia, and
  Wallman}}]{Harper20}
\bibinfo{author}{\bibfnamefont{R.}~\bibnamefont{Harper}},
  \bibinfo{author}{\bibfnamefont{S.~T.} \bibnamefont{Flammia}},
  \bibnamefont{and} \bibinfo{author}{\bibfnamefont{J.~J.}
  \bibnamefont{Wallman}}, \emph{\bibinfo{title}{{Efficient learning of quantum
  noise}}}, \bibinfo{journal}{Nature Phys.} \textbf{\bibinfo{volume}{332}},
  \bibinfo{pages}{1} (\bibinfo{year}{2020}).

\bibitem[{\citenamefont{Flammia and Wallman}(2020)}]{Flammia2020}
\bibinfo{author}{\bibfnamefont{S.~T.} \bibnamefont{Flammia}} \bibnamefont{and}
  \bibinfo{author}{\bibfnamefont{J.~J.} \bibnamefont{Wallman}},
  \emph{\bibinfo{title}{Efficient estimation of pauli channels}}
  (\bibinfo{year}{2020}), \eprint{arXiv:1907.12976}.

\bibitem[{\citenamefont{Aliferis and Preskill}(2008)}]{Aliferis2008}
\bibinfo{author}{\bibfnamefont{P.}~\bibnamefont{Aliferis}} \bibnamefont{and}
  \bibinfo{author}{\bibfnamefont{J.}~\bibnamefont{Preskill}},
  \emph{\bibinfo{title}{Fault-tolerant quantum computation against biased
  noise}}, \bibinfo{journal}{Phys. Rev. A} \textbf{\bibinfo{volume}{78}}
  (\bibinfo{year}{2008}).

\bibitem[{\citenamefont{Puri et~al.}(2020)\citenamefont{Puri, St-Jean, Gross,
  Grimm, Frattini, Iyer, Krishna, Touzard, Jiang, Blais et~al.}}]{Puri2020}
\bibinfo{author}{\bibfnamefont{S.}~\bibnamefont{Puri}},
  \bibinfo{author}{\bibfnamefont{L.}~\bibnamefont{St-Jean}},
  \bibinfo{author}{\bibfnamefont{J.~A.} \bibnamefont{Gross}},
  \bibinfo{author}{\bibfnamefont{A.}~\bibnamefont{Grimm}},
  \bibinfo{author}{\bibfnamefont{N.~E.} \bibnamefont{Frattini}},
  \bibinfo{author}{\bibfnamefont{P.~S.} \bibnamefont{Iyer}},
  \bibinfo{author}{\bibfnamefont{A.}~\bibnamefont{Krishna}},
  \bibinfo{author}{\bibfnamefont{S.}~\bibnamefont{Touzard}},
  \bibinfo{author}{\bibfnamefont{L.}~\bibnamefont{Jiang}},
  \bibinfo{author}{\bibfnamefont{A.}~\bibnamefont{Blais}},
  \bibnamefont{et~al.}, \emph{\bibinfo{title}{Bias-preserving gates with
  stabilized cat qubits}}, \bibinfo{journal}{Science Adv.}
  \textbf{\bibinfo{volume}{6}}, \bibinfo{pages}{eaay5901}
  (\bibinfo{year}{2020}).

\bibitem[{\citenamefont{Guillaud and Mirrahimi}(2019)}]{Guillaud2019}
\bibinfo{author}{\bibfnamefont{J.}~\bibnamefont{Guillaud}} \bibnamefont{and}
  \bibinfo{author}{\bibfnamefont{M.}~\bibnamefont{Mirrahimi}},
  \emph{\bibinfo{title}{Repetition cat qubits for fault-tolerant quantum
  computation}}, \bibinfo{journal}{Phys. Rev. X} \textbf{\bibinfo{volume}{9}}
  (\bibinfo{year}{2019}).

\bibitem[{\citenamefont{Tillich and Zemor}(2009)}]{Tillich09}
\bibinfo{author}{\bibfnamefont{J.-P.} \bibnamefont{Tillich}} \bibnamefont{and}
  \bibinfo{author}{\bibfnamefont{G.}~\bibnamefont{Zemor}},
  \emph{\bibinfo{title}{{Quantum LDPC codes with positive rate and minimum
  distance proportional to $n^{1/2}$}}} (\bibinfo{year}{2009}),
  \eprint{arXiv:0903.0566}.

\bibitem[{\citenamefont{Hastings}(2013)}]{Hastings13}
\bibinfo{author}{\bibfnamefont{M.~B.} \bibnamefont{Hastings}},
  \emph{\bibinfo{title}{{Decoding in Hyperbolic Spaces: LDPC Codes With Linear
  Rate and Efficient Error Correction}}} (\bibinfo{year}{2013}),
  \eprint{arXiv:1312.2546}.

\bibitem[{\citenamefont{Gottesman}(2013)}]{Gottesman13}
\bibinfo{author}{\bibfnamefont{D.}~\bibnamefont{Gottesman}},
  \emph{\bibinfo{title}{{Fault-Tolerant Quantum Computation with Constant
  Overhead}}} (\bibinfo{year}{2013}), \eprint{arXiv:1310.2984}.

\bibitem[{\citenamefont{Tuckett et~al.}(2018)\citenamefont{Tuckett, Bartlett,
  and Flammia}}]{Tuckett18}
\bibinfo{author}{\bibfnamefont{D.~K.} \bibnamefont{Tuckett}},
  \bibinfo{author}{\bibfnamefont{S.~D.} \bibnamefont{Bartlett}},
  \bibnamefont{and} \bibinfo{author}{\bibfnamefont{S.~T.}
  \bibnamefont{Flammia}}, \emph{\bibinfo{title}{Ultrahigh error threshold for
  surface codes with biased noise}}, \bibinfo{journal}{Phys. Rev. Lett.}
  \textbf{\bibinfo{volume}{120}}, \bibinfo{pages}{050505}
  (\bibinfo{year}{2018}).

\bibitem[{\citenamefont{Tuckett et~al.}(2019)\citenamefont{Tuckett, Darmawan,
  Chubb, Bravyi, Bartlett, and Flammia}}]{Tuckett19}
\bibinfo{author}{\bibfnamefont{D.~K.} \bibnamefont{Tuckett}},
  \bibinfo{author}{\bibfnamefont{A.~S.} \bibnamefont{Darmawan}},
  \bibinfo{author}{\bibfnamefont{C.~T.} \bibnamefont{Chubb}},
  \bibinfo{author}{\bibfnamefont{S.}~\bibnamefont{Bravyi}},
  \bibinfo{author}{\bibfnamefont{S.~D.} \bibnamefont{Bartlett}},
  \bibnamefont{and} \bibinfo{author}{\bibfnamefont{S.~T.}
  \bibnamefont{Flammia}}, \emph{\bibinfo{title}{Tailoring surface codes for
  highly biased noise}}, \bibinfo{journal}{Phys. Rev. X}
  \textbf{\bibinfo{volume}{9}}, \bibinfo{pages}{041031} (\bibinfo{year}{2019}).

\bibitem[{\citenamefont{Tuckett et~al.}(2020)\citenamefont{Tuckett, Bartlett,
  Flammia, and Brown}}]{Tuckett20}
\bibinfo{author}{\bibfnamefont{D.~K.} \bibnamefont{Tuckett}},
  \bibinfo{author}{\bibfnamefont{S.~D.} \bibnamefont{Bartlett}},
  \bibinfo{author}{\bibfnamefont{S.~T.} \bibnamefont{Flammia}},
  \bibnamefont{and} \bibinfo{author}{\bibfnamefont{B.~J.} \bibnamefont{Brown}},
  \emph{\bibinfo{title}{Fault-tolerant thresholds for the surface code in
  excess of $5\%$ under biased noise}}, \bibinfo{journal}{Phys. Rev. Lett.}
  \textbf{\bibinfo{volume}{124}}, \bibinfo{pages}{130501}
  (\bibinfo{year}{2020}).

\bibitem[{\citenamefont{\surname{Bonilla Ataides}
  et~al.}(2020)\citenamefont{\surname{Bonilla Ataides}, Tuckett, Bartlett,
  Flammia, and Brown}}]{BonillaAtaides20}
\bibinfo{author}{\bibfnamefont{J.~P.} \bibnamefont{\surname{Bonilla Ataides}}},
  \bibinfo{author}{\bibfnamefont{D.~K.} \bibnamefont{Tuckett}},
  \bibinfo{author}{\bibfnamefont{S.~D.} \bibnamefont{Bartlett}},
  \bibinfo{author}{\bibfnamefont{S.~T.} \bibnamefont{Flammia}},
  \bibnamefont{and} \bibinfo{author}{\bibfnamefont{B.~J.} \bibnamefont{Brown}},
  \emph{\bibinfo{title}{The {XZZX} surface code}} (\bibinfo{year}{2020}),
  \eprint{arXiv:2009.07851}.

\bibitem[{Sha()}]{Shannon48}
\bibinfo{note}{C. E. Shannon, \textit{A Mathematical Theory of Communication},
  Bell System Technical Journal. 27, 379 (1948)}.

\bibitem[{\citenamefont{Gallager}(1962)}]{Gallager62}
\bibinfo{author}{\bibfnamefont{R.~G.} \bibnamefont{Gallager}},
  \emph{\bibinfo{title}{Low-density parity-check codes}}, \bibinfo{journal}{IRE
  Trans. Info. Theory} \textbf{\bibinfo{volume}{8}}, \bibinfo{pages}{21}
  (\bibinfo{year}{1962}).

\bibitem[{\citenamefont{Gallager}(1973)}]{Gallager73}
\bibinfo{author}{\bibfnamefont{R.}~\bibnamefont{Gallager}},
  \emph{\bibinfo{title}{The random coding bound is tight for the average
  code}}, \bibinfo{journal}{IEEE Trans. Inform. Theory}
  \textbf{\bibinfo{volume}{IT-19}}, \bibinfo{pages}{244}
  (\bibinfo{year}{1973}).

\bibitem[{\citenamefont{Brown and Fawzi}(2015)}]{Brown12}
\bibinfo{author}{\bibfnamefont{W.}~\bibnamefont{Brown}} \bibnamefont{and}
  \bibinfo{author}{\bibfnamefont{O.}~\bibnamefont{Fawzi}},
  \emph{\bibinfo{title}{{Decoupling with Random Quantum Circuits}}},
  \bibinfo{journal}{Commun. Math. Phys.} \textbf{\bibinfo{volume}{340}},
  \bibinfo{pages}{867} (\bibinfo{year}{2015}), \eprint{arXiv:1210.6644}.

\bibitem[{\citenamefont{Brown and Fawzi}(2013)}]{Brown13}
\bibinfo{author}{\bibfnamefont{W.}~\bibnamefont{Brown}} \bibnamefont{and}
  \bibinfo{author}{\bibfnamefont{O.}~\bibnamefont{Fawzi}},
  \emph{\bibinfo{title}{{Short random circuits define good quantum error
  correcting codes}}}, \bibinfo{journal}{Proceedings of ISIT, pages 346 - 350}
  (\bibinfo{year}{2013}), \eprint{arXiv:1312.7646}.

\bibitem[{\citenamefont{Cleve et~al.}(2016)\citenamefont{Cleve, Leung, Liu, and
  Wang}}]{Cleve15}
\bibinfo{author}{\bibfnamefont{R.}~\bibnamefont{Cleve}},
  \bibinfo{author}{\bibfnamefont{D.}~\bibnamefont{Leung}},
  \bibinfo{author}{\bibfnamefont{L.}~\bibnamefont{Liu}}, \bibnamefont{and}
  \bibinfo{author}{\bibfnamefont{C.}~\bibnamefont{Wang}},
  \emph{\bibinfo{title}{{Near-linear constructions of exact unitary
  2-designs}}}, \bibinfo{journal}{Quantum Inf. Comp.}
  \textbf{\bibinfo{volume}{16}}, \bibinfo{pages}{0721} (\bibinfo{year}{2016}),
  \eprint{arXiv:1501.04592}.

\bibitem[{\citenamefont{Brandao et~al.}(2016)\citenamefont{Brandao, Harrow, and
  Horodecki}}]{Brandao16}
\bibinfo{author}{\bibfnamefont{F.~G. S.~L.} \bibnamefont{Brandao}},
  \bibinfo{author}{\bibfnamefont{A.~W.} \bibnamefont{Harrow}},
  \bibnamefont{and}
  \bibinfo{author}{\bibfnamefont{M.}~\bibnamefont{Horodecki}},
  \emph{\bibinfo{title}{{Local Random Quantum Circuits are Approximate
  Polynomial-Designs}}}, \bibinfo{journal}{Commun. Math. Phys.}
  \textbf{\bibinfo{volume}{346}}, \bibinfo{pages}{397} (\bibinfo{year}{2016}).

\bibitem[{\citenamefont{Harrow and Mehraban}(2018)}]{Harrow18}
\bibinfo{author}{\bibfnamefont{A.}~\bibnamefont{Harrow}} \bibnamefont{and}
  \bibinfo{author}{\bibfnamefont{S.}~\bibnamefont{Mehraban}},
  \emph{\bibinfo{title}{{Approximate unitary $t$-designs by short random
  quantum circuits using nearest-neighbor and long-range gates}}}
  (\bibinfo{year}{2018}), \eprint{arXiv:1809.06957}.

\bibitem[{\citenamefont{Delfosse et~al.}(2016)\citenamefont{Delfosse, Iyer, and
  Poulin}}]{Delfosse16}
\bibinfo{author}{\bibfnamefont{N.}~\bibnamefont{Delfosse}},
  \bibinfo{author}{\bibfnamefont{P.}~\bibnamefont{Iyer}}, \bibnamefont{and}
  \bibinfo{author}{\bibfnamefont{D.}~\bibnamefont{Poulin}},
  \emph{\bibinfo{title}{{A linear-time benchmarking tool for generalized
  surface codes}}} (\bibinfo{year}{2016}), \eprint{arXiv:1611.04256}.

\bibitem[{\citenamefont{Delfosse and Z\'emor}(2020)}]{Delfosse17}
\bibinfo{author}{\bibfnamefont{N.}~\bibnamefont{Delfosse}} \bibnamefont{and}
  \bibinfo{author}{\bibfnamefont{G.}~\bibnamefont{Z\'emor}},
  \emph{\bibinfo{title}{Linear-time maximum likelihood decoding of surface
  codes over the quantum erasure channel}}, \bibinfo{journal}{Phys. Rev.
  Research} \textbf{\bibinfo{volume}{2}}, \bibinfo{pages}{033042}
  (\bibinfo{year}{2020}), \eprint{arXiv:1703.01517}.

\bibitem[{\citenamefont{Imry and Ma}(1975)}]{Imry75}
\bibinfo{author}{\bibfnamefont{Y.}~\bibnamefont{Imry}} \bibnamefont{and}
  \bibinfo{author}{\bibfnamefont{S.-k.} \bibnamefont{Ma}},
  \emph{\bibinfo{title}{Random-field instability of the ordered state of
  continuous symmetry}}, \bibinfo{journal}{Phys. Rev. Lett.}
  \textbf{\bibinfo{volume}{35}}, \bibinfo{pages}{1399} (\bibinfo{year}{1975}).

\bibitem[{\citenamefont{Shor}(1996)}]{Shor96}
\bibinfo{author}{\bibfnamefont{P.~W.} \bibnamefont{Shor}},
  \emph{\bibinfo{title}{{Fault-tolerant quantum computation}}}
  (\bibinfo{year}{1996}), \eprint{arxiv:quant-ph/9605011}.

\bibitem[{\citenamefont{Aharonov and Ben-Or}(1999)}]{Aharonov99}
\bibinfo{author}{\bibfnamefont{D.}~\bibnamefont{Aharonov}} \bibnamefont{and}
  \bibinfo{author}{\bibfnamefont{M.}~\bibnamefont{Ben-Or}},
  \emph{\bibinfo{title}{{Fault-Tolerant Quantum Computation With Constant Error
  Rate}}} (\bibinfo{year}{1999}), \eprint{arXiv:quant-ph/9906129}.

\bibitem[{\citenamefont{Gottesman}(2000)}]{Gottesman00}
\bibinfo{author}{\bibfnamefont{D.}~\bibnamefont{Gottesman}},
  \emph{\bibinfo{title}{{Fault-tolerant quantum computation with local
  gates}}}, \bibinfo{journal}{J. Mod. Opt.} \textbf{\bibinfo{volume}{47}},
  \bibinfo{pages}{333} (\bibinfo{year}{2000}).

\bibitem[{\citenamefont{Kitaev}(1997)}]{Kitaev97}
\bibinfo{author}{\bibfnamefont{A.~Y.} \bibnamefont{Kitaev}},
  \emph{\bibinfo{title}{{Quantum computations: algorithms and error
  correction}}}, \bibinfo{journal}{Russ. Math. Surv.}
  \textbf{\bibinfo{volume}{52}}, \bibinfo{pages}{1191} (\bibinfo{year}{1997}).

\bibitem[{\citenamefont{Dennis et~al.}(2002)\citenamefont{Dennis, Kitaev,
  Landahl, and Preskill}}]{Dennis02}
\bibinfo{author}{\bibfnamefont{E.}~\bibnamefont{Dennis}},
  \bibinfo{author}{\bibfnamefont{A.}~\bibnamefont{Kitaev}},
  \bibinfo{author}{\bibfnamefont{A.}~\bibnamefont{Landahl}}, \bibnamefont{and}
  \bibinfo{author}{\bibfnamefont{J.}~\bibnamefont{Preskill}},
  \emph{\bibinfo{title}{{Topological quantum memory}}}, \bibinfo{journal}{J.
  Math. Phys.} \textbf{\bibinfo{volume}{43}}, \bibinfo{pages}{4452}
  (\bibinfo{year}{2002}).

\bibitem[{\citenamefont{Chubb and Flammia}(2018)}]{Chubb18}
\bibinfo{author}{\bibfnamefont{C.~T.} \bibnamefont{Chubb}} \bibnamefont{and}
  \bibinfo{author}{\bibfnamefont{S.~T.} \bibnamefont{Flammia}},
  \emph{\bibinfo{title}{{Statistical mechanical models for quantum codes with
  correlated noise}}} (\bibinfo{year}{2018}), \eprint{arXiv:1809.10704}.

\bibitem[{\citenamefont{Duclos-Cianci and Poulin}(2010)}]{Duclos10}
\bibinfo{author}{\bibfnamefont{G.}~\bibnamefont{Duclos-Cianci}}
  \bibnamefont{and} \bibinfo{author}{\bibfnamefont{D.}~\bibnamefont{Poulin}},
  \emph{\bibinfo{title}{Fast decoders for topological quantum codes}},
  \bibinfo{journal}{Phys. Rev. Lett.} \textbf{\bibinfo{volume}{104}},
  \bibinfo{pages}{050504} (\bibinfo{year}{2010}).

\bibitem[{\citenamefont{Wootton and Loss}(2012)}]{Wootton12}
\bibinfo{author}{\bibfnamefont{J.~R.} \bibnamefont{Wootton}} \bibnamefont{and}
  \bibinfo{author}{\bibfnamefont{D.}~\bibnamefont{Loss}},
  \emph{\bibinfo{title}{High threshold error correction for the surface code}},
  \bibinfo{journal}{Phys. Rev. Lett.} \textbf{\bibinfo{volume}{109}},
  \bibinfo{pages}{160503} (\bibinfo{year}{2012}).

\bibitem[{\citenamefont{Hutter et~al.}(2014)\citenamefont{Hutter, Wootton, and
  Loss}}]{Hutter14}
\bibinfo{author}{\bibfnamefont{A.}~\bibnamefont{Hutter}},
  \bibinfo{author}{\bibfnamefont{J.~R.} \bibnamefont{Wootton}},
  \bibnamefont{and} \bibinfo{author}{\bibfnamefont{D.}~\bibnamefont{Loss}},
  \emph{\bibinfo{title}{Efficient markov chain monte carlo algorithm for the
  surface code}}, \bibinfo{journal}{Phys. Rev. A}
  \textbf{\bibinfo{volume}{89}}, \bibinfo{pages}{022326}
  (\bibinfo{year}{2014}).

\bibitem[{\citenamefont{Delfosse and Nickerson}(2017)}]{Delfosse17b}
\bibinfo{author}{\bibfnamefont{N.}~\bibnamefont{Delfosse}} \bibnamefont{and}
  \bibinfo{author}{\bibfnamefont{N.~H.} \bibnamefont{Nickerson}},
  \emph{\bibinfo{title}{{Almost-linear time decoding algorithm for topological
  codes}}} (\bibinfo{year}{2017}), \eprint{arXiv:1709.06218}.

\bibitem[{\citenamefont{Bombin and Martin-Delgado}(2006)}]{Bombin06}
\bibinfo{author}{\bibfnamefont{H.}~\bibnamefont{Bombin}} \bibnamefont{and}
  \bibinfo{author}{\bibfnamefont{M.~A.} \bibnamefont{Martin-Delgado}},
  \emph{\bibinfo{title}{Topological quantum distillation}},
  \bibinfo{journal}{Phys. Rev. Lett.} \textbf{\bibinfo{volume}{97}},
  \bibinfo{pages}{180501} (\bibinfo{year}{2006}).

\bibitem[{\citenamefont{Bombin and Martin-Delgado}(2007)}]{Bombin07}
\bibinfo{author}{\bibfnamefont{H.}~\bibnamefont{Bombin}} \bibnamefont{and}
  \bibinfo{author}{\bibfnamefont{M.~A.} \bibnamefont{Martin-Delgado}},
  \emph{\bibinfo{title}{Topological computation without braiding}},
  \bibinfo{journal}{Phys. Rev. Lett.} \textbf{\bibinfo{volume}{98}},
  \bibinfo{pages}{160502} (\bibinfo{year}{2007}).

\bibitem[{\citenamefont{Bombin}(2011)}]{Bombin11}
\bibinfo{author}{\bibfnamefont{H.}~\bibnamefont{Bombin}},
  \emph{\bibinfo{title}{{Clifford gates by code deformation}}},
  \bibinfo{journal}{New J. Phys.} \textbf{\bibinfo{volume}{13}},
  \bibinfo{pages}{043005} (\bibinfo{year}{2011}).

\bibitem[{\citenamefont{Horsman et~al.}(2012)\citenamefont{Horsman, Fowler,
  Devitt, and Van~Meter}}]{Horsman12}
\bibinfo{author}{\bibfnamefont{C.}~\bibnamefont{Horsman}},
  \bibinfo{author}{\bibfnamefont{A.~G.} \bibnamefont{Fowler}},
  \bibinfo{author}{\bibfnamefont{S.}~\bibnamefont{Devitt}}, \bibnamefont{and}
  \bibinfo{author}{\bibfnamefont{R.}~\bibnamefont{Van~Meter}},
  \emph{\bibinfo{title}{{Surface code quantum computing by lattice surgery}}},
  \bibinfo{journal}{New J. Phys.} \textbf{\bibinfo{volume}{14}},
  \bibinfo{pages}{123011} (\bibinfo{year}{2012}).

\bibitem[{\citenamefont{Brown et~al.}(2017)\citenamefont{Brown, Laubscher,
  Kesselring, and Wootton}}]{Brown17}
\bibinfo{author}{\bibfnamefont{B.~J.} \bibnamefont{Brown}},
  \bibinfo{author}{\bibfnamefont{K.}~\bibnamefont{Laubscher}},
  \bibinfo{author}{\bibfnamefont{M.~S.} \bibnamefont{Kesselring}},
  \bibnamefont{and} \bibinfo{author}{\bibfnamefont{J.~R.}
  \bibnamefont{Wootton}}, \emph{\bibinfo{title}{Poking holes and cutting
  corners to achieve clifford gates with the surface code}},
  \bibinfo{journal}{Phys. Rev. X} \textbf{\bibinfo{volume}{7}},
  \bibinfo{pages}{021029} (\bibinfo{year}{2017}).

\bibitem[{\citenamefont{Webster and Bartlett}(2020)}]{Webster19}
\bibinfo{author}{\bibfnamefont{P.}~\bibnamefont{Webster}} \bibnamefont{and}
  \bibinfo{author}{\bibfnamefont{S.~D.} \bibnamefont{Bartlett}},
  \emph{\bibinfo{title}{Fault-tolerant quantum gates with defects in
  topological stabilizer codes}}, \bibinfo{journal}{Phys. Rev. A}
  \textbf{\bibinfo{volume}{102}}, \bibinfo{pages}{022403}
  (\bibinfo{year}{2020}).

\bibitem[{\citenamefont{Brown}(2020)}]{Brown20}
\bibinfo{author}{\bibfnamefont{B.~J.} \bibnamefont{Brown}},
  \emph{\bibinfo{title}{{A fault-tolerant non-Clifford gate for the surface
  code in two dimensions}}}, \bibinfo{journal}{Science Adv.}
  \textbf{\bibinfo{volume}{6}}, \bibinfo{pages}{eaay4929}
  (\bibinfo{year}{2020}).

\bibitem[{\citenamefont{Bravyi et~al.}(2010)\citenamefont{Bravyi, Poulin, and
  Terhal}}]{Bravyi10b}
\bibinfo{author}{\bibfnamefont{S.}~\bibnamefont{Bravyi}},
  \bibinfo{author}{\bibfnamefont{D.}~\bibnamefont{Poulin}}, \bibnamefont{and}
  \bibinfo{author}{\bibfnamefont{B.}~\bibnamefont{Terhal}},
  \emph{\bibinfo{title}{Tradeoffs for reliable quantum information storage in
  2d systems}}, \bibinfo{journal}{Phys. Rev. Lett.}
  \textbf{\bibinfo{volume}{104}}, \bibinfo{pages}{050503}
  (\bibinfo{year}{2010}).

\bibitem[{\citenamefont{Bacon}(2006)}]{Bacon06}
\bibinfo{author}{\bibfnamefont{D.}~\bibnamefont{Bacon}},
  \emph{\bibinfo{title}{Operator quantum error-correcting subsystems for
  self-correcting quantum memories}}, \bibinfo{journal}{Phys. Rev. A}
  \textbf{\bibinfo{volume}{73}}, \bibinfo{pages}{012340}
  (\bibinfo{year}{2006}).

\bibitem[{\citenamefont{Leverrier et~al.}(2015)\citenamefont{Leverrier,
  Tillich, and Zemor}}]{Leverrier15}
\bibinfo{author}{\bibfnamefont{A.}~\bibnamefont{Leverrier}},
  \bibinfo{author}{\bibfnamefont{J.-P.} \bibnamefont{Tillich}},
  \bibnamefont{and} \bibinfo{author}{\bibfnamefont{G.}~\bibnamefont{Zemor}},
  \emph{\bibinfo{title}{{Quantum Expander Codes}}}, \bibinfo{journal}{2015 IEEE
  56th Annual Symposium on Foundations of Computer Science (FOCS)} pp.
  \bibinfo{pages}{810--824} (\bibinfo{year}{2015}), \eprint{arXiv:1504.00822}.

\bibitem[{\citenamefont{Bolt et~al.}(2018)\citenamefont{Bolt, Poulin, and
  Stace}}]{Bolt18}
\bibinfo{author}{\bibfnamefont{A.}~\bibnamefont{Bolt}},
  \bibinfo{author}{\bibfnamefont{D.}~\bibnamefont{Poulin}}, \bibnamefont{and}
  \bibinfo{author}{\bibfnamefont{T.~M.} \bibnamefont{Stace}},
  \emph{\bibinfo{title}{Decoding schemes for foliated sparse quantum
  error-correcting codes}}, \bibinfo{journal}{Phys. Rev. A}
  \textbf{\bibinfo{volume}{98}}, \bibinfo{pages}{062302}
  (\bibinfo{year}{2018}).

\bibitem[{\citenamefont{Kovalev and Pryadko}(2013)}]{Kovalev13}
\bibinfo{author}{\bibfnamefont{A.~A.} \bibnamefont{Kovalev}} \bibnamefont{and}
  \bibinfo{author}{\bibfnamefont{L.~P.} \bibnamefont{Pryadko}},
  \emph{\bibinfo{title}{Fault tolerance of quantum low-density parity check
  codes with sublinear distance scaling}}, \bibinfo{journal}{Phys. Rev. A}
  \textbf{\bibinfo{volume}{87}}, \bibinfo{pages}{020304}
  (\bibinfo{year}{2013}).

\bibitem[{\citenamefont{Fawzi et~al.}(2018)\citenamefont{Fawzi, Grospellier,
  and Leverrier}}]{Fawzi18}
\bibinfo{author}{\bibfnamefont{O.}~\bibnamefont{Fawzi}},
  \bibinfo{author}{\bibfnamefont{A.}~\bibnamefont{Grospellier}},
  \bibnamefont{and}
  \bibinfo{author}{\bibfnamefont{A.}~\bibnamefont{Leverrier}},
  \emph{\bibinfo{title}{{Constant overhead quantum fault-tolerance with quantum
  expander codes}}} (\bibinfo{year}{2018}), \eprint{arXiv:1808.03821}.

\bibitem[{\citenamefont{Monroe et~al.}(2014)\citenamefont{Monroe, Raussendorf,
  Ruthven, Brown, Maunz, Duan, and Kim}}]{Monroe14}
\bibinfo{author}{\bibfnamefont{C.}~\bibnamefont{Monroe}},
  \bibinfo{author}{\bibfnamefont{R.}~\bibnamefont{Raussendorf}},
  \bibinfo{author}{\bibfnamefont{A.}~\bibnamefont{Ruthven}},
  \bibinfo{author}{\bibfnamefont{K.~R.} \bibnamefont{Brown}},
  \bibinfo{author}{\bibfnamefont{P.}~\bibnamefont{Maunz}},
  \bibinfo{author}{\bibfnamefont{L.-M.} \bibnamefont{Duan}}, \bibnamefont{and}
  \bibinfo{author}{\bibfnamefont{J.}~\bibnamefont{Kim}},
  \emph{\bibinfo{title}{Large-scale modular quantum-computer architecture with
  atomic memory and photonic interconnects}}, \bibinfo{journal}{Phys. Rev. A}
  \textbf{\bibinfo{volume}{89}}, \bibinfo{pages}{022317}
  (\bibinfo{year}{2014}).

\bibitem[{\citenamefont{Landsman et~al.}(2019)\citenamefont{Landsman, Wu,
  Leung, Zhu, Linke, Brown, Duan, and Monroe}}]{Landsman19}
\bibinfo{author}{\bibfnamefont{K.~A.} \bibnamefont{Landsman}},
  \bibinfo{author}{\bibfnamefont{Y.}~\bibnamefont{Wu}},
  \bibinfo{author}{\bibfnamefont{P.~H.} \bibnamefont{Leung}},
  \bibinfo{author}{\bibfnamefont{D.}~\bibnamefont{Zhu}},
  \bibinfo{author}{\bibfnamefont{N.~M.} \bibnamefont{Linke}},
  \bibinfo{author}{\bibfnamefont{K.~R.} \bibnamefont{Brown}},
  \bibinfo{author}{\bibfnamefont{L.}~\bibnamefont{Duan}}, \bibnamefont{and}
  \bibinfo{author}{\bibfnamefont{C.}~\bibnamefont{Monroe}},
  \emph{\bibinfo{title}{Two-qubit entangling gates within arbitrarily long
  chains of trapped ions}}, \bibinfo{journal}{Phys. Rev. A}
  \textbf{\bibinfo{volume}{100}}, \bibinfo{pages}{022332}
  (\bibinfo{year}{2019}).

\bibitem[{\citenamefont{Kimble}(2008)}]{Kimble08}
\bibinfo{author}{\bibfnamefont{H.~J.} \bibnamefont{Kimble}},
  \emph{\bibinfo{title}{{The quantum internet}}}, \bibinfo{journal}{Nature}
  \textbf{\bibinfo{volume}{453}}, \bibinfo{pages}{1023} (\bibinfo{year}{2008}).

\bibitem[{\citenamefont{Pastawski et~al.}(2015)\citenamefont{Pastawski,
  Yoshida, Harlow, and Preskill}}]{Pastawski15}
\bibinfo{author}{\bibfnamefont{F.}~\bibnamefont{Pastawski}},
  \bibinfo{author}{\bibfnamefont{B.}~\bibnamefont{Yoshida}},
  \bibinfo{author}{\bibfnamefont{D.}~\bibnamefont{Harlow}}, \bibnamefont{and}
  \bibinfo{author}{\bibfnamefont{J.}~\bibnamefont{Preskill}},
  \emph{\bibinfo{title}{{Holographic quantum error-correcting codes: toy models
  for the bulk/boundary correspondence}}}, \bibinfo{journal}{J. High Energ.
  Phys.} \textbf{\bibinfo{volume}{2015}}, \bibinfo{pages}{163}
  (\bibinfo{year}{2015}).

\bibitem[{\citenamefont{Harris et~al.}(2018)\citenamefont{Harris, McMahon,
  Brennen, and Stace}}]{Harris18}
\bibinfo{author}{\bibfnamefont{R.~J.} \bibnamefont{Harris}},
  \bibinfo{author}{\bibfnamefont{N.~A.} \bibnamefont{McMahon}},
  \bibinfo{author}{\bibfnamefont{G.~K.} \bibnamefont{Brennen}},
  \bibnamefont{and} \bibinfo{author}{\bibfnamefont{T.~M.} \bibnamefont{Stace}},
  \emph{\bibinfo{title}{Calderbank-shor-steane holographic quantum
  error-correcting codes}}, \bibinfo{journal}{Phys. Rev. A}
  \textbf{\bibinfo{volume}{98}}, \bibinfo{pages}{052301}
  (\bibinfo{year}{2018}).

\bibitem[{\citenamefont{Hayden and Preskill}(2007)}]{Hayden07}
\bibinfo{author}{\bibfnamefont{P.}~\bibnamefont{Hayden}} \bibnamefont{and}
  \bibinfo{author}{\bibfnamefont{J.}~\bibnamefont{Preskill}},
  \emph{\bibinfo{title}{{Black holes as mirrors: quantum information in random
  subsystems}}}, \bibinfo{journal}{J. High Energ. Phys.}
  \textbf{\bibinfo{volume}{2007}}, \bibinfo{pages}{120} (\bibinfo{year}{2007}),
  \eprint{arXiv:0708.4025}.

\bibitem[{\citenamefont{Lashkari et~al.}(2013)\citenamefont{Lashkari, Stanford,
  Hastings, Osborne, and Hayden}}]{Lashkari11}
\bibinfo{author}{\bibfnamefont{N.}~\bibnamefont{Lashkari}},
  \bibinfo{author}{\bibfnamefont{D.}~\bibnamefont{Stanford}},
  \bibinfo{author}{\bibfnamefont{M.}~\bibnamefont{Hastings}},
  \bibinfo{author}{\bibfnamefont{T.}~\bibnamefont{Osborne}}, \bibnamefont{and}
  \bibinfo{author}{\bibfnamefont{P.}~\bibnamefont{Hayden}},
  \emph{\bibinfo{title}{{Towards the fast scrambling conjecture}}},
  \bibinfo{journal}{J. High Energ. Phys.} \textbf{\bibinfo{volume}{2013}},
  \bibinfo{pages}{22} (\bibinfo{year}{2013}), \eprint{arXiv:1111.6580}.

\bibitem[{\citenamefont{Hosur et~al.}(2016)\citenamefont{Hosur, Qi, Roberts,
  and Yoshida}}]{Hosur16}
\bibinfo{author}{\bibfnamefont{P.}~\bibnamefont{Hosur}},
  \bibinfo{author}{\bibfnamefont{X.-L.} \bibnamefont{Qi}},
  \bibinfo{author}{\bibfnamefont{D.~A.} \bibnamefont{Roberts}},
  \bibnamefont{and} \bibinfo{author}{\bibfnamefont{B.}~\bibnamefont{Yoshida}},
  \emph{\bibinfo{title}{{Chaos in quantum channels}}}, \bibinfo{journal}{J.
  High Energ. Phys.} \textbf{\bibinfo{volume}{2016}}, \bibinfo{pages}{4}
  (\bibinfo{year}{2016}), \eprint{arXiv:1511.04021}.

\bibitem[{\citenamefont{Nahum et~al.}(2017)\citenamefont{Nahum, Ruhman, Vijay,
  and Haah}}]{Nahum16}
\bibinfo{author}{\bibfnamefont{A.}~\bibnamefont{Nahum}},
  \bibinfo{author}{\bibfnamefont{J.}~\bibnamefont{Ruhman}},
  \bibinfo{author}{\bibfnamefont{S.}~\bibnamefont{Vijay}}, \bibnamefont{and}
  \bibinfo{author}{\bibfnamefont{J.}~\bibnamefont{Haah}},
  \emph{\bibinfo{title}{Quantum entanglement growth under random unitary
  dynamics}}, \bibinfo{journal}{Phys. Rev. X} \textbf{\bibinfo{volume}{7}},
  \bibinfo{pages}{031016} (\bibinfo{year}{2017}).

\bibitem[{\citenamefont{Nahum et~al.}(2018{\natexlab{a}})\citenamefont{Nahum,
  Vijay, and Haah}}]{Nahum17}
\bibinfo{author}{\bibfnamefont{A.}~\bibnamefont{Nahum}},
  \bibinfo{author}{\bibfnamefont{S.}~\bibnamefont{Vijay}}, \bibnamefont{and}
  \bibinfo{author}{\bibfnamefont{J.}~\bibnamefont{Haah}},
  \emph{\bibinfo{title}{Operator spreading in random unitary circuits}},
  \bibinfo{journal}{Phys. Rev. X} \textbf{\bibinfo{volume}{8}},
  \bibinfo{pages}{021014} (\bibinfo{year}{2018}{\natexlab{a}}).

\bibitem[{\citenamefont{von Keyserlingk et~al.}(2018)\citenamefont{von
  Keyserlingk, Rakovszky, Pollmann, and Sondhi}}]{vonKeyserlingk17}
\bibinfo{author}{\bibfnamefont{C.~W.} \bibnamefont{von Keyserlingk}},
  \bibinfo{author}{\bibfnamefont{T.}~\bibnamefont{Rakovszky}},
  \bibinfo{author}{\bibfnamefont{F.}~\bibnamefont{Pollmann}}, \bibnamefont{and}
  \bibinfo{author}{\bibfnamefont{S.~L.} \bibnamefont{Sondhi}},
  \emph{\bibinfo{title}{Operator hydrodynamics, otocs, and entanglement growth
  in systems without conservation laws}}, \bibinfo{journal}{Phys. Rev. X}
  \textbf{\bibinfo{volume}{8}}, \bibinfo{pages}{021013} (\bibinfo{year}{2018}).

\bibitem[{\citenamefont{Nahum et~al.}(2018{\natexlab{b}})\citenamefont{Nahum,
  Ruhman, and Huse}}]{Nahum17b}
\bibinfo{author}{\bibfnamefont{A.}~\bibnamefont{Nahum}},
  \bibinfo{author}{\bibfnamefont{J.}~\bibnamefont{Ruhman}}, \bibnamefont{and}
  \bibinfo{author}{\bibfnamefont{D.~A.} \bibnamefont{Huse}},
  \emph{\bibinfo{title}{Dynamics of entanglement and transport in
  one-dimensional systems with quenched randomness}}, \bibinfo{journal}{Phys.
  Rev. B} \textbf{\bibinfo{volume}{98}}, \bibinfo{pages}{035118}
  (\bibinfo{year}{2018}{\natexlab{b}}).

\bibitem[{\citenamefont{Khemani et~al.}(2018)\citenamefont{Khemani, Vishwanath,
  and Huse}}]{Khemani17}
\bibinfo{author}{\bibfnamefont{V.}~\bibnamefont{Khemani}},
  \bibinfo{author}{\bibfnamefont{A.}~\bibnamefont{Vishwanath}},
  \bibnamefont{and} \bibinfo{author}{\bibfnamefont{D.~A.} \bibnamefont{Huse}},
  \emph{\bibinfo{title}{Operator spreading and the emergence of dissipative
  hydrodynamics under unitary evolution with conservation laws}},
  \bibinfo{journal}{Phys. Rev. X} \textbf{\bibinfo{volume}{8}},
  \bibinfo{pages}{031057} (\bibinfo{year}{2018}).

\bibitem[{\citenamefont{Rakovszky et~al.}(2018)\citenamefont{Rakovszky,
  Pollmann, and von Keyserlingk}}]{Tibor17}
\bibinfo{author}{\bibfnamefont{T.}~\bibnamefont{Rakovszky}},
  \bibinfo{author}{\bibfnamefont{F.}~\bibnamefont{Pollmann}}, \bibnamefont{and}
  \bibinfo{author}{\bibfnamefont{C.~W.} \bibnamefont{von Keyserlingk}},
  \emph{\bibinfo{title}{Diffusive hydrodynamics of out-of-time-ordered
  correlators with charge conservation}}, \bibinfo{journal}{Phys. Rev. X}
  \textbf{\bibinfo{volume}{8}}, \bibinfo{pages}{031058} (\bibinfo{year}{2018}).

\bibitem[{\citenamefont{Chan et~al.}(2018)\citenamefont{Chan, De~Luca, and
  Chalker}}]{Chan18}
\bibinfo{author}{\bibfnamefont{A.}~\bibnamefont{Chan}},
  \bibinfo{author}{\bibfnamefont{A.}~\bibnamefont{De~Luca}}, \bibnamefont{and}
  \bibinfo{author}{\bibfnamefont{J.~T.} \bibnamefont{Chalker}},
  \emph{\bibinfo{title}{Solution of a minimal model for many-body quantum
  chaos}}, \bibinfo{journal}{Phys. Rev. X} \textbf{\bibinfo{volume}{8}},
  \bibinfo{pages}{041019} (\bibinfo{year}{2018}).

\bibitem[{\citenamefont{Zhou and Nahum}(2019)}]{Zhou19}
\bibinfo{author}{\bibfnamefont{T.}~\bibnamefont{Zhou}} \bibnamefont{and}
  \bibinfo{author}{\bibfnamefont{A.}~\bibnamefont{Nahum}},
  \emph{\bibinfo{title}{Emergent statistical mechanics of entanglement in
  random unitary circuits}}, \bibinfo{journal}{Phys. Rev. B}
  \textbf{\bibinfo{volume}{99}}, \bibinfo{pages}{174205}
  (\bibinfo{year}{2019}).

\bibitem[{\citenamefont{Brandao and Horodecki}(2013)}]{Brandao13}
\bibinfo{author}{\bibfnamefont{F.~G. S.~L.} \bibnamefont{Brandao}}
  \bibnamefont{and}
  \bibinfo{author}{\bibfnamefont{M.}~\bibnamefont{Horodecki}},
  \emph{\bibinfo{title}{{Exponential Quantum Speed-ups are Generic}}},
  \bibinfo{journal}{Q. Inf. Comp.} \textbf{\bibinfo{volume}{13}},
  \bibinfo{pages}{0901} (\bibinfo{year}{2013}), \eprint{arXiv:1010.3654}.

\bibitem[{\citenamefont{Boixo et~al.}(2018)\citenamefont{Boixo, Isakov,
  Smelyanskiy, Babbush, Ding, Jiang, Bremner, Martinis, and Neven}}]{Boixo18}
\bibinfo{author}{\bibfnamefont{S.}~\bibnamefont{Boixo}},
  \bibinfo{author}{\bibfnamefont{S.~V.} \bibnamefont{Isakov}},
  \bibinfo{author}{\bibfnamefont{V.~N.} \bibnamefont{Smelyanskiy}},
  \bibinfo{author}{\bibfnamefont{R.}~\bibnamefont{Babbush}},
  \bibinfo{author}{\bibfnamefont{N.}~\bibnamefont{Ding}},
  \bibinfo{author}{\bibfnamefont{Z.}~\bibnamefont{Jiang}},
  \bibinfo{author}{\bibfnamefont{M.~J.} \bibnamefont{Bremner}},
  \bibinfo{author}{\bibfnamefont{J.~M.} \bibnamefont{Martinis}},
  \bibnamefont{and} \bibinfo{author}{\bibfnamefont{H.}~\bibnamefont{Neven}},
  \emph{\bibinfo{title}{{Characterizing quantum supremacy in near-term
  devices}}}, \bibinfo{journal}{Nature Phys.} \textbf{\bibinfo{volume}{14}},
  \bibinfo{pages}{595} (\bibinfo{year}{2018}).

\bibitem[{\citenamefont{Bouland et~al.}(2019)\citenamefont{Bouland, Fefferman,
  Nirkhe, and Vazirani}}]{Bouland18}
\bibinfo{author}{\bibfnamefont{A.}~\bibnamefont{Bouland}},
  \bibinfo{author}{\bibfnamefont{B.}~\bibnamefont{Fefferman}},
  \bibinfo{author}{\bibfnamefont{C.}~\bibnamefont{Nirkhe}}, \bibnamefont{and}
  \bibinfo{author}{\bibfnamefont{U.}~\bibnamefont{Vazirani}},
  \emph{\bibinfo{title}{{On the complexity and verification of quantum random
  circuit sampling}}}, \bibinfo{journal}{Nature Phys.}
  \textbf{\bibinfo{volume}{15}}, \bibinfo{pages}{159} (\bibinfo{year}{2019}).

\bibitem[{\citenamefont{Movassagh}(2019)}]{Movassagh19}
\bibinfo{author}{\bibfnamefont{R.}~\bibnamefont{Movassagh}},
  \emph{\bibinfo{title}{{Quantum supremacy and random circuits}}}
  (\bibinfo{year}{2019}), \eprint{arXiv:1909.06210}.

\bibitem[{\citenamefont{Deutsch}(1991)}]{Deutsch91}
\bibinfo{author}{\bibfnamefont{J.~M.} \bibnamefont{Deutsch}},
  \emph{\bibinfo{title}{Quantum statistical mechanics in a closed system}},
  \bibinfo{journal}{Phys. Rev. A} \textbf{\bibinfo{volume}{43}},
  \bibinfo{pages}{2046} (\bibinfo{year}{1991}).

\bibitem[{\citenamefont{Srednicki}(1993)}]{Srednicki93}
\bibinfo{author}{\bibfnamefont{M.}~\bibnamefont{Srednicki}},
  \emph{\bibinfo{title}{Entropy and area}}, \bibinfo{journal}{Phys. Rev. Lett.}
  \textbf{\bibinfo{volume}{71}}, \bibinfo{pages}{666} (\bibinfo{year}{1993}).

\bibitem[{\citenamefont{Nandkishore and Huse}(2015)}]{Nandkishore15}
\bibinfo{author}{\bibfnamefont{R.}~\bibnamefont{Nandkishore}} \bibnamefont{and}
  \bibinfo{author}{\bibfnamefont{D.~A.} \bibnamefont{Huse}},
  \emph{\bibinfo{title}{{Many-body localization and thermalization in quantum
  statistical mechanics}}}, \bibinfo{journal}{Annu. Rev. Condens. Matter Phys.}
  \textbf{\bibinfo{volume}{6}}, \bibinfo{pages}{15} (\bibinfo{year}{2015}).

\bibitem[{\citenamefont{D'Alessio et~al.}(2016)\citenamefont{D'Alessio, Kafri,
  Polkovnikov, and Rigol}}]{DAlessio16}
\bibinfo{author}{\bibfnamefont{L.}~\bibnamefont{D'Alessio}},
  \bibinfo{author}{\bibfnamefont{Y.}~\bibnamefont{Kafri}},
  \bibinfo{author}{\bibfnamefont{A.}~\bibnamefont{Polkovnikov}},
  \bibnamefont{and} \bibinfo{author}{\bibfnamefont{M.}~\bibnamefont{Rigol}},
  \emph{\bibinfo{title}{{From quantum chaos and eigenstate thermalization to
  statistical mechanics and thermodynamics}}}, \bibinfo{journal}{Adv. Phys.}
  \textbf{\bibinfo{volume}{65}}, \bibinfo{pages}{239} (\bibinfo{year}{2016}).

\bibitem[{\citenamefont{Li et~al.}(2018)\citenamefont{Li, Chen, and
  Fisher}}]{Li18}
\bibinfo{author}{\bibfnamefont{Y.}~\bibnamefont{Li}},
  \bibinfo{author}{\bibfnamefont{X.}~\bibnamefont{Chen}}, \bibnamefont{and}
  \bibinfo{author}{\bibfnamefont{M.~P.~A.} \bibnamefont{Fisher}},
  \emph{\bibinfo{title}{Quantum zeno effect and the many-body entanglement
  transition}}, \bibinfo{journal}{Phys. Rev. B} \textbf{\bibinfo{volume}{98}},
  \bibinfo{pages}{205136} (\bibinfo{year}{2018}).

\bibitem[{\citenamefont{Skinner et~al.}(2019)\citenamefont{Skinner, Ruhman, and
  Nahum}}]{Skinner18}
\bibinfo{author}{\bibfnamefont{B.}~\bibnamefont{Skinner}},
  \bibinfo{author}{\bibfnamefont{J.}~\bibnamefont{Ruhman}}, \bibnamefont{and}
  \bibinfo{author}{\bibfnamefont{A.}~\bibnamefont{Nahum}},
  \emph{\bibinfo{title}{Measurement-induced phase transitions in the dynamics
  of entanglement}}, \bibinfo{journal}{Phys. Rev. X}
  \textbf{\bibinfo{volume}{9}}, \bibinfo{pages}{031009} (\bibinfo{year}{2019}).

\bibitem[{\citenamefont{Li et~al.}(2019)\citenamefont{Li, Chen, and
  Fisher}}]{Li19}
\bibinfo{author}{\bibfnamefont{Y.}~\bibnamefont{Li}},
  \bibinfo{author}{\bibfnamefont{X.}~\bibnamefont{Chen}}, \bibnamefont{and}
  \bibinfo{author}{\bibfnamefont{M.~P.~A.} \bibnamefont{Fisher}},
  \emph{\bibinfo{title}{Measurement-driven entanglement transition in hybrid
  quantum circuits}}, \bibinfo{journal}{Phys. Rev. B}
  \textbf{\bibinfo{volume}{100}}, \bibinfo{pages}{134306}
  (\bibinfo{year}{2019}).

\bibitem[{\citenamefont{Chan et~al.}(2019)\citenamefont{Chan, Nandkishore,
  Pretko, and Smith}}]{Chan18b}
\bibinfo{author}{\bibfnamefont{A.}~\bibnamefont{Chan}},
  \bibinfo{author}{\bibfnamefont{R.~M.} \bibnamefont{Nandkishore}},
  \bibinfo{author}{\bibfnamefont{M.}~\bibnamefont{Pretko}}, \bibnamefont{and}
  \bibinfo{author}{\bibfnamefont{G.}~\bibnamefont{Smith}},
  \emph{\bibinfo{title}{Unitary-projective entanglement dynamics}},
  \bibinfo{journal}{Phys. Rev. B} \textbf{\bibinfo{volume}{99}},
  \bibinfo{pages}{224307} (\bibinfo{year}{2019}).

\bibitem[{\citenamefont{Gullans and Huse}(2020{\natexlab{a}})}]{Gullans19c}
\bibinfo{author}{\bibfnamefont{M.~J.} \bibnamefont{Gullans}} \bibnamefont{and}
  \bibinfo{author}{\bibfnamefont{D.~A.} \bibnamefont{Huse}},
  \emph{\bibinfo{title}{Dynamical purification phase transition induced by
  quantum measurements}}, \bibinfo{journal}{Phys. Rev. X}
  \textbf{\bibinfo{volume}{10}}, \bibinfo{pages}{041020}
  (\bibinfo{year}{2020}{\natexlab{a}}).

\bibitem[{\citenamefont{Choi et~al.}(2020)\citenamefont{Choi, Bao, Qi, and
  Altman}}]{Choi20}
\bibinfo{author}{\bibfnamefont{S.}~\bibnamefont{Choi}},
  \bibinfo{author}{\bibfnamefont{Y.}~\bibnamefont{Bao}},
  \bibinfo{author}{\bibfnamefont{X.-L.} \bibnamefont{Qi}}, \bibnamefont{and}
  \bibinfo{author}{\bibfnamefont{E.}~\bibnamefont{Altman}},
  \emph{\bibinfo{title}{Quantum error correction in scrambling dynamics and
  measurement-induced phase transition}}, \bibinfo{journal}{Phys. Rev. Lett.}
  \textbf{\bibinfo{volume}{125}}, \bibinfo{pages}{030505}
  (\bibinfo{year}{2020}).

\bibitem[{\citenamefont{Cao et~al.}(2019)\citenamefont{Cao, Tilloy, and
  Luca}}]{Cao19}
\bibinfo{author}{\bibfnamefont{X.}~\bibnamefont{Cao}},
  \bibinfo{author}{\bibfnamefont{A.}~\bibnamefont{Tilloy}}, \bibnamefont{and}
  \bibinfo{author}{\bibfnamefont{A.~D.} \bibnamefont{Luca}},
  \emph{\bibinfo{title}{{Entanglement in a fermion chain under continuous
  monitoring}}}, \bibinfo{journal}{SciPost Phys.} \textbf{\bibinfo{volume}{7}},
  \bibinfo{pages}{24} (\bibinfo{year}{2019}).

\bibitem[{\citenamefont{Szyniszewski et~al.}(2019)\citenamefont{Szyniszewski,
  Romito, and Schomerus}}]{Szyniszewski19}
\bibinfo{author}{\bibfnamefont{M.}~\bibnamefont{Szyniszewski}},
  \bibinfo{author}{\bibfnamefont{A.}~\bibnamefont{Romito}}, \bibnamefont{and}
  \bibinfo{author}{\bibfnamefont{H.}~\bibnamefont{Schomerus}},
  \emph{\bibinfo{title}{Entanglement transition from variable-strength weak
  measurements}}, \bibinfo{journal}{Phys. Rev. B}
  \textbf{\bibinfo{volume}{100}}, \bibinfo{pages}{064204}
  (\bibinfo{year}{2019}).

\bibitem[{\citenamefont{Bao et~al.}(2020)\citenamefont{Bao, Choi, and
  Altman}}]{Bao20}
\bibinfo{author}{\bibfnamefont{Y.}~\bibnamefont{Bao}},
  \bibinfo{author}{\bibfnamefont{S.}~\bibnamefont{Choi}}, \bibnamefont{and}
  \bibinfo{author}{\bibfnamefont{E.}~\bibnamefont{Altman}},
  \emph{\bibinfo{title}{Theory of the phase transition in random unitary
  circuits with measurements}}, \bibinfo{journal}{Phys. Rev. B}
  \textbf{\bibinfo{volume}{101}}, \bibinfo{pages}{104301}
  (\bibinfo{year}{2020}).

\bibitem[{\citenamefont{Jian et~al.}(2020)\citenamefont{Jian, You, Vasseur, and
  Ludwig}}]{Jian19}
\bibinfo{author}{\bibfnamefont{C.-M.} \bibnamefont{Jian}},
  \bibinfo{author}{\bibfnamefont{Y.-Z.} \bibnamefont{You}},
  \bibinfo{author}{\bibfnamefont{R.}~\bibnamefont{Vasseur}}, \bibnamefont{and}
  \bibinfo{author}{\bibfnamefont{A.~W.~W.} \bibnamefont{Ludwig}},
  \emph{\bibinfo{title}{Measurement-induced criticality in random quantum
  circuits}}, \bibinfo{journal}{Phys. Rev. B} \textbf{\bibinfo{volume}{101}},
  \bibinfo{pages}{104302} (\bibinfo{year}{2020}).

\bibitem[{\citenamefont{Gullans and Huse}(2020{\natexlab{b}})}]{Gullans19d}
\bibinfo{author}{\bibfnamefont{M.~J.} \bibnamefont{Gullans}} \bibnamefont{and}
  \bibinfo{author}{\bibfnamefont{D.~A.} \bibnamefont{Huse}},
  \emph{\bibinfo{title}{Scalable probes of measurement-induced criticality}},
  \bibinfo{journal}{Phys. Rev. Lett.} \textbf{\bibinfo{volume}{125}},
  \bibinfo{pages}{070606} (\bibinfo{year}{2020}{\natexlab{b}}).

\bibitem[{\citenamefont{Zabalo et~al.}(2020)\citenamefont{Zabalo, Gullans,
  Wilson, Gopalakrishnan, Huse, and Pixley}}]{Zabalo20}
\bibinfo{author}{\bibfnamefont{A.}~\bibnamefont{Zabalo}},
  \bibinfo{author}{\bibfnamefont{M.~J.} \bibnamefont{Gullans}},
  \bibinfo{author}{\bibfnamefont{J.~H.} \bibnamefont{Wilson}},
  \bibinfo{author}{\bibfnamefont{S.}~\bibnamefont{Gopalakrishnan}},
  \bibinfo{author}{\bibfnamefont{D.~A.} \bibnamefont{Huse}}, \bibnamefont{and}
  \bibinfo{author}{\bibfnamefont{J.~H.} \bibnamefont{Pixley}},
  \emph{\bibinfo{title}{Critical properties of the measurement-induced
  transition in random quantum circuits}}, \bibinfo{journal}{Phys. Rev. B}
  \textbf{\bibinfo{volume}{101}}, \bibinfo{pages}{060301}
  (\bibinfo{year}{2020}).

\bibitem[{\citenamefont{Zhang et~al.}(2020)\citenamefont{Zhang, Reyes, Kourtis,
  Chamon, Mucciolo, and Ruckenstein}}]{Zhang20}
\bibinfo{author}{\bibfnamefont{L.}~\bibnamefont{Zhang}},
  \bibinfo{author}{\bibfnamefont{J.~A.} \bibnamefont{Reyes}},
  \bibinfo{author}{\bibfnamefont{S.}~\bibnamefont{Kourtis}},
  \bibinfo{author}{\bibfnamefont{C.}~\bibnamefont{Chamon}},
  \bibinfo{author}{\bibfnamefont{E.~R.} \bibnamefont{Mucciolo}},
  \bibnamefont{and} \bibinfo{author}{\bibfnamefont{A.~E.}
  \bibnamefont{Ruckenstein}}, \emph{\bibinfo{title}{Nonuniversal entanglement
  level statistics in projection-driven quantum circuits}},
  \bibinfo{journal}{Phys. Rev. B} \textbf{\bibinfo{volume}{101}},
  \bibinfo{pages}{235104} (\bibinfo{year}{2020}).

\bibitem[{\citenamefont{Tang and Zhu}(2020)}]{Tang20}
\bibinfo{author}{\bibfnamefont{Q.}~\bibnamefont{Tang}} \bibnamefont{and}
  \bibinfo{author}{\bibfnamefont{W.}~\bibnamefont{Zhu}},
  \emph{\bibinfo{title}{Measurement-induced phase transition: A case study in
  the nonintegrable model by density-matrix renormalization group
  calculations}}, \bibinfo{journal}{Phys. Rev. Research}
  \textbf{\bibinfo{volume}{2}}, \bibinfo{pages}{013022} (\bibinfo{year}{2020}).

\bibitem[{\citenamefont{Li et~al.}(2020)\citenamefont{Li, Chen, Ludwig, and
  Fisher}}]{Li20}
\bibinfo{author}{\bibfnamefont{Y.}~\bibnamefont{Li}},
  \bibinfo{author}{\bibfnamefont{X.}~\bibnamefont{Chen}},
  \bibinfo{author}{\bibfnamefont{A.~W.~W.} \bibnamefont{Ludwig}},
  \bibnamefont{and} \bibinfo{author}{\bibfnamefont{M.~P.~A.}
  \bibnamefont{Fisher}}, \emph{\bibinfo{title}{{Conformal invariance and
  quantum non-locality in hybrid quantum circuits}}} (\bibinfo{year}{2020}),
  \eprint{arXiv:2003.12721}.

\bibitem[{\citenamefont{Fuji and Ashida}(2020)}]{Fuji20}
\bibinfo{author}{\bibfnamefont{Y.}~\bibnamefont{Fuji}} \bibnamefont{and}
  \bibinfo{author}{\bibfnamefont{Y.}~\bibnamefont{Ashida}},
  \emph{\bibinfo{title}{Measurement-induced quantum criticality under
  continuous monitoring}}, \bibinfo{journal}{Phys. Rev. B}
  \textbf{\bibinfo{volume}{102}}, \bibinfo{pages}{054302}
  (\bibinfo{year}{2020}).

\bibitem[{\citenamefont{Lavasani et~al.}(2020)\citenamefont{Lavasani, Alavirad,
  and Barkeshli}}]{Lavasani20}
\bibinfo{author}{\bibfnamefont{A.}~\bibnamefont{Lavasani}},
  \bibinfo{author}{\bibfnamefont{Y.}~\bibnamefont{Alavirad}}, \bibnamefont{and}
  \bibinfo{author}{\bibfnamefont{M.}~\bibnamefont{Barkeshli}},
  \emph{\bibinfo{title}{{Measurement-induced topological entanglement
  transitions in symmetric random quantum circuits}}} (\bibinfo{year}{2020}),
  \eprint{arXiv:2004.07243}.

\bibitem[{\citenamefont{Sang and Hsieh}(2020)}]{Sang20}
\bibinfo{author}{\bibfnamefont{S.}~\bibnamefont{Sang}} \bibnamefont{and}
  \bibinfo{author}{\bibfnamefont{T.~H.} \bibnamefont{Hsieh}},
  \emph{\bibinfo{title}{{Measurement Protected Quantum Phases}}}
  (\bibinfo{year}{2020}), \eprint{arXiv:2004.09509}.

\bibitem[{\citenamefont{Ippoliti et~al.}(2020)\citenamefont{Ippoliti, Gullans,
  Gopalakrishnan, Huse, and Khemani}}]{Ippoliti20}
\bibinfo{author}{\bibfnamefont{M.}~\bibnamefont{Ippoliti}},
  \bibinfo{author}{\bibfnamefont{M.~J.} \bibnamefont{Gullans}},
  \bibinfo{author}{\bibfnamefont{S.}~\bibnamefont{Gopalakrishnan}},
  \bibinfo{author}{\bibfnamefont{D.~A.} \bibnamefont{Huse}}, \bibnamefont{and}
  \bibinfo{author}{\bibfnamefont{V.}~\bibnamefont{Khemani}},
  \emph{\bibinfo{title}{{Entanglement phase transitions in measurement-only
  dynamics}}} (\bibinfo{year}{2020}), \eprint{arXiv:2004.09560}.

\bibitem[{\citenamefont{Alberton et~al.}(2020)\citenamefont{Alberton, Buchhold,
  and Diehl}}]{Alberton20}
\bibinfo{author}{\bibfnamefont{O.}~\bibnamefont{Alberton}},
  \bibinfo{author}{\bibfnamefont{M.}~\bibnamefont{Buchhold}}, \bibnamefont{and}
  \bibinfo{author}{\bibfnamefont{S.}~\bibnamefont{Diehl}},
  \emph{\bibinfo{title}{{Trajectory dependent entanglement transition in a free
  fermion chain -- from extended criticality to area law}}}
  (\bibinfo{year}{2020}), \eprint{arXiv:2005.09722}.

\bibitem[{\citenamefont{Lunt and Pal}(2020)}]{Lunt20}
\bibinfo{author}{\bibfnamefont{O.}~\bibnamefont{Lunt}} \bibnamefont{and}
  \bibinfo{author}{\bibfnamefont{A.}~\bibnamefont{Pal}},
  \emph{\bibinfo{title}{Measurement-induced entanglement transitions in
  many-body localized systems}}, \bibinfo{journal}{Phys. Rev. Research}
  \textbf{\bibinfo{volume}{2}}, \bibinfo{pages}{043072} (\bibinfo{year}{2020}).

\bibitem[{\citenamefont{Rossini and Vicari}(2020)}]{Rossini20}
\bibinfo{author}{\bibfnamefont{D.}~\bibnamefont{Rossini}} \bibnamefont{and}
  \bibinfo{author}{\bibfnamefont{E.}~\bibnamefont{Vicari}},
  \emph{\bibinfo{title}{Measurement-induced dynamics of many-body systems at
  quantum criticality}}, \bibinfo{journal}{Phys. Rev. B}
  \textbf{\bibinfo{volume}{102}}, \bibinfo{pages}{035119}
  (\bibinfo{year}{2020}).

\bibitem[{\citenamefont{Goto and Danshita}(2020)}]{Goto20}
\bibinfo{author}{\bibfnamefont{S.}~\bibnamefont{Goto}} \bibnamefont{and}
  \bibinfo{author}{\bibfnamefont{I.}~\bibnamefont{Danshita}},
  \emph{\bibinfo{title}{Measurement-induced transitions of the entanglement
  scaling law in ultracold gases with controllable dissipation}},
  \bibinfo{journal}{Phys. Rev. A} \textbf{\bibinfo{volume}{102}},
  \bibinfo{pages}{033316} (\bibinfo{year}{2020}).

\bibitem[{\citenamefont{Lopez-Piqueres
  et~al.}(2020)\citenamefont{Lopez-Piqueres, Ware, and
  Vasseur}}]{LopezPiqueres20}
\bibinfo{author}{\bibfnamefont{J.}~\bibnamefont{Lopez-Piqueres}},
  \bibinfo{author}{\bibfnamefont{B.}~\bibnamefont{Ware}}, \bibnamefont{and}
  \bibinfo{author}{\bibfnamefont{R.}~\bibnamefont{Vasseur}},
  \emph{\bibinfo{title}{Mean-field entanglement transitions in random tree
  tensor networks}}, \bibinfo{journal}{Phys. Rev. B}
  \textbf{\bibinfo{volume}{102}}, \bibinfo{pages}{064202}
  (\bibinfo{year}{2020}).

\bibitem[{\citenamefont{Shtanko et~al.}(2020)\citenamefont{Shtanko, Kharkov,
  Garc{\'\i}a-Pintos, and Gorshkov}}]{Shtanko20}
\bibinfo{author}{\bibfnamefont{O.}~\bibnamefont{Shtanko}},
  \bibinfo{author}{\bibfnamefont{Y.~A.} \bibnamefont{Kharkov}},
  \bibinfo{author}{\bibfnamefont{L.~P.} \bibnamefont{Garc{\'\i}a-Pintos}},
  \bibnamefont{and} \bibinfo{author}{\bibfnamefont{A.~V.}
  \bibnamefont{Gorshkov}}, \emph{\bibinfo{title}{{Classical Models of
  Entanglement in Monitored Random Circuits}}} (\bibinfo{year}{2020}),
  \eprint{arXiv:2004.06736}.

\bibitem[{\citenamefont{Vijay}(2020)}]{Vijay20}
\bibinfo{author}{\bibfnamefont{S.}~\bibnamefont{Vijay}},
  \emph{\bibinfo{title}{{Measurement-Driven Phase Transition within a
  Volume-Law Entangled Phase}}} (\bibinfo{year}{2020}),
  \eprint{arXiv:2005.03052}.

\bibitem[{\citenamefont{Lang and B\"uchler}(2020)}]{Lang20}
\bibinfo{author}{\bibfnamefont{N.}~\bibnamefont{Lang}} \bibnamefont{and}
  \bibinfo{author}{\bibfnamefont{H.~P.} \bibnamefont{B\"uchler}},
  \emph{\bibinfo{title}{Entanglement transition in the projective transverse
  field ising model}}, \bibinfo{journal}{Phys. Rev. B}
  \textbf{\bibinfo{volume}{102}}, \bibinfo{pages}{094204}
  (\bibinfo{year}{2020}).

\bibitem[{\citenamefont{Nahum et~al.}(2020)\citenamefont{Nahum, Roy, Skinner,
  and Ruhman}}]{Nahum20}
\bibinfo{author}{\bibfnamefont{A.}~\bibnamefont{Nahum}},
  \bibinfo{author}{\bibfnamefont{S.}~\bibnamefont{Roy}},
  \bibinfo{author}{\bibfnamefont{B.}~\bibnamefont{Skinner}}, \bibnamefont{and}
  \bibinfo{author}{\bibfnamefont{J.}~\bibnamefont{Ruhman}},
  \emph{\bibinfo{title}{{Measurement and entanglement phase transitions in
  all-to-all quantum circuits, on quantum trees, and in Landau-Ginsburg
  theory}}} (\bibinfo{year}{2020}), \eprint{arXiv:2009.11311}.

\bibitem[{\citenamefont{Fan et~al.}(2020)\citenamefont{Fan, Vijay, Vishwanath,
  and You}}]{Fan20}
\bibinfo{author}{\bibfnamefont{R.}~\bibnamefont{Fan}},
  \bibinfo{author}{\bibfnamefont{S.}~\bibnamefont{Vijay}},
  \bibinfo{author}{\bibfnamefont{A.}~\bibnamefont{Vishwanath}},
  \bibnamefont{and} \bibinfo{author}{\bibfnamefont{Y.-Z.} \bibnamefont{You}},
  \emph{\bibinfo{title}{{Self-Organized Error Correction in Random Unitary
  Circuits with Measurement}}} (\bibinfo{year}{2020}),
  \eprint{arXiv:2002.12385}.

\bibitem[{\citenamefont{Li and Fisher}(2020)}]{Li20b}
\bibinfo{author}{\bibfnamefont{Y.}~\bibnamefont{Li}} \bibnamefont{and}
  \bibinfo{author}{\bibfnamefont{M.~P.~A.} \bibnamefont{Fisher}},
  \emph{\bibinfo{title}{{Statistical Mechanics of Quantum Error-Correcting
  Codes}}} (\bibinfo{year}{2020}), \eprint{arXiv:2007.03822}.

\bibitem[{\citenamefont{Fidkowski et~al.}(2020)\citenamefont{Fidkowski, Haah,
  and Hastings}}]{Fidkowski20}
\bibinfo{author}{\bibfnamefont{L.}~\bibnamefont{Fidkowski}},
  \bibinfo{author}{\bibfnamefont{J.}~\bibnamefont{Haah}}, \bibnamefont{and}
  \bibinfo{author}{\bibfnamefont{M.~B.} \bibnamefont{Hastings}},
  \emph{\bibinfo{title}{{How Dynamical Quantum Memories Forget}}}
  (\bibinfo{year}{2020}), \eprint{arXiv:2008.10611}.

\bibitem[{\citenamefont{Napp et~al.}(2019)\citenamefont{Napp, La~Placa,
  Dalzell, Brandao, and Harrow}}]{Napp19}
\bibinfo{author}{\bibfnamefont{J.}~\bibnamefont{Napp}},
  \bibinfo{author}{\bibfnamefont{R.~L.} \bibnamefont{La~Placa}},
  \bibinfo{author}{\bibfnamefont{A.~M.} \bibnamefont{Dalzell}},
  \bibinfo{author}{\bibfnamefont{F.~G. S.~L.} \bibnamefont{Brandao}},
  \bibnamefont{and} \bibinfo{author}{\bibfnamefont{A.~W.}
  \bibnamefont{Harrow}}, \emph{\bibinfo{title}{{Efficient classical simulation
  of random shallow 2D quantum circuits}}} (\bibinfo{year}{2019}),
  \eprint{arXiv:2001.00021}.

\bibitem[{\citenamefont{Hayden et~al.}(2007)\citenamefont{Hayden, Horodecki,
  Winter, and Yard}}]{Hayden07b}
\bibinfo{author}{\bibfnamefont{P.}~\bibnamefont{Hayden}},
  \bibinfo{author}{\bibfnamefont{M.}~\bibnamefont{Horodecki}},
  \bibinfo{author}{\bibfnamefont{A.}~\bibnamefont{Winter}}, \bibnamefont{and}
  \bibinfo{author}{\bibfnamefont{J.}~\bibnamefont{Yard}},
  \emph{\bibinfo{title}{{A decoupling approach to the quantum capacity}}},
  \bibinfo{journal}{Open Syst. Inf. Dyn.} \textbf{\bibinfo{volume}{15}},
  \bibinfo{pages}{7} (\bibinfo{year}{2007}).

\bibitem[{\citenamefont{Chubb et~al.}(2017)\citenamefont{Chubb, Tan, and
  Tomamichel}}]{Chubb17}
\bibinfo{author}{\bibfnamefont{C.~T.} \bibnamefont{Chubb}},
  \bibinfo{author}{\bibfnamefont{V.~Y.~F.} \bibnamefont{Tan}},
  \bibnamefont{and}
  \bibinfo{author}{\bibfnamefont{M.}~\bibnamefont{Tomamichel}},
  \emph{\bibinfo{title}{{Moderate Deviation Analysis for Classical
  Communication over Quantum Channels}}}, \bibinfo{journal}{Commun. Math.
  Phys.} \textbf{\bibinfo{volume}{355}}, \bibinfo{pages}{1283}
  (\bibinfo{year}{2017}).

\bibitem[{\citenamefont{Tomamichel et~al.}(2016)\citenamefont{Tomamichel,
  Berta, and Renes}}]{Tomamichel16}
\bibinfo{author}{\bibfnamefont{M.}~\bibnamefont{Tomamichel}},
  \bibinfo{author}{\bibfnamefont{M.}~\bibnamefont{Berta}}, \bibnamefont{and}
  \bibinfo{author}{\bibfnamefont{J.~M.} \bibnamefont{Renes}},
  \emph{\bibinfo{title}{{Quantum coding with finite resources}}},
  \bibinfo{journal}{Nature Commun.} \textbf{\bibinfo{volume}{7}},
  \bibinfo{pages}{1} (\bibinfo{year}{2016}).

\bibitem[{\citenamefont{DiVincenzo et~al.}(1998)\citenamefont{DiVincenzo, Shor,
  and Smolin}}]{DiVincenzo98}
\bibinfo{author}{\bibfnamefont{D.~P.} \bibnamefont{DiVincenzo}},
  \bibinfo{author}{\bibfnamefont{P.~W.} \bibnamefont{Shor}}, \bibnamefont{and}
  \bibinfo{author}{\bibfnamefont{J.~A.} \bibnamefont{Smolin}},
  \emph{\bibinfo{title}{Quantum-channel capacity of very noisy channels}},
  \bibinfo{journal}{Phys. Rev. A} \textbf{\bibinfo{volume}{57}},
  \bibinfo{pages}{830} (\bibinfo{year}{1998}).

\bibitem[{\citenamefont{Smith and Smolin}(2007)}]{Smith07}
\bibinfo{author}{\bibfnamefont{G.}~\bibnamefont{Smith}} \bibnamefont{and}
  \bibinfo{author}{\bibfnamefont{J.~A.} \bibnamefont{Smolin}},
  \emph{\bibinfo{title}{Degenerate quantum codes for pauli channels}},
  \bibinfo{journal}{Phys. Rev. Lett.} \textbf{\bibinfo{volume}{98}},
  \bibinfo{pages}{030501} (\bibinfo{year}{2007}).

\bibitem[{\citenamefont{Bennett et~al.}(1997)\citenamefont{Bennett, DiVincenzo,
  and Smolin}}]{Bennett97}
\bibinfo{author}{\bibfnamefont{C.~H.} \bibnamefont{Bennett}},
  \bibinfo{author}{\bibfnamefont{D.~P.} \bibnamefont{DiVincenzo}},
  \bibnamefont{and} \bibinfo{author}{\bibfnamefont{J.~A.}
  \bibnamefont{Smolin}}, \emph{\bibinfo{title}{Capacities of quantum erasure
  channels}}, \bibinfo{journal}{Phys. Rev. Lett.}
  \textbf{\bibinfo{volume}{78}}, \bibinfo{pages}{3217} (\bibinfo{year}{1997}).

\bibitem[{\citenamefont{Harrow and Low}(2009)}]{Harrow09}
\bibinfo{author}{\bibfnamefont{A.~W.} \bibnamefont{Harrow}} \bibnamefont{and}
  \bibinfo{author}{\bibfnamefont{R.~A.} \bibnamefont{Low}},
  \emph{\bibinfo{title}{{Random Quantum Circuits are Approximate 2-designs}}},
  \bibinfo{journal}{Commun. Math. Phys.} \textbf{\bibinfo{volume}{291}},
  \bibinfo{pages}{257} (\bibinfo{year}{2009}).

\bibitem[{\citenamefont{Nielsen and Chuang}(2011)}]{NielsenChuang}
\bibinfo{author}{\bibfnamefont{M.~A.} \bibnamefont{Nielsen}} \bibnamefont{and}
  \bibinfo{author}{\bibfnamefont{I.~L.} \bibnamefont{Chuang}},
  \emph{\bibinfo{title}{Quantum Computation and Quantum Information}}
  (\bibinfo{publisher}{Cambridge University Press}, \bibinfo{address}{New York,
  NY, USA}, \bibinfo{year}{2011}), \bibinfo{edition}{10th} ed.

\bibitem[{\citenamefont{Schumacher}(1996)}]{Schumacher96b}
\bibinfo{author}{\bibfnamefont{B.}~\bibnamefont{Schumacher}},
  \emph{\bibinfo{title}{{Sending entanglement through noisy quantum
  channels}}}, \bibinfo{journal}{Phys. Rev. A} \textbf{\bibinfo{volume}{54}},
  \bibinfo{pages}{2614} (\bibinfo{year}{1996}).

\bibitem[{\citenamefont{Horodecki et~al.}(1999)\citenamefont{Horodecki,
  Horodecki, and Horodecki}}]{Horodecki99}
\bibinfo{author}{\bibfnamefont{M.}~\bibnamefont{Horodecki}},
  \bibinfo{author}{\bibfnamefont{P.}~\bibnamefont{Horodecki}},
  \bibnamefont{and}
  \bibinfo{author}{\bibfnamefont{R.}~\bibnamefont{Horodecki}},
  \emph{\bibinfo{title}{General teleportation channel, singlet fraction, and
  quasidistillation}}, \bibinfo{journal}{Phys. Rev. A}
  \textbf{\bibinfo{volume}{60}}, \bibinfo{pages}{1888} (\bibinfo{year}{1999}).

\bibitem[{\citenamefont{Nielsen}(2002)}]{Nielsen02}
\bibinfo{author}{\bibfnamefont{M.~A.} \bibnamefont{Nielsen}},
  \emph{\bibinfo{title}{{A simple formula for the average gate fidelity of a
  quantum dynamical operation}}}, \bibinfo{journal}{Phys. Lett. A}
  \textbf{\bibinfo{volume}{303}}, \bibinfo{pages}{249} (\bibinfo{year}{2002}).

\bibitem[{\citenamefont{Schumacher and Nielsen}(1996)}]{Schumacher96}
\bibinfo{author}{\bibfnamefont{B.}~\bibnamefont{Schumacher}} \bibnamefont{and}
  \bibinfo{author}{\bibfnamefont{M.~A.} \bibnamefont{Nielsen}},
  \emph{\bibinfo{title}{{Quantum data processing and error correction}}},
  \bibinfo{journal}{Phys. Rev. A} \textbf{\bibinfo{volume}{54}},
  \bibinfo{pages}{2629} (\bibinfo{year}{1996}).

\bibitem[{\citenamefont{Schumacher and Westmoreland}(2001)}]{Schumacher01}
\bibinfo{author}{\bibfnamefont{B.}~\bibnamefont{Schumacher}} \bibnamefont{and}
  \bibinfo{author}{\bibfnamefont{M.~D.} \bibnamefont{Westmoreland}},
  \emph{\bibinfo{title}{{Approximate quantum error correction}}}
  (\bibinfo{year}{2001}), \eprint{arXiv:quant-ph/0112106}.

\bibitem[{\citenamefont{Lloyd}(1997)}]{Lloyd97}
\bibinfo{author}{\bibfnamefont{S.}~\bibnamefont{Lloyd}},
  \emph{\bibinfo{title}{Capacity of the noisy quantum channel}},
  \bibinfo{journal}{Phys. Rev. A} \textbf{\bibinfo{volume}{55}},
  \bibinfo{pages}{1613} (\bibinfo{year}{1997}).

\bibitem[{\citenamefont{Devetak and Shor}(2005)}]{Devetak05}
\bibinfo{author}{\bibfnamefont{I.}~\bibnamefont{Devetak}} \bibnamefont{and}
  \bibinfo{author}{\bibfnamefont{P.~W.} \bibnamefont{Shor}},
  \emph{\bibinfo{title}{{The Capacity of a Quantum Channel for Simultaneous
  Transmission of Classical and Quantum Information}}},
  \bibinfo{journal}{Commun. Math. Phys.} \textbf{\bibinfo{volume}{256}},
  \bibinfo{pages}{287} (\bibinfo{year}{2005}).

\bibitem[{\citenamefont{Wilde}(2017)}]{Wildebook}
\bibinfo{author}{\bibfnamefont{M.~M.} \bibnamefont{Wilde}},
  \emph{\bibinfo{title}{Quantum Information Theory}}
  (\bibinfo{publisher}{Cambridge University Press},
  \bibinfo{address}{Cambridge, UK}, \bibinfo{year}{2017}).

\bibitem[{\citenamefont{Gottesman}(1998)}]{Gottesman98}
\bibinfo{author}{\bibfnamefont{D.}~\bibnamefont{Gottesman}},
  \emph{\bibinfo{title}{{The Heisenberg Representation of Quantum Computers}}}
  (\bibinfo{year}{1998}), \eprint{arXiv:quant-ph/9807006}.

\bibitem[{\citenamefont{Aaronson and Gottesman}(2004)}]{Aaronson04}
\bibinfo{author}{\bibfnamefont{S.}~\bibnamefont{Aaronson}} \bibnamefont{and}
  \bibinfo{author}{\bibfnamefont{D.}~\bibnamefont{Gottesman}},
  \emph{\bibinfo{title}{Improved simulation of stabilizer circuits}},
  \bibinfo{journal}{Phys. Rev. A} \textbf{\bibinfo{volume}{70}},
  \bibinfo{pages}{052328} (\bibinfo{year}{2004}).

\bibitem[{Pre()}]{Preskill}
\bibinfo{note}{See Ch.\ 7 of lecture notes by J. Preskill}.

\bibitem[{\citenamefont{Abdel-Ghaffar}(2012)}]{Abdel12}
\bibinfo{author}{\bibfnamefont{K.~A.~S.} \bibnamefont{Abdel-Ghaffar}},
  \emph{\bibinfo{title}{{Counting matrices over finite fields having a given
  number of rows of unit weight}}}, \bibinfo{journal}{Linear Algebra and its
  Applications} \textbf{\bibinfo{volume}{436}}, \bibinfo{pages}{2665}
  (\bibinfo{year}{2012}).

\bibitem[{\citenamefont{Dalzell et~al.}(2020)\citenamefont{Dalzell,
  Hunter-Jones, and Brandao}}]{Dalzell20}
\bibinfo{author}{\bibfnamefont{A.~M.} \bibnamefont{Dalzell}},
  \bibinfo{author}{\bibfnamefont{N.}~\bibnamefont{Hunter-Jones}},
  \bibnamefont{and} \bibinfo{author}{\bibfnamefont{F.~G. S.~L.}
  \bibnamefont{Brandao}}, \emph{\bibinfo{title}{{Random quantum circuits
  anti-concentrate in log depth}}} (\bibinfo{year}{2020}), \eprint{2011.12277}.

\bibitem[{\citenamefont{Fisher}(1984)}]{Fisher84}
\bibinfo{author}{\bibfnamefont{M.~E.} \bibnamefont{Fisher}},
  \emph{\bibinfo{title}{{Walks, walls, wetting, and melting}}},
  \bibinfo{journal}{J. Stat. Phys.} \textbf{\bibinfo{volume}{34}},
  \bibinfo{pages}{667} (\bibinfo{year}{1984}).

\bibitem[{\citenamefont{Raussendorf et~al.}(2005)\citenamefont{Raussendorf,
  Bravyi, and Harrington}}]{Raussendorf05}
\bibinfo{author}{\bibfnamefont{R.}~\bibnamefont{Raussendorf}},
  \bibinfo{author}{\bibfnamefont{S.}~\bibnamefont{Bravyi}}, \bibnamefont{and}
  \bibinfo{author}{\bibfnamefont{J.}~\bibnamefont{Harrington}},
  \emph{\bibinfo{title}{Long-range quantum entanglement in noisy cluster
  states}}, \bibinfo{journal}{Phys. Rev. A} \textbf{\bibinfo{volume}{71}},
  \bibinfo{pages}{062313} (\bibinfo{year}{2005}).

\bibitem[{\citenamefont{Han et~al.}(2007)\citenamefont{Han, Raussendorf, and
  Duan}}]{Han07}
\bibinfo{author}{\bibfnamefont{Y.-J.} \bibnamefont{Han}},
  \bibinfo{author}{\bibfnamefont{R.}~\bibnamefont{Raussendorf}},
  \bibnamefont{and} \bibinfo{author}{\bibfnamefont{L.-M.} \bibnamefont{Duan}},
  \emph{\bibinfo{title}{Scheme for demonstration of fractional statistics of
  anyons in an exactly solvable model}}, \bibinfo{journal}{Phys. Rev. Lett.}
  \textbf{\bibinfo{volume}{98}}, \bibinfo{pages}{150404}
  (\bibinfo{year}{2007}).

\end{thebibliography}

\end{document}